\documentclass[manuscript,screen,nonacm]{acmart} 

\renewcommand\footnotetextcopyrightpermission[1]{} 
\makeatletter
\renewcommand\@formatdoi[1]{\ignorespaces}
\makeatother

\acmConference[]{}{}{}
\acmBooktitle{}
\acmPrice{}
\acmISBN{}

\usepackage{subcaption}
\usepackage{rotating, graphicx}
\usepackage{booktabs}
\usepackage{tabularx}
\usepackage[normalem]{ulem}
\useunder{\uline}{\ul}{}
\usepackage[export]{adjustbox}

\usepackage[ruled,vlined]{algorithm2e}

\usepackage{url}

\usepackage{pifont}

\usepackage{url}
\urldef{\urlCaseStudyDocs}\url{https://docs.oracle.com/javase/7/docs/api/java/io/Reader.html#read(char[],%20int,%20int)} 

\newcommand{\etal}{\textit{et al. }}

\newcommand{\revise}[1]{\textcolor{black}{#1}}  

\newcommand{\RQ}[1]{\textbf{RQ#1}}

\usepackage{mdframed}

\newmdenv[innerlinewidth=0.5pt, roundcorner=4pt,linecolor=gray,innerleftmargin=4pt,
innerrightmargin=4pt,innertopmargin=4pt,innerbottommargin=4pt]{note}

\newenvironment{result}%
{\medskip\begin{note}\centering\em}%
{\end{note}\medskip}

\begin{document}

\title{Mutation Testing in Evolving Systems: Studying the relevance of mutants to code evolution}

\author{Milos Ojdanic}
\email{milos.ojdanic@uni.lu}
\affiliation{
  \institution{SnT, University of Luxembourg}
  \country{Luxembourg}
}
\author{Ezekiel Soremekun}
\email{ezekiel.soremekun@uni.lu}
\affiliation{
  \institution{SnT, University of Luxembourg}
  \country{Luxembourg}
}
\author{Renzo Degiovanni}
\email{renzo.degiovanni@uni.lu}
\affiliation{
  \institution{SnT, University of Luxembourg}
  \country{Luxembourg}
}
\author{Mike Papadakis}
\email{michail.papadakis@uni.lu}
\affiliation{
  \institution{SnT, University of Luxembourg}
  \country{Luxembourg}
}
\author{Yves Le Traon}
\email{yves.letraon@uni.lu}
\affiliation{%
  \institution{SnT, University of Luxembourg}
  \country{Luxembourg}
}

\renewcommand{\shortauthors}{Ojdanić, et al.}


\begin{abstract}
\textbf{Context:}
When software evolves, opportunities for introducing faults appear. Therefore, it is important to test the evolved program behaviors during each evolution cycle. However, while software evolves, its complexity is also evolving, introducing challenges to the testing process. To deal with this issue, testing techniques should be adapted to target the effect of the program changes instead of the entire program functionality. To this end, \textit{commit-aware mutation testing}, a powerful testing technique, has been proposed. Unfortunately, commit-aware mutation testing is challenging due to the complex program semantics involved. Hence, it is pertinent to understand the characteristics, predictability, and potential of the technique.

\textbf{Objective:}
We conduct an exploratory study to investigate the properties of \textit{commit-relevant mutants}, i.e., the test elements of commit-aware mutation testing, by proposing a general definition and an experimental approach to identify them. We thus, aim at investigating the prevalence, location, and comparative advantages of commit-aware mutation testing over time (i.e., the program evolution).  We also investigate the predictive power of several commit-related features in identifying and selecting commit-relevant mutants to understand the essential properties for its best-effort application case.

\textbf{Method:}
Our commit-relevant definition relies on the notion of observational slicing, approximated by higher-order mutation. Specifically, our approach utilizes the impact of mutants, effects of one mutant on another in capturing and analyzing the implicit interactions between the changed and unchanged code parts. The study analyses millions of mutants (over 10 million), 288 commits, five (5) different open-source software projects involving over 68,213 CPU days of computation and sets a ground truth where we perform our analysis. 

\textbf{Results:}
Our analysis shows that commit-relevant mutants are \textit{located mainly outside of program commit change} (81\%), suggesting a limitation in previous work. We also note that effective selection of commit-relevant mutants has the potential of reducing \revise{the number of mutants} by up to 93\%. In addition, we demonstrate that commit relevant mutation testing is significantly more effective and efficient than state-of-the-art baselines, i.e., random mutant selection and 
analysis of only mutants within the program change. In our analysis of the predictive power of mutants and commit-related features (e.g., number of mutants within a change, mutant type, and commit size) in predicting commit-relevant mutants, we found that most \textit{proxy features do not reliably predict commit-relevant mutants}. 

\textbf{Conclusion:} 
This empirical study highlights the properties of commit-relevant mutants and demonstrates the importance of identifying and selecting commit-relevant mutants when testing evolving software systems. 

\end{abstract}


\maketitle

%
%
%

\section{Introduction}\label{introduction}


Software systems evolve and are typically developed through program evolution cycles that involve frequent code modifications  \cite{handsOnDevOps}. Therefore, when software evolves, the program modifications need to be tested to avoid introducing faults and ensure the expected program behavior. To test an evolving program, developers need to perform 
\textit{regression testing}, i.e., assessing the impact of the change on the program by generating additional test cases targeting the change and its dependencies~\cite{YooH12}. Typically, developers have to write or generate test cases that exercise the changes, stress their dependencies, and check that the program changes behave as intended \cite{ApiwattanapongSCOH06}.

Mutation testing is an established software testing technique~\cite{PapadakisK00TH19}. It is typically applied to reveal faults in a program by modifying the program (aka injecting mutants) and generating tests to reveal the faults (i.e., kill the mutants) in the modified program. Mutation testing is an effective approach to improve the test suite's \revise{strengths} by ensuring that it is adequate and diverse enough to kill all injected mutants. In the last decade, mutation testing has focused on selecting or reducing the number of executed mutants to ensure that mutation testing is feasible and scales in practice. To this end, researchers have proposed mutation testing with a specific type of mutants \cite{offut1993}, mutant reduction by detecting equivalent mutants \cite{KintisPJMTH18, ChekamPBTS20} or by focusing on a particular category of mutants \revise{such as subsuming mutants\footnote{Subsuming mutants \cite{jia2009} or disjoint mutants \cite{KintisPM10} is a set of mutants that has no mutant that is killed by a proper subset of tests that kill another mutant.}} or hard-to-kill mutants \cite{PapadakisCT18, KurtzAODKG16, KintisPM10}. 

Traditional mutation testing involves injecting mutants into the entire code base of the software. 
However, mutation testing of evolving programs is challenging due to the \textit{scale of the required mutation analysis, the complexity of the program, and the difficulty of determining the impact of the dependencies of the program changes}. The sub-field of mutation testing addressing these issues by targeting the mutation testing program changes is referred to as \textit{commit-aware mutation testing} \cite{ma2020commit}.

A few commit-aware mutation testing approaches have been proposed to tackle the challenges of mutation testing of evolving software systems~\cite{petrovic2018, ma2020commit, ma2021mudelta, CACHIA2013}. These approaches suggest that mutation testing of evolving systems should focus on the program changes rather than the entire program. Recent studies have also indicated that commit-relevant mutants can be found on unchanged code due to unforeseen interactions between changed and unchanged code~\cite{ma2020commit, ma2021mudelta}. However, these studies do not provide scientific insights into the nature and properties of commit-relevant mutants and their utility over time. For instance, it is necessary to understand the distribution and program location of commit-relevant mutants to effectively identify, select, or predict commit-relevant mutants.   

In this paper, we address this challenge by conducting an exploratory empirical study to investigate the properties of commit-relevant mutants. 
Specifically, we examined the distribution, location, prevalence, predictability, and utility of commit-relevant mutants, as well as subsuming commit-relevant mutants\footnote{\revise{Subsuming commit-relevant mutants is a set of commit-relevant mutants that has no commit-relevant mutant killed by a proper subset of tests that kill another commit-relevant mutant}}. To achieve this, we propose an experimental approach for identifying commit-relevant mutants using the notion of observational slicing \cite{BinkleyGHIKY14}, i.e., the relevance of an instruction to a program point of interest (such as a program state or variable(s)) can be determined by mutating instructions and observing their impact to the point of interest (changes on the target program state or variable). Since we aim to identify mutants relevant to changed instructions, we check the impact of mutants located on the changed code, as performed by observational slicing, on mutants located on unchanged code. In essence, with this approach, we measure the impact of second-order mutants on the first-order ones \cite{KintisPM12, KintisPM15}, which captures the existence of implicit interactions between the changed and unchanged code parts. 

Overall, our formulation of the commit-aware mutation testing addresses the limitations and challenges of the state of the art~\cite{ma2020commit, ma2021mudelta}, in particular, making it more general and applicable for most evolving systems (\textit{see Section \ref{exp-goals}}). Using this approach, we elicit the properties of commit-relevant mutants and study the advantage of commit-relevant mutant selection in comparison to random mutant selection or mutants located on program changes.

To the best of our knowledge, this is the most extensive empirical study of commit-relevant mutants. Specifically, our evaluation setup contains 10,071,875 mutants and 288 commits extracted from five (5) mature open-source software repositories. Our experiments took over 68,213 CPU days of computation. The main objective of this work is to provide scientific insights concerning the application of mutation analysis in testing evolving software systems. The main findings of this paper are summarized as follows:
 \begin{itemize}
 \item \textit{Commit-relevant mutants are prevalent}. In our evaluation, \textit{30\%} of mutants are commit-relevant, on average. Hence, \revise{by reducing the number of mutants (by around 70\%) and concentrating merely on those representing change-aware test requirements, considerable cost reductions can be achieved}.

 \item \textit{Selecting subsuming commit-relevant mutants significantly reduces \revise{the number of mutants}}. \revise{Selection of \textit{subsuming commit-relevant mutants} reduces even further the number of mutants, by about 93\%, on average.} 
 
\item \textit{A large proportion of commit-relevant mutants are located outside of the program changes}. The majority of the commit-relevant mutants are located outside the changed methods (\textit{69\%}). 

\item \textit{Several evaluated commit or mutant related features can not reliably predict (subsuming) commit-relevant mutants}. For instance, (the number of) commit-relevant mutants cannot be reliably predicted by features such as the commit size or mutant operator types.

\item \textit{State of the art mutant selection approaches miss a large portion of commit-relevant mutants.}  As an example, random mutant selection techniques miss approximately 45\% of subsuming commit-relevant mutants when analyzing the scope of 20 mutants.

\item \textit{Commit-relevant mutation testing significantly reduces the text executions in comparison to the state of the art mutant selection methods}. \revise{Specifically, commit-relevant mutation testing reduces the number of test executions by about 16 times compared to random mutant selection.}
\end{itemize}

\section{Background}\label{mutation_testing}


\subsection{Mutation Testing} 
Mutation is a test adequacy criterion in which test requirements are characterized by mean of \emph{mutants} obtained by performing slight syntactic modifications to the original program (for instance, the relational expression  $\texttt{a > b}$ can be mutated into $\texttt{a < b}$). 
Intuitively, these mutants aim at representing artificially injected faults that can be used to assess the effectiveness and thoroughness of a test suite in detecting these seeded faults. 
Then, the tester starts by analyzing the mutants and proceeds to design test cases to \emph{kill} them, i.e., to distinguish the observable behavior between the mutant and the original program. 
Hence, the adequacy of a test suite concerning the mutation criterion, called \emph{mutation score}, is computed as the ratio of killed mutants over the total number of mutants. 

Notice that the number of mutants not necessarily represent the number of test cases required to cover all of them since several mutants can be redundant. 
On the one hand, there may exist mutants that cannot be killed by any test since they are functionally \emph{equivalent} to the original program. 
On the other hand, one test may kill other mutants at the same time. 
Thus, the effort put into analyzing and executing redundant mutants is wasted; hence it is desirable to analyze only the mutants that add value. 


\subsection{Subsuming Mutants}
\label{sec:subsuming_mutants}

Subsuming relations aims at finding the minimal set of mutants required to cover all (killable) mutants~\cite{ammann_establishing_2014}. Intuitively, this set of mutants has minimal redundancies and represents a nearly optimal mutation testing process with respect to cost \cite{PapadakisHHJT16, PapadakisCT18}. More formally, let us consider that $M_1$, $M_2$, and $T$ be two mutants and a test suite, respectively. 
Consider also that $T_1 \subseteq T$ and $T_2 \subseteq T$ are the set of tests from $T$ that kill mutants $M_1$ and $M_2$, respectively, where $T_1 \neq \emptyset$ and $T_2 \neq \emptyset$ indicating that both $M_1$ and $M_2$ are killable mutants. 
We say that mutant $M_1$ subsumes mutant $M_2$, if and only if,  $T_1 \subseteq T_2$.
In case $T_1=T_2$, we say that mutants $M_1$ and $M_2$ are indistinguishable. 
The set of mutants that are both killable and subsumed only by indistinguishable mutants are called \emph{subsuming mutants}. 
For instance, assuming that $T_1=\{t_1,t_2\}$ and $T_2=\{t_1,t_2,t_3\}$, one can notice that every time we run a test to kill mutant $M_1$ (i.e., $t_1$ or $t_2$) we will also kill mutant $M_2$. 
While the vice versa does not hold since if we kill mutant $M_2$ by $t_3$, we will not kill mutant $M_1$. 
In this case we say that $M_1$ subsumes $M_2$.

\revise{Several researchers have studied the impact and prevalence of subsuming mutants for traditional mutation testing~\cite{PapadakisHHJT16, PapadakisCT18, alipour2016evaluating, guimaraes2020optimizing}. 
For instance, \citet{alipour2016evaluating} demonstrated that subsuming mutants can reduce traditional mutation testing effort by up to 80\%. In particular, in their empirical study on mutation test reduction, found that subsuming mutants can reduce the number of mutants requiring analysis by up to 80\%. Their study demonstrated the importance of subsuming mutants in traditional mutation testing, emphasizing that there is strong inter-dependency among mutants. In their empirical evaluation of \textit{traditional mutation testing} (involving four C projects and thousands of mutants), the paper found that test case reduction based on a single mutant can reduce mutation testing effort (in terms of the number of mutant test executions) by 33 to 80\%~\cite{alipour2016evaluating}.
Likewise, \citet{guimaraes2020optimizing} empirically demonstrated that identifying dynamic subsumption relations among mutants reduces traditional mutation test execution time by 53\%. \citet{delamaro2001interface} also demonstrated that identifying the inter-procedural relation among mutants in two program units helps to identify interface mutants, i.e., the mutants that are 
relevant for 
mutation testing during system integration. 
} 
\revise{However, despite the evidence of the impact of subsuming mutants on traditional mutation testing and integration testing, their impact 
on commit-aware mutation testing remains unknown. Thus, \textit{in this paper, we study the prevalence and distribution of mutants relevant for a committed change and the extent to which subsuming relations are maintained.} }


\subsection{High-Order Mutants} 

Depending on the number of mutation operators we apply to the original program, we can categorize the obtained mutants by the number of simple changes one has to introduce to form them.  
That is, \emph{first-order mutants} (FOM) is obtained by making only one simple syntactic change to the original program. 
Second-order mutants (SOM) are obtained by making two syntactic changes to the original program (or applying one mutation to first-order mutants). In the general case, higher-order mutants (HOM)~\cite{jia2009} are produced after the successful application of \textit{n} mutations to the original program.  

At the very beginning, using higher-order mutants in mutation testing was not considered viable because of the \emph{Coupling Effect} proposed by DeMillo et al. \cite{demillo1978}. It stated that ``Test data that distinguishes all programs differing from a correct one by only simple errors is so sensitive that it also implicitly distinguishes more complex errors''. 
However, later on, \citet{offut1992} defined first-order mutants as simple faults while characterizing higher-order mutants as complex artificial defects. 

In this study, we plan to use second-order mutants as the means for studying whether mutants located outside the commit change interacts with the mutants located within the change. Then we use this information to determine if mutants are relevant or not for given commit changes. Details are presented later in Section~\ref{sec:approach}.

\begin{figure}[bt]
    \centering
    \includegraphics[scale=0.30,trim={0cm 3.5cm 0cm 3.5cm},clip]{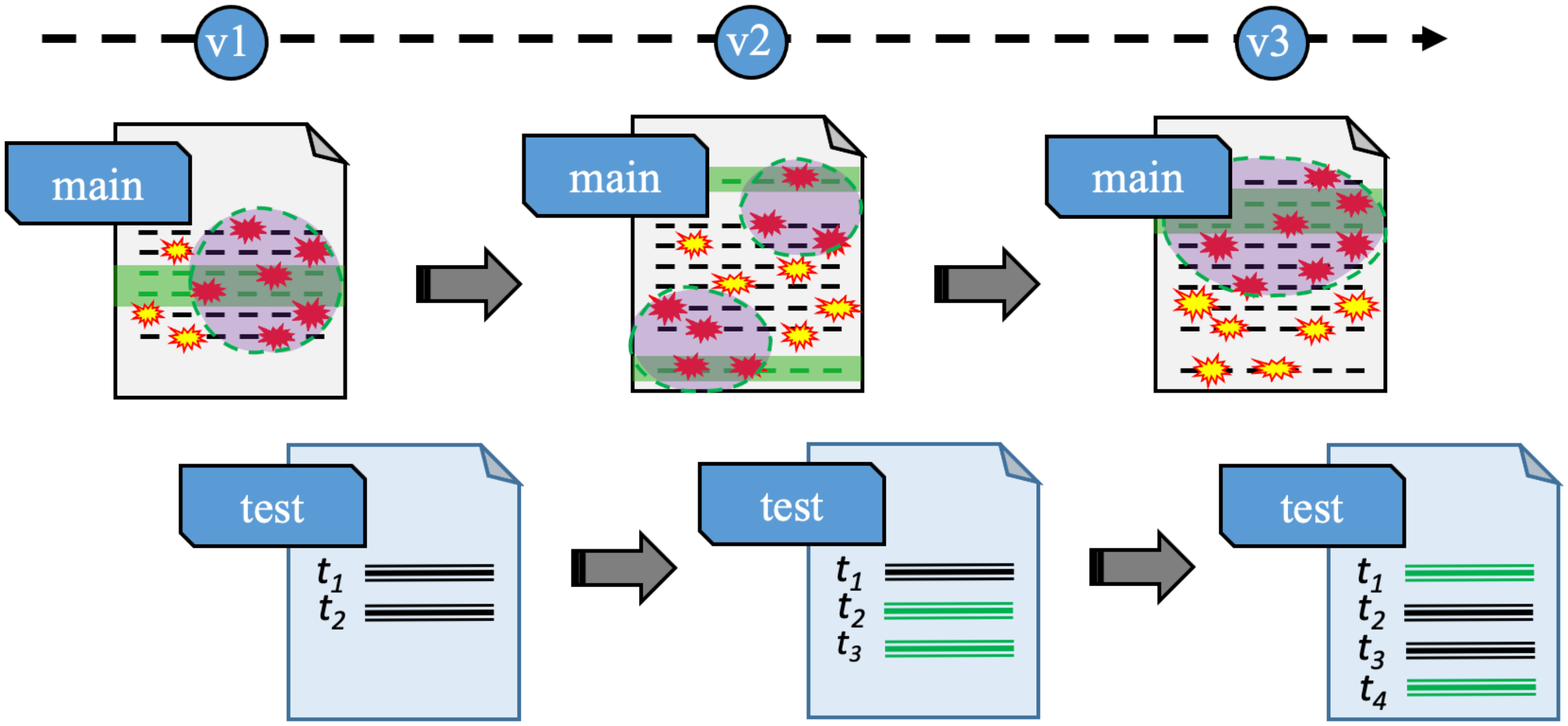}
    \caption{\revise{Typical  evolution of a software and its test suite showing three versions ($v_1$ to $v_3$) of a program (\texttt{main}) and and its test suite (\texttt{test}). The {\color{green} green} portions of the program (\texttt{main}) symbolize the \textit{program changes} (e.g., a commit), and the explosions symbolize the \textit{mutants} injected into the program. 
In the test suite (\texttt{test}), $t_i$ symbolizes a \textit{test case} $i$, and the {\color{green} green} rectangles represent changes in the test suite (i.e., addition and modification of tests). The test suite and source code evolve as the program evolves through versions. As the size of the program increases, we can observe that the number of mutants increases as well. This eventually leads to a substantial number of irrelevant mutants that result in waste of efforts.  In the figure, with {\color{red} red} the mutants that are commit-relevant and with {\color{yellow} yellow} the irrelevant ones. Focusing only on commit-relevant mutants reduces the number of mutants requiring attention and leads to significant cost reductions. Additionally, the set of commit-relevant mutants quantifies the extent to which practitioners have tested the program behaviors affected by the change. } }
    \label{fig:testrig}
\end{figure}

\subsection{Testing Evolving Systems}

Software systems evolve frequently, hence, it is pertinent to provide methods and tools to analyze the impact of the program changes. 
\emph{Regression testing} helps in this respect by re-running the test suite on the new version of the code to ensure that the previously developed functionality behaves as expected. 
Software evolves for many reasons (e.g., due to bug fixes, code refactoring or new features). 
Therefore it is important to understand how to 
test the program change, if it is enough to test only the changed lines, as well as how many test requirements and test cases will need to be analyzed. \revise{Notably, developers are burdened with the challenge of testing evolving systems, specifically, 
how to effectively analyse 
the difference in the program behaviors induced by their changes. These are the main challenges of regression testing, and in this work, we aim to study these challenges via the lens of mutation testing.} 


\revise{Generally, as the software evolves, the test suite also evolves. Concretely, as the program changes (e.g., due to new features or bug fixes), new tests are added or old tests may be modified to exercise those changes. 
Figure \ref{fig:testrig} illustrates the evolution of a program and its test suite during a typical software development process,  
showing changes in four versions of the program (\texttt{main}) and the test suite (\texttt{test}).
In this example, we illustrate that analyzing all mutants is costly, as the number of mutants (both red and yellow in \autoref{fig:testrig}) is independent of the program changes (is actually depended on the size of the programs)  and increases as the program size increases. 
Hence, traditional mutation testing will be costly, since it uses more mutants than required. More importantly, by doing so, developers will have to analyze mutants that are not relevant to what they actually committed. In this work, we study how to address these challenges using commit-aware mutation testing. 
}

A common approach to address these challenges is to leverage code coverage information, i.e., analyzing the test coverage of a particular change, to decide if the change needs further test cases or not. 
However, previous works~\cite{Fowler,KurtzAODKG16} have shown that many severe integration issues arise from unforeseen interactions triggered between introduced change and the rest of the software.  
\revise{Therefore, there is a need for change-aware test metrics to guide effective regression testing 
and allow developers to quantify the extent to which they 
tested the error-prone program behaviors affected by their changes.} 
We plan to use mutation testing to address these interactions by targeting suitable mutants that demonstrate an (implicit) interaction between the changed lines and the unmodified part of the program (i.e., the code outside the change). 
\revise{These mutants form the change-relevant requirements and should be used to determine whether test suites are adequate and provide guidance in improving the test suite.}



\section{Commit-aware Mutation Testing} 
\label{sec:approach}





\subsection{Definition}
\revise{
Intuitively, \textit{commit-relevant} mutants are those that are linked with (capture) changed program behaviour, by the committed changes. 
    }
\revise{
These mutants are those that a) are killable and are located on the changed lines, because they capture behaviour relevant to the committed changes, and b) those that are killable, are located on unchanged lines and affect the changed, by the commit, program behaviour, because they capture the interaction of the changed and unchanged code. This is approximated by a special form of observational slicing  that uses higher order mutants. The idea is that mutants, located on unmodified code, that impact the behavior of mutants located on modified code, are commit-relevant because they interact/depend with the changed code. Consider two first-order mutants $M_X$ and $M_Y$, such that $M_X$ is located on changed code and $M_Y$ is located within the changed code. Then, the higher order mutant ($M_{XY}$) is the one created by combining $M_X$ and $M_Y$. We say that $M_X$ is \textit{commit-relevant } if the higher order mutant ($M_{XY}$) has a different program behaviour from the first order mutants $M_X$ and $M_Y$. That is, $M_X$ is commit-relevant if ($M_{XY}$ != $M_{Y}$) and ($M_{XY}$ != $M_{X}$). 
%
Formally, the definition of \textit{commit-relevant mutant} can be formed as: 
}

\begin{definition}[Commit-relevant mutants]\label{def:relevant-mutant}
\revise{
    A mutant \textit{$M_X$} is relevant to a commit-change, if a) it is killable and is located on the changed code, or b) there is a second order mutant \textit{$M_{XY}$} (formed by the mutant pair of  \textit{$M_X$}, located outside the change, and \textit{$M_Y$}, located on the change) that has different behaviour from the two first-order mutants \textit{$M_X$} and \textit{$M_Y$} that it is composed of. 
    }
\end{definition}

\begin{figure*}[bt]
    \begin{center}
    \includegraphics[width=1.0\textwidth,trim={0cm 4cm 0cm 4cm},clip]{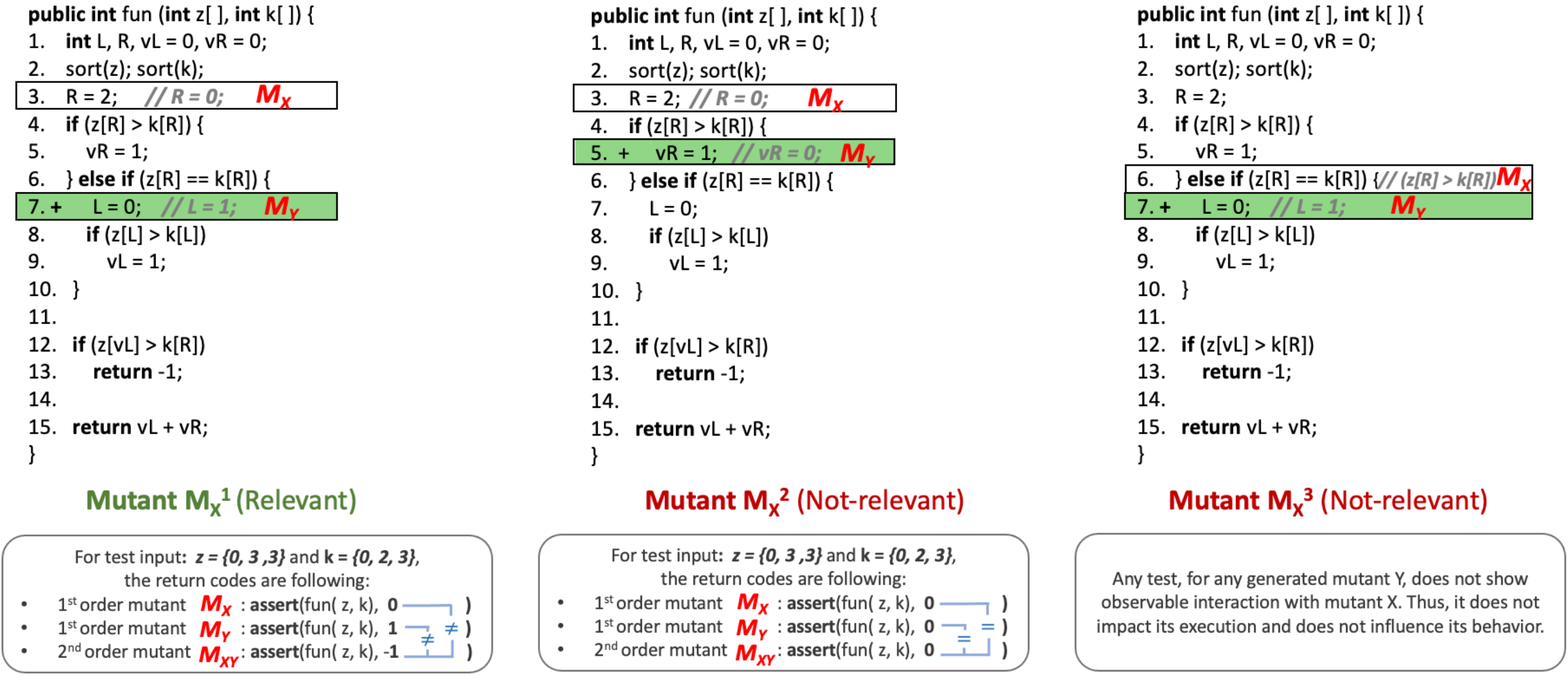}
    \caption{Example of relevant and not-relevant mutants. Left Sub-Figure: Mutant $M_X^1$ is relevant as mutant \textit{$M_Y$} impacts its behavior. Center Sub-Figure: Mutant $M_X^2$ is non-relevant as mutant \textit{$M_Y$} does not impacts its behavior. Right Sub-Figure: mutant $M_X^3$ is not relevant since there is no behavioral difference for every possible $M_Y$.}
    \label{fig:Demonstrating_example}
    \end{center}
\end{figure*}



\subsection{\revise{Motivating Examples}}

\subsubsection*{\textbf{\revise{Simple Example}}}
\revise{\autoref{fig:Demonstrating_example} describes three simple scenarios illustrating commit-relevant mutants on a toy code example.} 
In the code snippet on the left, we observe the example function \textit{fun} that takes two arguments (integer arrays of size \textit{3}).
It starts by sorting the arrays' elements, then makes computations, and returns an integer as a result. 

The {\color{green}green} rectangle on line seven (\textit{7}) represents the line that has been modified in the code.
Using Java comments (symbols ``\textit{//}'') on line three (\textit{3}) we represent mutant outside the change $M_X$, and the mutant on the change $M_Y$ on line seven (\textit{7}). 
Mutant $M_X$ changes the value of variable ``\textit{R}'' to zero (\textit{0}), while the mutant \textit{$M_Y$} changes the value of variable ``\text{L}'' to one (\textit{1}). 

Consider that our test suite is confirmed just by one test that invokes function \textit{fun} with the following arguments: \textit{z = \{0, 3, 3\} } and \textit{k = \{0, 2, 3\}}. 
The output of the corresponding input value is observed from the inside of an atomic assertion as the input's actual value.
We can see that after comparing obtained values after running each mutant in isolation, given the same test input, the mutant's behavior is different.
Following our definition, this suggests that \textit{$M_X$} is relevant to the modification, since the actual execution value output (\textit{fun(z, k)}) for mutant \textit{$M_X$} is \textit{0}, which is different from mutant \textit{$M_Y$} whereas  \textit{fun(z, k) = 1}, and \textit{$M_{XY}$} \textit{fun(z, k) = -1}.

The code snippet in the middle of the figure presents the scenario in which a mutant is not relevant to the modification.
Precisely, let us consider the same mutant $M_Y$ on the change on line \textit{5.+} as before, but now mutant $M_X$ located outside the change is on line \textit{3}. 
Mutant $M_X$ modifies the assignment statement into \textit{R = 0}. 
Given the same test input as before (i.e.,  \textit{z = \{0, 3, 3\} } and \textit{k = \{0, 2, 3\}}), and following the mutants execution behavior, we can 
observe that mutants show no observable interaction. 
Therefore, mutant \textit{$M_X$} is not considered relevant for this particular change.

The code snippet on the right side shows an additional example of a non-relevant mutant. 
However, in this example, we observe two mutants that are unreachable from each other. 
These two mutants, for any test input, do not show observable differential interaction. 
Therefore, mutant \textit{$M_X$} is considered to be non-relevant to test the corresponding change.

\begin{minipage}{\textwidth}
    \begin{minipage}[b]{\textwidth}
        \begin{figure}[H]
            \centering
            \includegraphics[scale=0.5,trim={0cm 3.4cm 0cm 3.3cm},clip]{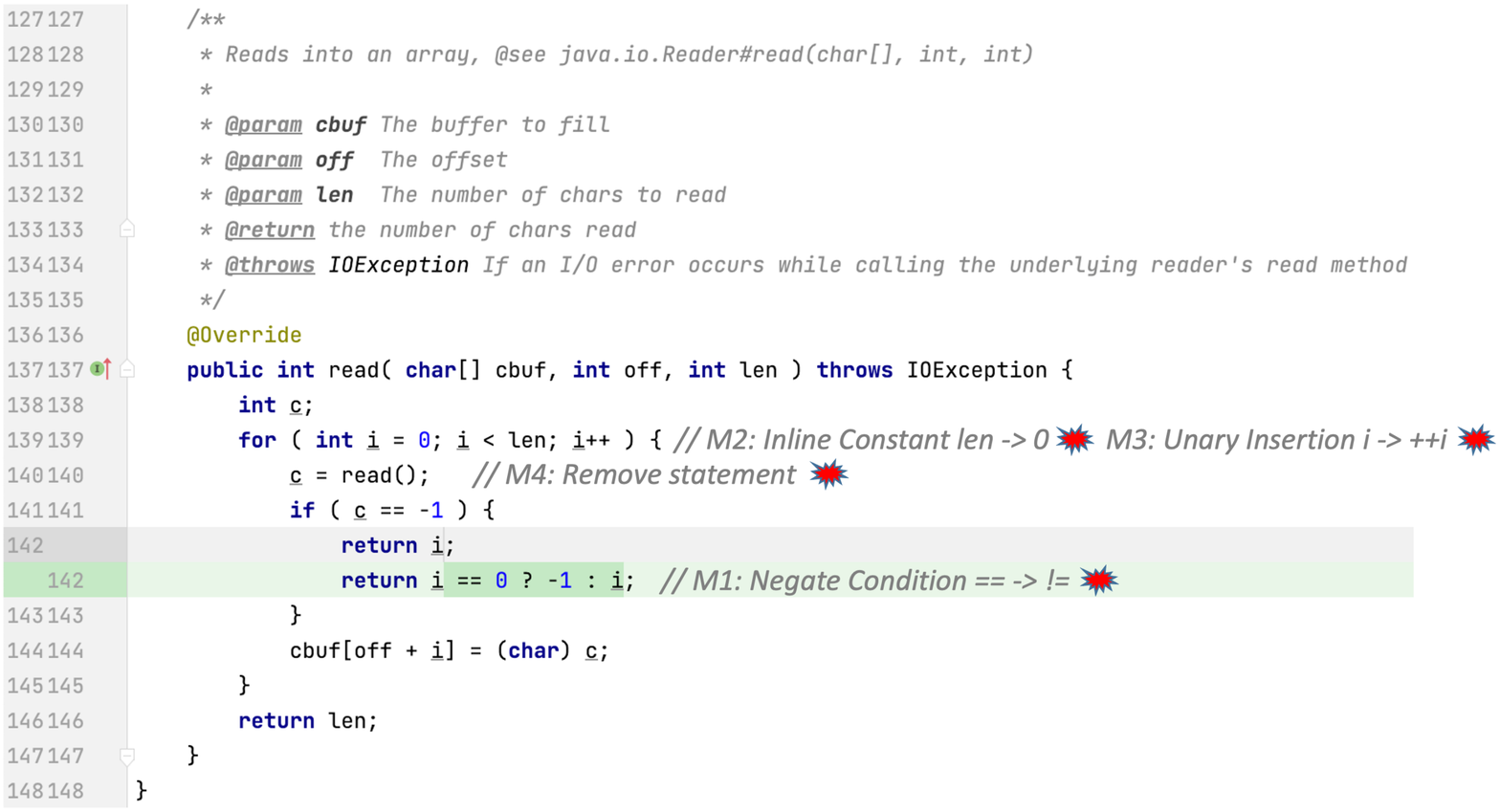}
            \caption{\revise{Method (\texttt{read()}) excerpted from the \texttt{BoundedReader.java} program in the 
            Apache commons-io project  (version \texttt{81210eb})}} 
            \label{fig:Real_case_example}
        \end{figure}
    \end{minipage}
   \end{minipage}



\subsubsection*{\textbf{\revise{Real Example}}}
\revise{
\autoref{fig:Real_case_example} presents an excerpt of a program from the Apache commons-io\footnote{https://github.com/apache/commons-io} project, version \texttt{81210eb}. The figure shows an evolution of the program in which the function \texttt{read} was modified in line 142 (from \texttt{org.apache.commons.io.input.BoundedReader.java}). The program change adds a constraints on the function's return value, suggesting that it should return negative one (-1) in case the buffer does not contain any more values, and otherwise it should return the index of the current iteration (i.e., \texttt{i}). Specifically, the previous version of the program always returned \texttt{i} in line 142, but the modified version either returns -1, when \texttt{i} is equals to 0, or \texttt{i} otherwise.}

\revise{
Function \texttt{read} takes three parameters, namely, an array of chars \texttt{cbuf}, and two  integers \texttt{off} and \texttt{len}. Intuitively, function \texttt{read} aims at modifying a certain number of characters (\texttt{len}) of array \texttt{cbuf}, starting from the given offset position \texttt{off}. 
The function starts by reading a new character from a different buffer (see built-in \texttt{read()} invocation in line 140), then it proceeds to update \texttt{cbuf} array with the new character, and finally it returns the number of updated characters\footnote{For more information about the implementation of this method, please refer to official implementation documentation page:\cite{junitAPIlink}} 
Notice that the \texttt{read()} invocation (line 140) returns the fed character as an integer in the range between zero (0) to 65535 (0x00-0xffff), or it returns negative one (-1) if the end of the buffer has been reached.
}

\revise{
As an example, consider a testing scenario that executes function \texttt{read} with the following inputs:
}

\noindent
\revise{
\hspace*{2cm}\texttt{read([`X', `X', `X', `X'], 1, 2);} 
}

\revise{
\noindent
and the buffer accessed by the \texttt{read()} call in line 140 is as:
}

\noindent
\revise{
\hspace*{2cm}\texttt{[`0', `1', `2', `3', `4', `5', `6', `7', `8', `9', `0'];}
}

\noindent
\revise{
For this test case, both versions of the program, the previous one and the recently changed version, will return the same output (i.e., \texttt{len = 2}). 
Moreover, both versions of the program will produce the same modifications into array \texttt{cbuf} given as input, resulting in:
\begin{center}
 \texttt{[`X', `0', `1', `X']} 
 \end{center}
This is an example of a test case that does not exercise the program changes, since the change (line 142) is never executed for this test case. Hence, the test does not show any behavioral difference between the previous version of the program and the current modified version of it.
}

\begin{minipage}{\textwidth}
    \begin{minipage}[b]{\textwidth}
        \begin{table}[H]
        \caption{\revise{
            Test output observation for \autoref{fig:Real_case_example} showing the program behavior (outputs) of original program, the changed program, and the first and second order mutants of the program. The test observations are performed using input \texttt{read([`X',`X',`X',`X'], 1, 2)} and an empty buffer (\texttt{[]}) fed to the built-in function \texttt{read} (in line 140)}} 
            \label{tab:case-study-table}
            \centering
            \scalebox{0.9}{
                \begin{tabular}{l c c c c c}
                \toprule
                & \textbf{\scriptsize{Program Versions}} & \textbf{\scriptsize{Program Changes}} & \textbf{\scriptsize{Code line}} & \textbf{\scriptsize{Test output}} &  \textbf{\scriptsize{Commit-Relevance}}\tabularnewline
                \midrule
                \scriptsize{Pre-commit} & \scriptsize{Old version \texttt{51f13c84}} & \scriptsize{\texttt{i}} & \scriptsize{\textit{142}} & \scriptsize{0} & \scriptsize{N/A} \tabularnewline
                \scriptsize{Post-commit} & \scriptsize{New version \texttt{81210eb}} & \scriptsize{\texttt{i == 0 ? -1 : i}} & \scriptsize{\textit{142}} & \scriptsize{-1} & \scriptsize{N/A} \tabularnewline
                \scriptsize{First-order mutant} & \scriptsize{M1} & \scriptsize{\texttt{== $\Rightarrow$ !=}} & \scriptsize{\textit{142}} & \scriptsize{0} & \scriptsize{\textit{Relevant}} \tabularnewline
                \scriptsize{First-order mutant} & \scriptsize{M2} & \scriptsize{\texttt{len $\Rightarrow$ 0}} & \scriptsize{\textit{139}} & \scriptsize{2} & \scriptsize{\textit{Non-Relevant}} \tabularnewline
                \scriptsize{First-order mutant} & \scriptsize{M3} & \scriptsize{i $\Rightarrow$ ++i} & \scriptsize{\textit{139}} & \scriptsize{1} & \scriptsize{\textit{Relevant}} \tabularnewline
                \scriptsize{First-order mutant} & \scriptsize{M4} & \scriptsize{\texttt{delete statement}} & \scriptsize{\textit{140}} & \scriptsize{2} & \scriptsize{\textit{Non-Relevant}} \tabularnewline
                \scriptsize{Second-order mutant} & \scriptsize{M12} & \scriptsize{\texttt{== $\Rightarrow$ != $\land$ len $\Rightarrow$ 0}} & \scriptsize{\textit{142 $\land$ 139}} & \scriptsize{2} & \scriptsize{N/A} \tabularnewline
                \scriptsize{Second-order mutant} & \scriptsize{M13} & \scriptsize{\texttt{== $\Rightarrow$ != $\land$ i $\Rightarrow$ ++i}} & \scriptsize{\textit{142 $\land$ 139}} & \scriptsize{-1} & \scriptsize{N/A} \tabularnewline
                \scriptsize{Second-order mutant} & \scriptsize{M14} & \scriptsize{\texttt{== $\Rightarrow$ != $\land$ delete statement}} & \scriptsize{\textit{142 $\land$ 140}} & \scriptsize{2} & \scriptsize{N/A} \tabularnewline
                \bottomrule
                \end{tabular}
            }
             \end{table}
    \end{minipage}
\end{minipage}

\subsubsection*{Commit-aware Mutation Testing}

\revise{Now, consider that during the mutation testing analysis, four mutants ($M1$ to $M4$) are injected into the function\texttt{read}, as it is shown in \autoref{fig:Real_case_example} via Java comments (``\texttt{//}''). 
Particularly, \textit{Mutant $M1$} is located on the modified statement in line 142, i.e., it is a mutant withing the program change, and it replaces the condition \texttt{==} (equal) with \texttt{!=} (not equal). 
\textit{Mutant $M2$} is located on line 139 (outside the change) and replaces the variable \texttt{len} with a constant value zero (0), mutating the condition of the \texttt{for} loop \texttt{i < len} to \texttt{i < 0}. 
\textit{Mutant $M3$} is also injected on the same statement (line 139) 
but it uses an unary insertion  (\texttt{++i}) to update variable $i$ within the condition check of the \texttt{for} loop, such that condition (\texttt{i < len}) is mutated to (\texttt{++i < len}). 
Finally, \textit{Mutant $M4$} removes the statement located on line 140 (i.e., \texttt{c = read()}).
}

\revise{Then, by using our HOM-based approach, we can create higher-order mutants by pairing all four mutants. Precisely, we pair the mutants located outside the change with the mutants on the commit-change (line 142), and we obtain three higher-order mutants $M12$, $M13$, and $M14$. \autoref{tab:case-study-table} illustrates the behavior (outputs) of function \texttt{read} under its previous version, its current changed version, and all the mutants.
}


\revise{
Consider now a different testing scenario in which the input buffer accessed by the built-in function \texttt{read} in line 140 is empty (i.e., \texttt{[]}). 
This testing scenario shows the behavioral difference between the previous version and the modified version of the program, since it executes the change (in line 142). We observe that, while the execution of the previous version of the program returns zero (0),  the modified version returns negative one (-1). 
}

\revise{
\autoref{tab:case-study-table} highlights the behavior (i.e., output) of each mutant. First, according to the traditional definition of commit-relevant mutants, $M1$ is a commit-relevant mutant, since it is located on the program change~\cite{CACHIA2013}. Additionally, according to our extension of the definition of commit-aware mutation (\textit{see} \autoref{sec:approach}), we compare the output of the second-order mutants and their isolated first-order mutants. We observe that the second-order mutant $M13$ is also a commit relevant mutant. 
This is because the second order mutant ($M3$) has a different behavior from the isolated first-order mutants (i.e., $M13$ != $M3$, and $M13$ != $M1$). Meanwhile, the other second-order mutants, i.e., $M12$ and $M14$ are not commit-relevant because they have similar behaviors as the isolated first-order mutants (i.e., $M12$ == $M2$, and $M14$ == $M4$). 
}

\subsubsection*{Commit-aware Criteria}
\revise{
Let us illustrate the importance of the strict constraint employed in our approach to compare the behaviors of first-order and second-order mutants (e.g., the need to ensure $M13$ != $M3$ \textit{AND} $M13$ != $M1$) using a counter-example. 
Specifically, we will discuss the rationale for this constraint and why considering a less strict constraint does not suffice (and mutants like, for instance, $M2$ are not commit-relevant, despite the fact that $M12$ != $M1$ but $M12$ == $M2$). 
Let us consider the example program in \autoref{fig:Real_case_example} and the output behavior observed in \autoref{tab:case-study-table}. Even though $M1$ and $M12$ have different output, inspecting the behavior of $M12$ on the program change using the provided test cases, we observe 
that the behavior of the second-order mutant $M12$ is not  different from that of $M2$. In fact, $M2$ does not execute the program change nor the mutant within the change. Indeed there is no input that can force mutant $M2$ to execute the changed line (in line 142), since the \texttt{for} loop condition will always evaluate to false. 
This implies that using a less strict constraint (e.g., an ``OR'' operation) in our check, will lead to such mis-identification of commit-relevant mutants, implying that mutants that can never lead to the execution of the program change (e.,g., $M2$) can be mis-classified as relevant to the program change. Thus, it is important to ensure that the behavior of the second-order mutant and that of the isolated first-order mutants are indeed different. 
}

\subsubsection*{Subsumption Relation}
\revise{Let us illustrate the subsumption relation of commit-relevant mutants. 
Consider the two commit-relevant mutants in our example, i.e., 
\textit{mutant $M3$} that we have identified as commit-relevant, and 
\textit{mutant $M1$} 
located on the changed statement (commit-relevant by default). 
In our example (\autoref{fig:Real_case_example}), both mutants are killed by the initial test input (in \autoref{tab:case-study-table}). 
Let us consider that the test suite has one additional test, in particular the following test: 
\begin{center}
\texttt{read([`X',`X',`X',`X'], 1, 1)}
\end{center}
\noindent
Given this new test input, we observe that \textit{mutant $M1$} is killed by this test input (output zero (0)), 
but \textit{mutant $M3$} is not killed by the test (output one (1)).  
Following the definition of mutant subsumption \cite{ammann_establishing_2014}, we can observe that a test that distinguishes \textit{mutant $M3$}, will also distinguish \textit{mutant $M1$}, but 
\textit{mutant $M1$} can be distinguished by a test that cannot distinguish \textit{mutant $M3$}. 
In this situation, we say that \textit{$M3$ subsumes $M1$}, making $M3$ a \textit{subsuming commit-relevant mutant}. 
This example illustrates a scenario where a subsuming commit-relevant mutants is located outside the program change. 
The mutant residing outside the change subsumes a mutant residing on the program change, which makes the test requirement of the mutant on the change redundant. 
We can satisfy both mutants ($M1$ and $M3$) by writing test requirements to identify the subsuming commit-relevant mutant located outside of the committed change (i.e., $M3$), which is the commit-relevant mutant also identified by our approach.}


\subsection{State of the Art} 

Let us provide background on the state of the art in commit-aware mutation testing. There are very few research papers on commit-aware mutation testing. In particular, we highlight the papers on the formalization of the concept as well as techniques for selecting commit-aware mutants. 

\begin{itemize}
    \item \textbf{\textit{Formalization of Commit-aware Mutation Testing:}}
    Ma \etal~\cite{ma2020commit} defines and formalizes the concept of commit-aware mutation testing. 
\revise{
The paper defines commit-relevant mutants as a set of mutants capturing the interactions that affect the changed program behaviors. 
The paper further shows 
that there is only a weak correlation between commit-relevant mutation scores and traditional mutation scores, thus, highlighting the need for a commit-aware test assessment metric. 
In its evaluation, the authors demonstrate
the strength of commit-aware mutation in revealing commit-related faults by showing that traditional mutation 
has about 30\% less chances of revealing commit-introducing faults, in comparison to commit-aware mutation testing.}
\revise{
In this work, we propose an alternative approach to identify commit-relevant mutants.
In comparison to \cite{ma2020commit}, our approach 
removes the strict test contract assumption 
that requires test suites being executable across program versions, i.e., there are not contract changes. 
%
In our setup, we address this concern by employing only the post-commit version and the committed change for commit-aware mutation testing, with no restriction on the evolution of the test suite (\textit{see Section \ref{exp-goals}}). 
 }
    
    \item \textbf{\textit{Diff-based Commit-aware Mutant Selection:}} Cachia et al.~\cite{CACHIA2013} proposed an incremental mutation testing that limits the scope of mutant generation to strictly changed code regions since the last mutation run. Similarly, Petrovic et al.~\cite{petrovic2018} proposed a diff-based probabilistic mutant selection technique that focuses only on the mutants located within the program changes. These approaches ignore the program dependencies between the committed changes and the unmodified code by design. Hence, in this work, we present and employ an experimental approach that accounts for the dependencies between program changes and unmodified code when identifying commit-relevant mutants to demonstrate the extent to which using mutant from modified code can help. Our approach analyses the effect of second-order mutants, one within the program change and the other in the unmodified code, on the evolving program behavior. 
    
    \item \textbf{\textit{Machine Learning based Commit-aware Mutant Selection:}}
        \texttt{Mudelta}~\cite{ma2021mudelta} presents a machine learning and static analysis-based approach for predicting commit-relevant mutants.  
\revise{This paper illustrates the importance of commit-aware mutation testing, particularly its ability to reduce mutation testing effort and reveal commit-related faults. 
In comparison to random mutant selection, \texttt{Mudelta} reveals 45\% more commit-relevant mutants, and
achieves 27\% higher fault revealing ability in fault introducing commits. 
In this work, we propose a complementary dynamic analysis approach for commit-aware mutation testing. Specifically, we propose a new HOM-based observation slicing approach for commit-aware mutation testing, that is applicable in the absence of static code features or (large) training data. 
} 

     
\item \textbf{\textit{\revise{Interface Mutation:}}} \revise{
\citet{delamaro2001interface} proposed an inter-procedural mutation analysis approach for generating mutants that are relevant for integration testing, i.e., suitable for testing the interactions between program units. In particular, since interface mutation aims at testing component integrations, it  injects mutants on the component contracts (interfaces) and pairs of them at the component call sites and related uses of the interface parameter inside the components bodies in order to capture potential interactions between the caller and called components. This mechanism, of capturing dependences through pairs of mutants, is somehow similar to our approach but more restrictive as it targets interfaces (call sites and method parameter uses). In contrast commit-aware mutation testing aims at identifying relevant dependencies between changed and unchanged code and not between components.    
	    		}
    		
    \end{itemize}
    
    \subsection{Design Requirements}
    \label{exp-goals} 


        Our commit-relevant mutation approach aims to fulfill certain requirements to ensure we gather and study a vast number of commits and commit-relevant mutants. These design requirements address some of the limitations and challenges of state of the art ~\cite{ma2020commit, ma2021mudelta, petrovic2018}. In particular, we address the following: 
        
    \begin{itemize}
    
    \item \textbf{\textit{Location of Commit-relevant Mutants:}}
        In this work, we focus on identifying commit-relevant mutants within and outside the program changes, i.e., within the commit-change as done in prior work~\cite{petrovic2018}, and the unmodified program code. In particular, we are interested in revealing behavioral interactions induced by the program changes on the rest of the unmodified program code. We achieve this by identifying the commit-relevant mutants outside of the program changes. In particular, we employ second-order mutants; we analyze the impact of second-order mutants on the behavior of the evolving program (\textit{see Section \ref{sec:approach-overview}}). 
    
    \item \textbf{\textit{Test Contract:}}
        Our experimental approach employs only the test suite from the post-commit program version. In previous work~\cite{ma2020commit}, the experimental design requires the execution of test suites across the pre-commit and post-commit versions of the program. Plus, it implies that the number of tests does not increase or decreases across versions, i.e., the test contract is intact. Therefore, the approach observes the delta between versions by comparing mutants test suites from pre- and post-change commit. This assumption is impractical in several cases since the test suite also evolves as the program evolves, e.g., when implementing new features or fixing bugs. In our work, we observed that this assumption is not common in practice. In particular, in our study, the proportion of commits where the test contract is preserved is less than 40\%.
     
     \item \textbf{\textit{\revise{Commit Patches and Hunks}}}: 
        \revise{In our approach for commit-aware mutation testing, we require the commit patches and commit hunks for empirical evaluation and analysis. 
        Indeed, commit properties are vital for commit-aware mutation testing and commonly used 
by the state of the art techniques~\cite{ma2020commit,ma2021mudelta}. In our work, we employ commit properties (i.e., patches and commit hunks) for both commit-relevant mutant detection and 
experimental analysis. Particularly, our approach requires the commit patches (i.e., the delta between pre-commit and post-commit versions) to identify the interaction of the mutants within the change and the mutants outside the  change. 
We also employ commit hunks in our experimental analysis, to identify 
the number of individual code-chunks structure present in a commit. 
Applying commit hunks in our analysis sheds light on the relationship between altered statements in the commit hunk and the mutants residing outside the commit hunk.  
}
     
     
    \item \textbf{\textit{Post-Commit Version:}} Our experimental approach requires \textit{the post-commit version of the program to be compilable, executable, and testable}. \revise{These requirements are vital for the dynamic analysis of our approach and they only apply to the \textit{post-commit version} of the program. 
In contrast, previous work~\cite{ma2020commit} requires two program versions (pre-commit and post-commit) and \textit{assumes a green test suite, no build failures and no compilation errors for both program versions}. 
In our evaluation setup, we find that these conditions are uncommon (less than 40\% of the cases). 
To address this concern, we ensure our approach requires the post-commit version of the program, without need for the pre-commit version. This allows collecting significantly more commits for our study, and allows to evaluate a vast amount of commit-relevant mutants. }
    
    
    \item \textbf{\textit{Test Oracle: }} \revise{In this work, we employ test assertions of system units as our test oracle. This is a fine-grained test oracle used by unit tests. Here it is  important to note that when we refer to assertions these are \textit{test assertions}, and \textit{not program assertions}  and are used for checking the observable behaviour of the program units, as mandated by strong mutation.
%
In practice,  we use test assertions to define the mutant behaviour  
study the impact of mutants and changes on program behaviour.  
}

    \item \textbf{\textit{Number of Commits:}} Our empirical study characterized commit-relevant mutants and required a substantial number of commits and commit-relevant mutants. Dues to the flexibility of our experimental approach, in this study, we analyzed significantly (10x) more commits and mutants than previous work~\cite{ma2020commit}. We addressed the significant limitations and assumptions of prior work, which could prevent gathering a sufficient number of commits and commit-relevant mutants. For instance, as stated in the paper \revise{Ma et al.}~\cite{ma2020commit}, it is challenging to find commits in open-source projects that do not break the test contract, i.e., keep the test suite intact. This challenge further inhibits our goal of automatically studying the characteristics of relevant mutants. Addressing the concerns above allows us to gather and study more commits than previous studies.


\end{itemize}

Our experimental approach aims to target the requirements above to ensure that we gather many commits and cover several realistic corner cases for evolving software systems. Overall, fulfilling these requirements and addressing these concerns enables us to collect significantly more commits and identify significantly more commit-relevant mutants for our study. In particular, our study involved 10x more commits and 6x more commit-relevant mutants than previous studies. 
    

\subsection{Approach Overview}
\label{sec:approach-overview}
Our study aims at investigating the existence and distribution of commit-relevant mutants in evolving software systems. 
Specifically, we study the relationship between the lines of code changed in a commit hunk and the mutants residing on program locations outside the commit hunk under consideration. 
Intuitively, we want to study the interaction between two program locations, where one location is part of the commit hunk, and the other is outside the change. 
We plan to employ high-order mutants (second-order to be more precise) and simulate potential changes in a commit hunk and the mutants outside the commit hunk. 
This study aims at providing scientific evidence of the relationship and relevance of mutants (test requirements) outside commit hunks that need to be taken into account when testing evolving systems. 

To determine if a mutant is relevant for a commit hunk, we plan to observe whether the commit changes affect mutants' behavior. 
Intuitively, suppose a change in a location in the commit hunk (produced by a mutation) affects the outcome of the mutant outside the commit. In that case, we have evidence that there exists an interaction between these two locations, indicating that the mutant is \emph{relevant} for the commit. The absence of interactions indicates either the existence of equivalent mutants \cite{KintisPM12, KintisPM15} or the absence of dependence/relevance. To account for the case of equivalent mutants and ensure the relevance of observations, we sample multiple mutants per statement. 

Mutants' behavior is (partially) determined by observing their covering test set. 
Implying that if we want to observe the interaction between mutants on different locations, the test set should make any difference in the mutant's behavior whether they are run in isolation or combined both. 
More precisely, if we can observe that the behavior of two mutants $M_X$ and $M_Y$ run in isolation differs from the behavior of the second-order mutant $M_{XY}$ (obtained by combining both mutations $M_X$ and $M_Y$), then we can conclude that mutants $M_X$ and $M_Y$ influence each other. 
Figure \ref{fig:Approach_intro} depicts such a situation, where the test set $\{t_0, t_1\}$ is able to observe that $M_{XY}$'s behavior differs from $M_X$ and $M_Y$'s behavior. For instance, test $t_0$ passes on mutants $M_X$ and $M_Y$ but fails on mutant $M_{XY}$. 
Thus, we can conclude that locations in which mutations $M_X$ and $M_Y$ were applied to interact with each other.




Following a similar idea, consider generating one of the mutants outside the change ($M_X$) and the other one on the change ($M_Y$), and their combination makes a second-order mutant ($M_{XY}$) suitable for observing if there exists an interaction between them.
To determine if mutant $M_X$ is relevant for the commit change, we can iterate this process by exploring different high-order mutants $M_{XY}$ by varying mutant on the change $M_Y$, with the aim at finding one combination that evidence their interaction.

To compare mutants' behaviors, first, we need an intersection set of tests covering mutants $M_X$, $M_Y$, and $M_{XY}$. 
Second, we proceed to run these tests to observe a difference between the mutants. 
Instead of considering only passing and failing output as a standard unit level testing oracle, we instrument tests and contained assertions to obtain and compare actual assertion values. 
For instance, an assertion like \texttt{assertEquals(0, Z)} can be violated by a (potentially) infinite number of values for \texttt{Z}, all of them violating the assertion. Suppose after executing mutants $M_X$, $M_Y$ and $M_{XY}$, the value of \texttt{Z} is different. In that case, we can observe a difference in their execution, allowing us to determine if there exists an interaction between these mutants, concluding that mutant $M_X$ is relevant for the commit change. 
Section~\ref{sec:pitest-assert-tool} describes the implementation details on how we instrument test executions to obtain actual assertion values.

%
%
%

\begin{figure*}[htp!]
    \begin{center}
    \includegraphics[width=0.8\textwidth,trim={0cm 5cm 0cm 4cm},clip]{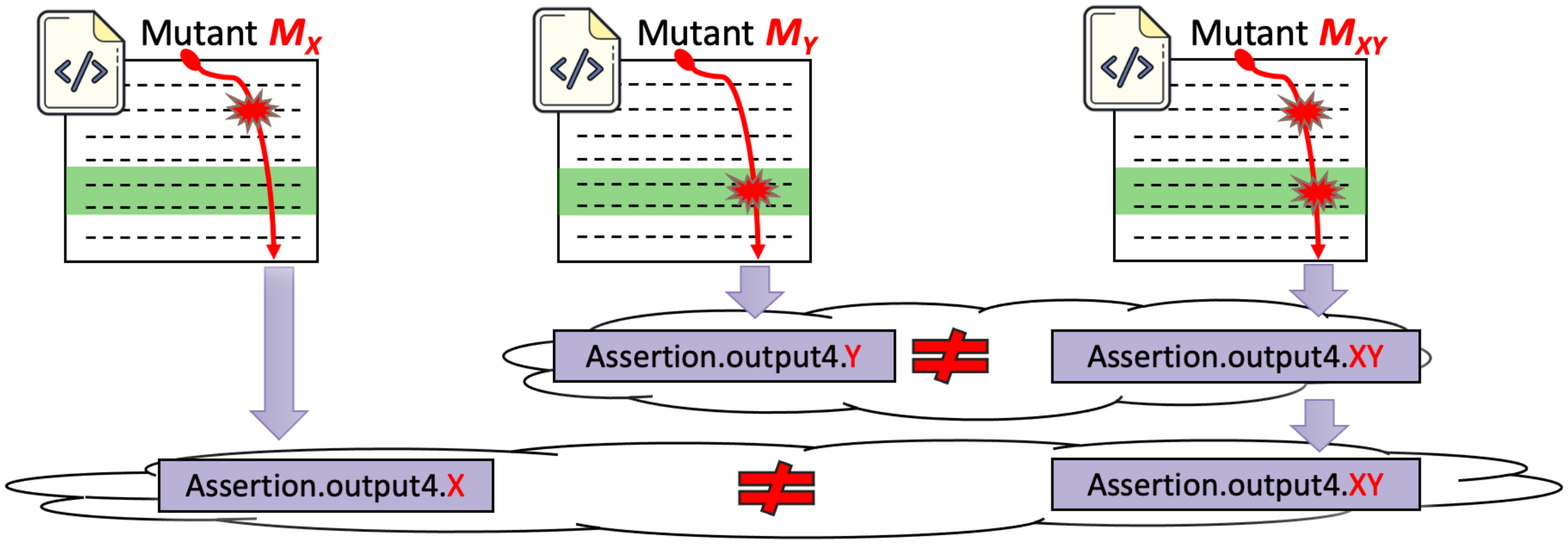}
    \caption{A mutant \textit{$M_X$} is relevant to a commit-change, if any higher order mutant \textit{$M_{XY}$}, shows different behavior from \textit{$M_X$} and \textit{$M_Y$} executed in isolation}
    \label{fig:Approach_figure_example}
    \end{center}
\end{figure*}

Figure \ref{fig:Approach_figure_example} illustrates our approach to detect interactions between mutants by comparing their behavioral assertion values. 
It depicts that after executing each first-order mutant $M_X$ and $M_Y$ in isolation (assertions that cover them), we compare the output values with the value obtained after running second-order mutant $M_{XY}$.
If running $M_X$ and $M_Y$ in isolation differs from running $M_{XY}$, we determine that mutant $M_X$ is relevant for the commit change. 

\subsection{Algorithm}  
\label{sec:approach-algorithm}


To perform an empirical study toward distinguishing relevant mutants, we generate the first order mutants located around and on the commit change (i.e., $M_X$ and $M_Y$, respectively). The second-order mutants (i.e., $M_{XY}$) are a combination of the previous two. 
The \emph{mutant-assertion matrices} were obtained by executing the mutants against developer-written and automatically generated test pools. Note that test run status is pass/fall for Java programs; therefore, to observe behavioral differences produced by mutants, we need to focus on test assertions and record assertion execution actual value output of each test on every mutant. 
Precisely, for every mutant and every test assertion, a mutant-assertion matrix stores the assertion values obtained after running a mutant against a test.
As noted, this study performs mutation analysis on commits from Java programs, using Pitest\footnote{http://pitest.org/} as the Java mutation testing tool, and EvoSuite\footnote{https://www.evosuite.org/} as the state of the art test case generation tool.
Section \ref{analysis_procedure} provides further details regarding mutants test case generation and test assertions instrumentation. 

After computing mutant-assertion matrices, we proceed to approximate which mutants are relevant to the change, according to our Definition~\ref{def:relevant-mutant} following the steps incorporated in Algorithm ~\ref{algo:relevant}.
The algorithm summarises previously described process, where functions \emph{MutantsonChangeMutantOuput}, \emph{aroundChangeMutantOutput}, \emph{highOrderMutantOutput} return the output of a specific test \textit{assertion} execution per specific \textit{mutant}. 
Finally, Algorithm~\ref{algo:relevant} returns a set of relevant mutants for a particular commit change.

\revise{
This algorithm has a worst-case polynomial time complexity of $O(n^4)$, due to the four nested \texttt{for} loops ($O(n*n*n*n)$). For each of the three inputs fed to the algorithm (\emph{TestSuite}, \emph{MutantsOnChange} and \emph{MutantsAroundChange}), 
there is a linear-time complexity ($O(n)$).  Additionally, there is a linear-time complexity  ($O(n)$) for evaluating each test assertion corresponding to the test cases. 
Overall, the performance of the algorithm depends on the number of mutants in the change, the number of mutants injected in the modified code, the size of the test suite and the number of assertions in each test. Specifically, to derive higher-order mutants, we consider every pair of mutants within and outside the change, we also execute all test cases corresponding to these mutants, and evaluate all test assertions in each test case. This algorithm can be optimized by improving the number of evaluated tests, assertions or pairs of mutants.} 
\revise{The complexity of this algorithm can be reduced to  ($O(log(n) * n^3)$ via a binary search on the pair of mutants (outside the change) that exposes a behavioral difference. Likewise, the complexity can be reduced to cubic complexity ($O(n^3)$) by executing a constant number of test cases/assertions ($O(1)$). For instance, an improvement is achievable by selecting and executing only the most relevant tests for the changes, e.g., from historical test executions in the CI. A reduction is also achievable if only one test assertion is evaluated for each test case, e.g., executing only the assertion that captures the interaction between the pair of mutants 
has a constant time complexity ($O(1)$). 
}

%

\setlength{\algomargin}{0.99em}
\setlength{\textfloatsep}{13pt}  
\begin{algorithm}
\SetAlgoLined
\LinesNumbered
\KwData{TestSuite, MutantsOnChange, MutantsAroundChange}
\KwResult{Relevant Mutants}
  $RelevantMuts \leftarrow \emptyset$\;
\For{$X \in MutantsAroundChange$}{
    \For{$Y \in MutantsOnChange$}{
        \For{$test \in TestSuite$}{
            \For{$assertion \in test$}{
                $Yval \leftarrow onChangeMutantOuput(assertion, Y)$\;
                $Xval \leftarrow aroundChangeMutantOutput(assertion, X)$\;
                $XYval \leftarrow highOrderMutantOutput(assertion, Y, X)$\;
                \If{$Yval \neq XYval \land Xval \neq XYval$}{
                    $RelevantMuts \leftarrow RelevantMuts \cup \{X\}$\;
                    \textbf{jump to line 2 and take next mutant $X$}\; 
                }
            }
        }
    }
}
 \Return $RelevantMuts$ \;
 \caption{Approximate Commit-relevant Mutants Set}
 \label{algo:relevant}
 \end{algorithm}

\section{Experimental Setup}\label{Setup}



\subsection{\revise{Goals}}\label{exp-goals}
\revise{
The main goal of our study is 
to investigate the prevalence and characteristics of commit-relevant mutants in evolving software systems in terms of their program location and relationship to \textit{commit hunks} and \textit{mutant types}. We also study their effectiveness and efficiency in testing evolving systems in comparison to the state-of-the-art. Specifically, our empirical goal is to achieve the following three main goals:
\begin{enumerate}
\item study the \textit{properties of commit-relevant mutants}, in terms of their prevalence, mutant types, location and proportions, as well as the\textit{ subsumption relation of commit-relevant mutants} (RQ1, RQ2 and RQ4);
\item examine the \textit{relationship between commit-relevant mutants and commit properties} (e.g., commit size) (RQ3);
\item investigate the benefit of commit-relevant mutation testing, in terms of their \textit{effectiveness} and \textit{efficiency} in comparison to the baselines (RQ5 and RQ6).
\end{enumerate}
}

\revise{
Overall, our study aims at providing insights on the properties of commit-relevant mutants and at demonstrating their importance and effectiveness in testing evolving systems. 
}

\subsection{Research Questions}\label{rqs}
As we aim at assessing the potential of mutation testing in evolving systems, we investigate the following research questions (\RQ s). 

\begin{description}

    \item[\RQ1] \textbf{Commit-Relevant Mutants:} What is the \emph{prevalence} of ``commit-relevant mutants'' among the whole set of mutants?
        \revise{
            \begin{description}
                \item[\RQ1.1] How are commit-relevant mutants distributed in the program?
                \item[\RQ1.2] Are commit-relevant mutants located within or outside the developers' committed changes?
                \item[\RQ1.3] Is there any correlation between the number of commit-relevant mutants located within program changes and the number of commit-relevant mutants outside the changes?
            \end{description}
        }   
    
    
%
%
%

    \item[\RQ2] \textbf{Subsuming Commit-Relevant Mutants:} What is the \emph{proportion} of ``subsuming commit-relevant mutants'', i.e., the number of commit-relevant mutants that subsumes other commit-relevant mutants, such that testing only these subsuming mutants is sufficient to test all other commit-relevant mutants?
    

%

    \item[\RQ3] \textbf{Commit Size:} Is there a relationship between the \textit{size of the commit} (i.e., number of commit hunks) and the number of (subsuming) commit-relevant mutants?
    

%
    
    \item[\RQ4] \textbf{Commit-Relevant Mutant Types:} What is the \emph{distribution of mutant types} in commit-relevant mutants? 
%
%
        
    \item[\RQ5] \textbf{Comparative Effectiveness: } How \textit{effective} are (subsuming) commit-relevant mutants, in comparison to the \textit{baselines} (i.e., random mutation and ``commit-only mutation'')? 
    
%
%
%

    \item[\RQ6] \textbf{Test Executions: } What is the \textit{performance} of (subsuming) commit-relevant mutants in comparison to the \textit{baselines}, in terms of the number of \textit{required test executions}?
%
%
    
\end{description}

RQ1 aims at improving our understanding of the locations, prevalence, and number of relevant mutants in relation to committed changes. The answer to the question allows having a rough view of the relevant mutant's distribution within and outside committed code. Answering RQ2 will show the extent to which the relevant mutant sets have redundancies. Previous work \cite{PapadakisHHJT16} has shown that redundant mutants inflate mutation scores with the unfortunate effect of obscuring its utility. We, therefore, would like to validate whether relevant mutant sets also suffer from such inflation effects. RQs 3 and 4 analyze the relation between commit size and prevalence of mutant types in relevant mutant sets to check whether there are levels/thresholds at which relevant mutants do not yield much benefits. Finally, RQs 5 and 6 aim at quantifying the potential benefits of using relevant mutants during project evolutions concerning cost and effectiveness.

\subsection{Analysis Procedure}\label{analysis_procedure}

We focus our empirical study on commits of Java programs as selected subjects. 
To perform the mutation analysis, we employ Pitest\footnote{http://pitest.org/}~\cite{pitest}, one of the state-of-the-art Java mutation testing tools. 
We approximate the set of commit-relevant mutants by following the algorithm introduced in Section~\ref{sec:approach}. 
Besides the approximated set of commit-relevant mutants located outside of commit-change, we also record and consider as commit-relevant all those mutants residing on the location of commit-change (in our approach, \textit{$M_Y$} mutants). 
This corresponds to work done by \cite{petrovic2018}, whereas the commit-relevant mutants set is made out of mutants located on the commit diff, i.e., statements modified or added by commit.

To make our approximation robust, we follow previous studies process steps \cite{KurtzAODKG16,ammann_establishing_2014,PapadakisK00TH19}. 
Our approach uses mutant-assertion matrices to identify mutants interactions that constitute, up to our knowledge, the first study conducted on test assertion level for Java programming language (bypassing standard tests passing/failing mutation behavior for Java programs). 
Mutant-assertion matrices were computed by running large test pools built by considering developer tests and adding automatically generated tests using EvoSuite\footnote{https://www.evosuite.org/}~\cite{FraserZ12}, a state of the art test case generation tool.  
From the computed mutant-assertion matrices,  we obtain three sets of mutants: \textit{mutants on a change}, \textit{mutants relevant to a change} and \textit{mutants not relevant to a change}. 

To answer \RQ1 we study the prevalence and location of commit-relevant mutants in every commit by analyzing the average number of relevant/non-relevant mutants and their distribution. 
We address \RQ2 by studying the proportion of subsuming commit-relevant mutants among all commit-relevant mutants and all subsuming mutants. This will estimate an extra possible reduction we can achieve if we focus only on subsuming mutants. 
We consider traditional passing/failing test behavior to compute the set of subsuming mutants per subject (notice that this information is also captured when mutant-assertion matrices were built). 

To address \RQ3 and \RQ4, we perform a similar statistical analysis. Still, we study any correlation between the number of commit-relevant mutants and the size of commit hunks, and the type of mutants. In \RQ5 and \RQ6, we simulate a mutation testing scenario where the tester starts by picking a mutant for analysis for which a test to kill it is developed.  
During this simulation, for each analyzed mutant, we randomly pick the test to kill it from the pool and compute which other mutants are collaterally killed by the same test.  
The process proceeds by picking a survived mutant until every mutant has been killed. 
We consider a mutant as equivalent if there is no test in the pool that kills it. 

This kind of simulation has been used in various related works to assess the effectiveness of mutation testing techniques~\cite{KurtzAODKG16,ammann_establishing_2014,PapadakisK00TH19,ma2020commit}. 
We consider four different mutant selection techniques when answering \RQ5 and \RQ6. Two of them we use as \emph{baselines}, where one consists of \emph{randomly} selecting \revise{from the set of all} mutants, and the other one consists of selecting only the mutants on the change. 
Another selection technique consists of selecting from the pool of commit-relevant mutants, while the last technique consists of selecting subsuming commit-relevant mutants. 
We aim to obtain the best-effort evaluation by maximizing effectiveness and minimizing the effort. We focus on the first 20 mutants picked by a tester to test commit changes, while we measure effectiveness in terms of the \emph{commit-relevant mutation score} reached by the selected mutants that guide the testing process. 
Simultaneously, we measure the computational effort in terms of the number of \emph{test executions} required to accomplish the same effect over the different baselines (different mutants pools). 
\revise{In this simulation, we are interested in the test executions with the tests derived by the analysed mutants. The dependent variable is the test sets, while the independent variable is the test executions.}
We iterate the process (killing all selected mutants) 100 times and compute the relevant mutation score and computation effort. 

\subsection{Subject Programs and Commits}\label{subject_programs}
We focus our empirical study on commits of a set of well-known, well-tested, and matured Java open-source projects taken from Apache Commons Proper repository\footnote{https://commons.apache.org}. 
The process of mining repositories, data analysis, and collection, was performed as follows:
\begin{enumerate}
\item Our study focuses on the following projects: \texttt{commons-collections, commons-lang, comm\\ons-net, commons-io, commons-csv}. 
These projects differ in size while having the most extended history of evolution. 
We extracted commits from the year 2005 to 2020. \revise{To extract commit patches and hunks in our setup, we employ \texttt{PyDriller}\footnote{\url{https://pydriller.readthedocs.io/en/latest/intro.html}} (\texttt{V1.15}) to mine commits from the selected projects\footnote{\texttt{PyDriller} is an open-source Python framework that helps developers mine software repositories and extract information given the GIT URL of the repository of interest.}. We applied \texttt{PyDriller} to query the project's information such as commits hash id, modifications date, modified source code, modification operation, and hunks of the commits and quickly exported such information into a JSON file.}

\item We kept only commits that use JUnit4+\footnote{https://junit.org/junit4/} as a framework to write repeatable tests since it is required by
EvoSuite~\cite{FraserZ12}, the test generation tool we use for automatically augmenting test suites. 
   
\item We filtered out those commits that do not compile, do not have a green test suite (i.e., some of the tests are failing), or do not affect a program's source code (i.e., commits that only change configuration files). 
Some commits with failing tests are filtered out since Pitest requires a green test suite to perform mutation testing analysis. 

\item Due to the significant execution time for commits containing several files, we set a limit for 72h of execution on a High-Performance Computer to generate and execute mutants per commit. 
Please note that the test suites contain developer-written and automatically generated tests, where both are used to create mutation matrices. 
All experiments were conducted on two nodes with 20 physical cores and 256GB of RAM. Specifically on Intel Skylake Xeon Gold 2.6GHz processors, running on Linux Ubuntu OS across four threads.  
\end{enumerate}

Overall, we generated 9,368,052 high-order mutants and 260,051 first-order mutants, over 288 commits, that required 68,213 CPUs days of execution.
Table \ref{tab:subjects} summarises the details of the mined commits. Column ``\# Commits'' reports the number of commits mined per project, column ``\# LOC'' (Lines Of Code) indicates a subject scope in terms of lines of code, ``Maturity" reports on the date of first commit, column ``\# FOM'' (First-Order Mutant) indicates the total number of First Order Mutants generated for those commits, ``\#Mutants on Change'' indicates the number of First Order Mutants generated on the changed lines, column ``\#HOM'' (High-Order Mutant) indicates the total number of High Order Mutants generated, column ``\# Dev. Tests'' (Developer written Tests) reports on the number of developer written test cases, and column ``\# Evosuite Tests'' reports on the number of automatically generated tests.

{\Large
\begin{table}[!bt]
  \centering
  \caption{Details of Subjects Programs and Studied Commits}
  \label{tab:subjects}
  \scalebox{.9}{
  \begin{tabular}{l r r r r r r r r r r}
  \toprule
  \textbf{\scriptsize{Commons Projects}} & \textbf{\scriptsize{\# LOC}} & \textbf{\scriptsize{\# Maturity}} & \textbf{\scriptsize{\# Commits}} & \textbf{\scriptsize{\# FOM}} & \textbf{\scriptsize{\# Mutants on Change}} & \textbf{\scriptsize{\# HOM}}& \textbf{\scriptsize{\# Dev. Tests}} & \textbf{\scriptsize{\# EvoSuite Tests}} \tabularnewline
  \midrule
  \scriptsize{collections} & \scriptsize{74,170} & \scriptsize{14/04/2001} & \scriptsize{45} & \scriptsize{27,417} & \scriptsize{2,026} & \scriptsize{1,192,188} & \scriptsize{4,797} & \scriptsize{1,285} \tabularnewline\midrule
  \scriptsize{io} & \scriptsize{29,193} & \scriptsize{25/01/2002} & \scriptsize{30} & \scriptsize{24,970} & \scriptsize{1,115} & \scriptsize{668,448} & \scriptsize{914} & \scriptsize{286} \tabularnewline\midrule 
  \scriptsize{text} & \scriptsize{22,933} & \scriptsize{11/11/2014} & \scriptsize{46} & \scriptsize{47,847} & \scriptsize{4,155} & \scriptsize{2,073,829} & \scriptsize{1,084} & \scriptsize{322} \tabularnewline\midrule
  \scriptsize{csv} & \scriptsize{4,844} & \scriptsize{25/01/2002} & \scriptsize{101} & \scriptsize{66,862} & \scriptsize{3,577} & \scriptsize{1,968,137} & \scriptsize{6,144} & \scriptsize{2,833} \tabularnewline\midrule
  \scriptsize{lang} & \scriptsize{85,709} & \scriptsize{19/07/2002} & \scriptsize{66} & \scriptsize{102,072} & \scriptsize{3,891} & \scriptsize{3,885,341} & \scriptsize{7,574} & \scriptsize{959} \tabularnewline\midrule
  \textbf{\scriptsize{Total}} & \scriptsize{216,489} & \scriptsize{N/A} & \textbf{\scriptsize{288}} & \textbf{\scriptsize{269,168}} & \textbf{\scriptsize{14,764}} & \textbf{\scriptsize{9,787,943}} & \textbf{\scriptsize{20,513}} & \textbf{\scriptsize{5,685}} \tabularnewline\midrule
  \bottomrule
  \multicolumn{9}{l}{\footnotesize \textit{ ``\# LOC'' - Lines Of Code, ``\# FOM'' - First Order Mutants, ``\# HOM'' - Higher Order Mutants, ``\# Dev. Tests'' - Developer written Tests} }\\
  \end{tabular}
  }
\end{table}
}

\subsection{Metrics and Measurements}\label{metrics}
\textbf{Statistical Analysis:} To answer our research questions, we performed several statistical analyses to evaluate correlations among several variables. For instance, in \RQ1, we analyzed whether the number of commit-relevant mutants correlates with the number of mutants residing on a change and whether the number of subsuming commit-relevant mutants correlates with the number of subsuming mutants. 


In this study, we employ \revise{two} correlation metrics, namely \textit{Kendall rank coefficient ($\tau$) (Tau-a)}, \revise{and \textit{Spearman's rank correlation coefficient ($\rho$ - (rho)) }}, with the level of statistical significance set-up to $p-value$ 0.05. The Kendall rank coefficient ($\tau$), measures the similarity in the ordering of studied scores, while \revise{Spearman's $\rho$ (rho) measures how well the relationship between two variables can be described using a monotonic function \cite{myers2004spearman}.} 
\revise{The correlation metrics calculate values between -1 to 1, where a value close to 1 or -1 indicates strong correlation, while a value close to zero indicates no correlation at all. }
Additionally, to facilitate comprehension of our figures, we employed \textit{coefficient of determination} (R$^2$ trendline) as a statistical measure that describes the proportion of the variance in the dependent variable that is predictable from the independent variable(s).

\smallskip\noindent  
\textbf{Mutation-specific Measures:} We also employed mutation-specific metrics such as \textit{the commit-relevant mutation score} and \textit{subsuming commit-relevant mutation score}, to measure the effectiveness and efficiency of the selected mutants that guide the testing process. 
\revise{We measure how the test suite effectiveness progresses when we analyze mutants from the different mutant sets (e.g., all mutants, relevant mutants, subsuming relevant, etc.). Similarly, we measure efficiency by counting the number of test executions involved (to identify which mutants are killed by the test suites) when test suite progresses.}

\subsection{Implementation Details} \label{subsection:implementation_details}
Our commit-relevant mutant identification approach is implemented in approximately 5 KLOC of Python code, ~600 LOC in Shell scripts and 3 KLOC of Java. It employs several external tools and libraries including \texttt{Evosuite}, \texttt{git-diff }and \texttt{PitTest}. We have also implemented additional infrastructure on \texttt{PitTest} to ensure analysis of evolving software and extract assertion information. In the following, we describe each of these tools. 



\subsubsection{\texttt{EvoSuite \revise{(V1.1.0)}}} 
To obtain a rich test suite for our study, we collected developer-written tests and automatically generated tests. For our mutation testing analysis, we augment developers' test suites with test cases automatically generated with \texttt{EvoSuite}~\cite{FraserZ12}. \texttt{EvoSuite} is an evolutionary testing tool that generates unit tests for Java software. In our analysis, we run \texttt{EvoSuite} against all several coverage criteria (e.g., line, branch, mutation, method, etc.); we also executed \texttt{EvoSuite} with default configurations, especially concerning running time.

\subsubsection{\texttt{Pitest (\revise{V1.5.1})} and \texttt{git-diff}}
Pitest does not have built-in functionality to satisfy the requirements of our experiment. Therefore, we extended the framework for High Order Mutants~\cite{Laurent} on top of Pitest that takes as an input the \texttt{gitdiff} output\footnote{https://git-scm.com/docs/git-diff}. 
Based on the statement difference between the versions, the framework extends the mutants generation functionality by generating, i.e., mapping, mutants on the change, with the mutants around the change. Thus, creating second-order mutants for that particular commit file. Our framework is configured to generate the extended set of mutants available in Pitest, introduced by Laurent \etal \cite{LaurentPKHTV17}. Kintis \etal \cite{KintisPPVMT18} has also shown that this extended set of mutants is more powerful than the mutant sets produced by other mutation testing tools.   

\subsubsection{\texttt{Pitest Assert}} \label{sec:pitest-assert-tool}
\texttt{Pitest (V1.5.1)} creates killing matrices and identifies whether a mutant is killed or not, based on test case oracle prediction (test fails or passes). 
These matrices were not suitable for our experimental procedure. 
Therefore, we built a framework on top of Pitest to extract additional information concerning each test case assertions (from tests that cover mutants). 
Our framework performs bytecode instrumentation of each test executed on a specific mutant, using ASM\footnote{https://asm.ow2.io/} as an all-purpose Java bytecode manipulation and analysis framework. 
By instrumenting each test case assertion, we can obtain execution information. 
More precisely, each assertion has a unique test name where it locates, assertion function name, assertion line number, and assertion actual execution value. 
If an assertion triggers an exception, we keep track of the stack-trace execution. 
However, for this study's purpose, we disregard the assertions that trigger the exception from our relevant mutants calculation (please refer to Algorithm \ref{algo:relevant}) since we only aim at actual mutants' observable behavioral output. 
Hence, the mutant-assertion matrix is a weighted matrix. For each (mutant, test-assertion) pair, the value corresponds to the actual assertion value obtained by running the test on the mutant, or the exception stack trace if an assertion throws an exception. 
\revise{
Concretely, we employ the \textit{JUnit4}\footnote{https://junit.org/junit4/} testing framework which contains a public class (called \texttt{Assert}) that provides 
a set of assertion methods to specify test conditions. 
Typically, these methods (e.g., \texttt{Assert.assertEquals(expected value, actual value)}) directly evaluate the assertion's conditions, then returns 
the final assertion's output (e.g., conditions not satisfied, pass, or fail). 
To obtain the value of parameters within the assert statement, in our framework, we use \texttt{Pitest Assert} to instrument each assertion method. Such that we serialize the provided input values in the assert statement before they propagate to conditional checks, i.e., before the conditional check is reached in \textit{org.junit.Assert\footnote{https://junit.org/junit4/javadoc/4.13/org/junit/Assert.html}} and the output values are fed to \textit{org.hamcrest.Matcher\footnote{http://hamcrest.org/JavaHamcrest/javadoc/1.3/org/hamcrest/Matchers.html}} for evaluation. 
Specifically, we serialize both the expected and actual values after they propagate as 
input parameters of the assert statement. This allows to 
assess the input parameters of the \texttt{assert} statement (e.g., an expression or a method call (\texttt{assertEqual(foo(), bar())})) for concrete values. 
Hence, in our setup, we compare the output values of both the expected and actual values present in each assertion. However, our experimental framework does not directly account for the 
potential dependencies within assertions and test cases, we address this concern in the \textit{threats to validity} (\textit{see} \autoref{ValidityThreats}). 
Our test assertion framework is built on top of \texttt{PiTest} and is publicly available. 
}

\subsection{Research Protocol}\label{protocol}
\begin{figure}[bt!]
    \begin{center}
    \includegraphics[width=0.95\textwidth,trim={0cm 8cm 0cm 7cm},clip]{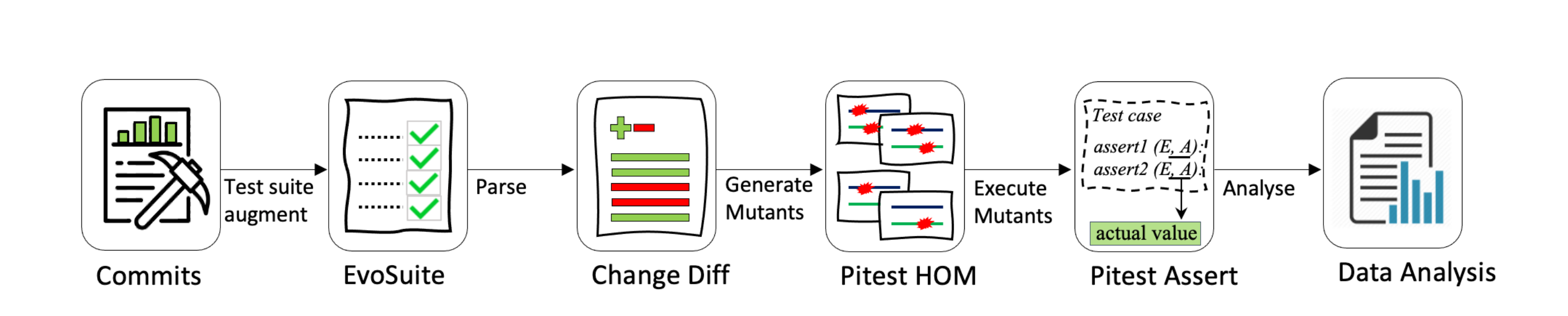}
    \caption{Research Protocol}
    \label{fig:Process_pipeline}
    \end{center}
\end{figure}

\autoref{fig:Process_pipeline} highlights our experimental protocol which proceeds as follows: For each project (e.g., \textit{commons-collections}) and each mined commit (e.g., hash: \textit{03543e5f9}, 
we first \textit{augment the developers' test suite} with automatically generated tests using EvoSuite~\cite{FraserZ12}. Next, we \textit{obtain the commit changes} (a.k.a hunks) of the commit using the \texttt{git-diff} tool, in order to identify the changed and unchanged program statements. We then \textit{generate both first-order and second-order mutants} for the program, using Pitest Assert as our extension of Pitest Mutation Testing tool~\cite{pitest}. After mutant generation, we \textit{execute every mutant to obtain the mutant-assertion matrices}, which provides information about test assertion type, position and value. Finally, we execute our relevant mutant detection algorithm~\ref{algo:relevant} to identify commit-relevant mutants. 

Our result analysis proceeds after computing mutant-assertion matrices and identifying commit-relevant mutants. 
We then perform the data gathering and analysis required to answer every research question (\RQ s). 
In particular, we compute subsuming mutant relations necessary to answer \RQ2, and perform the mutation testing simulation needed to answer \RQ5 and \RQ6.

\section{Experimental Results}\label{Results}



\smallskip\noindent
\subsection{\RQ1: Commit Relevant Mutants} 


We start by studying the proportion of \emph{commit-relevant mutants} that affect the commit changes out of \emph{all mutants} by using the pipeline just introduced in Section~\ref{sec:approach}. 
Thus, in this RQ, 
we consider as commit-relevant mutants all mutants identified by our approach, including the set of killable mutants residing on modified statements. We distinguish commit-relevant mutants in the categories of those located on changed and unchanged code to demonstrate

This allows us to estimate the potential reduction in terms of the number of mutants requiring analysis and the number of test executions required to cover them if the tester focuses testing only on commit-relevant mutants instead of the whole set mutants, \revise{or on the mutant set consisting of all mutants residing on the modification}.

Additionally, we evaluate the properties of commit-relevant mutants that can inform their selection among all mutants. Thus, we examine the location of commit-relevant mutants, whether they are mostly located within the commit or outside the committed changes. We also assess whether there is a correlation between the number of identified commit-relevant mutants and the number of commit-relevant mutants within the committed change to determine if the number of mutants within a  commit can serve as a proxy to determine the number of commit-relevant mutants. 

{\Large
\begin{table}[htp]
  \centering
  \caption{Details of the Prevalence of Commit-relevant Mutants. }
  \label{tab:prevalence-commit-relevant-mutants}
  \scalebox{.9}{
  \begin{tabular}{l r r r r r r r}
  \toprule
  \textbf{\scriptsize{Project}} & \textbf{\scriptsize{\# Commits (C)}}  & \textbf{\scriptsize{\# C. All R. M.}} & \textbf{\scriptsize{\# C. No R. M.}} &  \textbf{\scriptsize{\# Relevant}} & \textbf{\scriptsize{\# Not Relevant}} & \textbf{\scriptsize{Ratio}} & \textbf{\scriptsize{Reduction Ratio}} \tabularnewline
  \midrule
  \scriptsize{commons-collections} & \scriptsize{45} & \scriptsize{2} & \scriptsize{4} & \scriptsize{6,833} & \scriptsize{18,558} & \scriptsize{32,31\%} & \scriptsize{67,69\%} \tabularnewline\midrule
  \scriptsize{commons-io} & \scriptsize{30} & \scriptsize{0} & \scriptsize{3} & \scriptsize{6,052} & \scriptsize{17,803} & \scriptsize{28,70\%} & \scriptsize{71,30\%} \tabularnewline\midrule 
  \scriptsize{commons-text} & \scriptsize{46} & \scriptsize{1} & \scriptsize{4} & \scriptsize{8,810} & \scriptsize{34,882} & \scriptsize{27,10\%} & \scriptsize{72,90\%} \tabularnewline\midrule
  \scriptsize{commons-csv} & \scriptsize{101} & \scriptsize{4} & \scriptsize{0} &  \scriptsize{27,441} & \scriptsize{35,844} & \scriptsize{47,39\%} & \scriptsize{53,61\%} \tabularnewline\midrule
  \scriptsize{commons-lang} & \scriptsize{66} & \scriptsize{1} & \scriptsize{2} &\scriptsize{15,724} & \scriptsize{82,457} & \scriptsize{19,22\%} & \scriptsize{80,78\%} \tabularnewline\midrule
  \textbf{\scriptsize{Total}} & \textbf{\scriptsize{288}} & \textbf{\scriptsize{8}} & \textbf{\scriptsize{13}} & \textbf{\scriptsize{64,860}} & \textbf{\scriptsize{189,544}} & \textbf{\scriptsize{N/A}} & \textbf{\scriptsize{N/A}} \tabularnewline\midrule
  \textbf{\scriptsize{Average}} &  \textbf{\scriptsize{58}} & \textbf{\scriptsize{N/A}}& \textbf{\scriptsize{N/A}} & \textbf{\scriptsize{225}}  & \textbf{\scriptsize{658}}  & \textbf{\scriptsize{29,58\%}} & \textbf{\scriptsize{70,42\%}}  \\ \hline 
  \bottomrule
  \multicolumn{8}{l}{\footnotesize \textit{ ``\# C. All R. M.'' - Number of Commits with all relevant mutants, ``\# C. No R. M.'' - Number of Commits with no relevant mutants} }\\
\end{tabular}
  }
\end{table}
}

    \begin{figure*}[htp]
        \begin{center}
		\includegraphics[width=\textwidth]{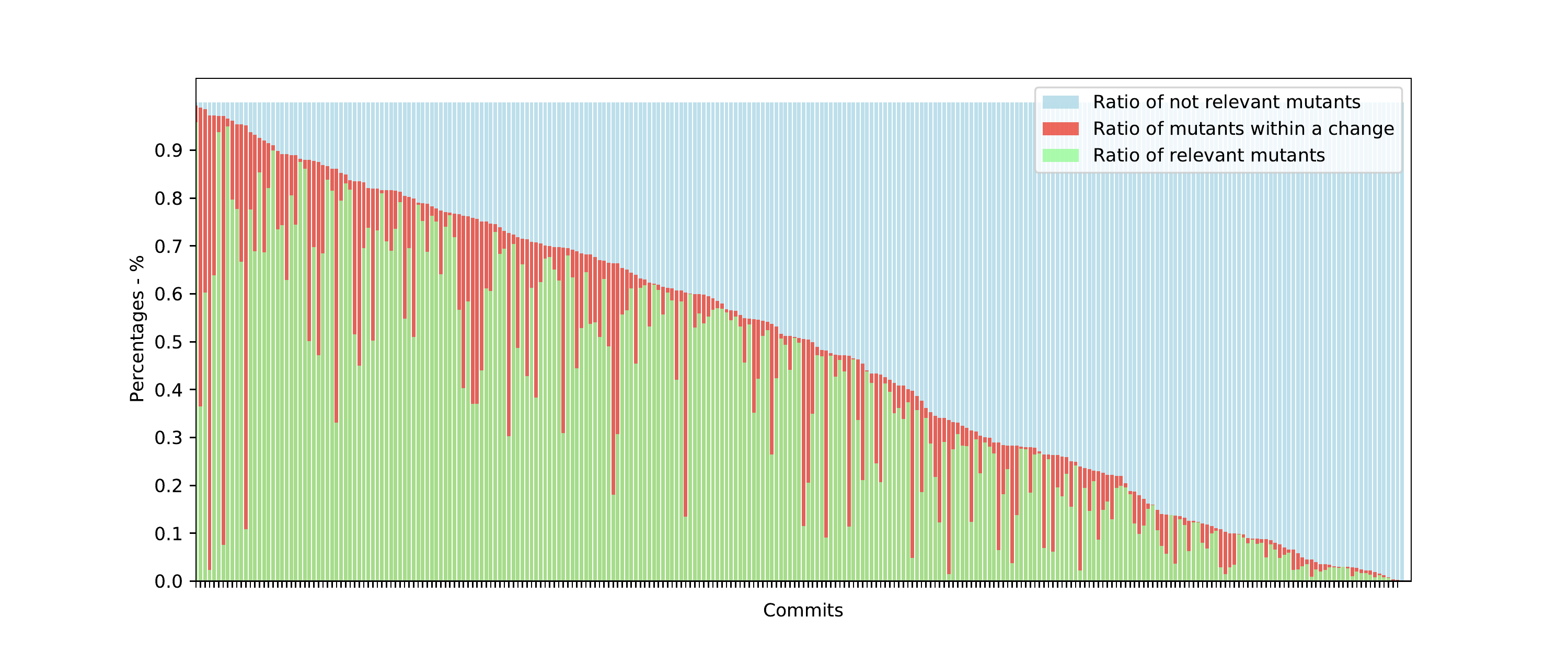}
        \end{center}
        \caption{Distribution of mutants across all commits showing the proportion of non-relevant mutants (in \textit{ \color{blue} blue}) as well as commit-relevant mutants within committed changes (in \textit{\color{red} red}) and outside committed changes (in \textit{\color{green} green})
		}
        \label{fig:RQ1-Distribution_plot}
    \end{figure*}
    
\begin{figure}[bt!]
    \centering
    \begin{minipage}{.45\textwidth}
      \centering
      \vspace{-0.4cm}
	\includegraphics[width=1.0\linewidth]{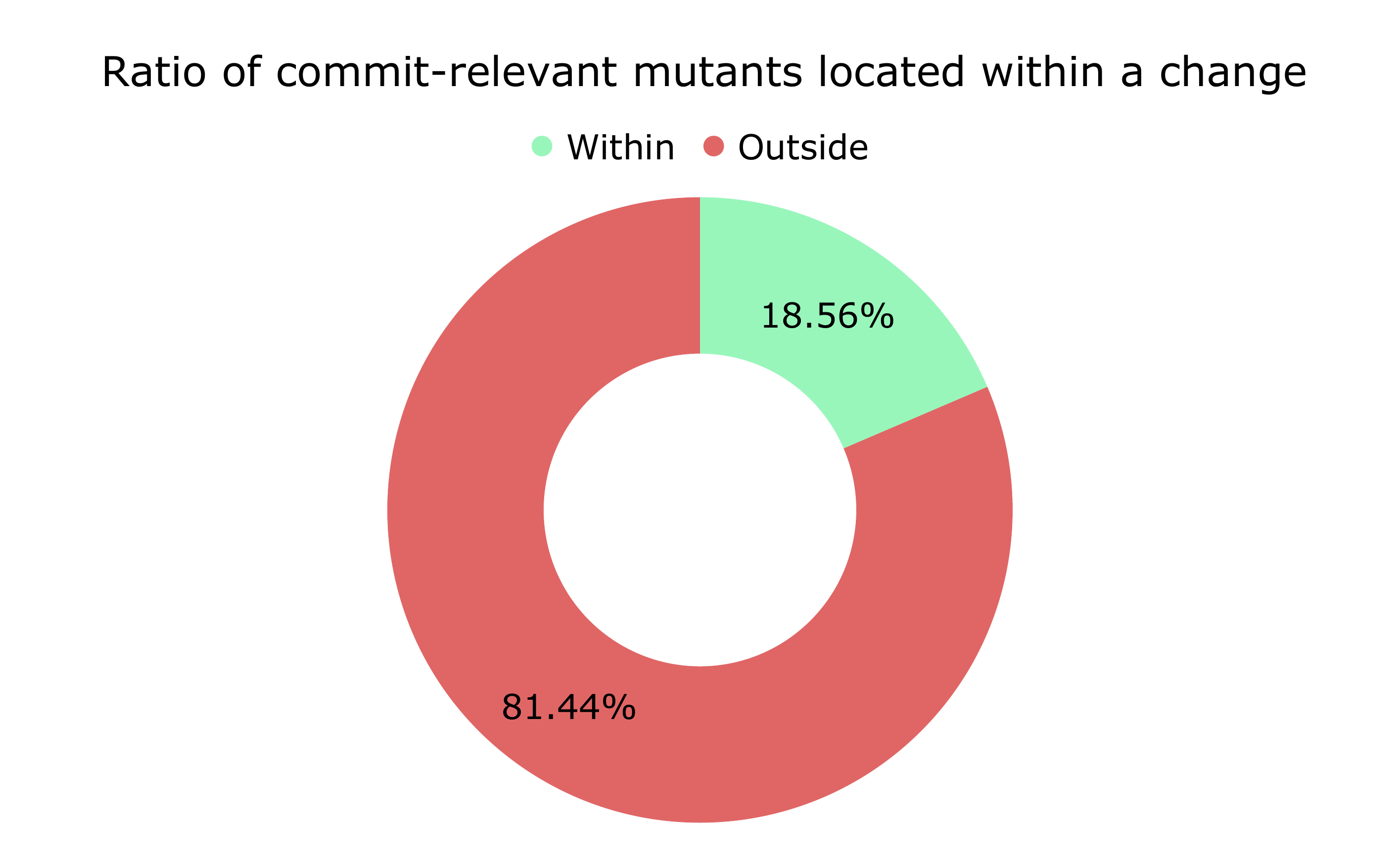}
      \captionof{figure}{Proportion of commit-relevant mutants within the commit (18.56\%) and outside the commit (81.44\%)}
      \label{fig:RQ1-ratio_ouside_methods_i}
    \end{minipage}%
    \hspace{0.4cm}
    \begin{minipage}{.45\textwidth}
      \centering
      \vspace{-0.4cm}
      \includegraphics[width=1.0\linewidth]{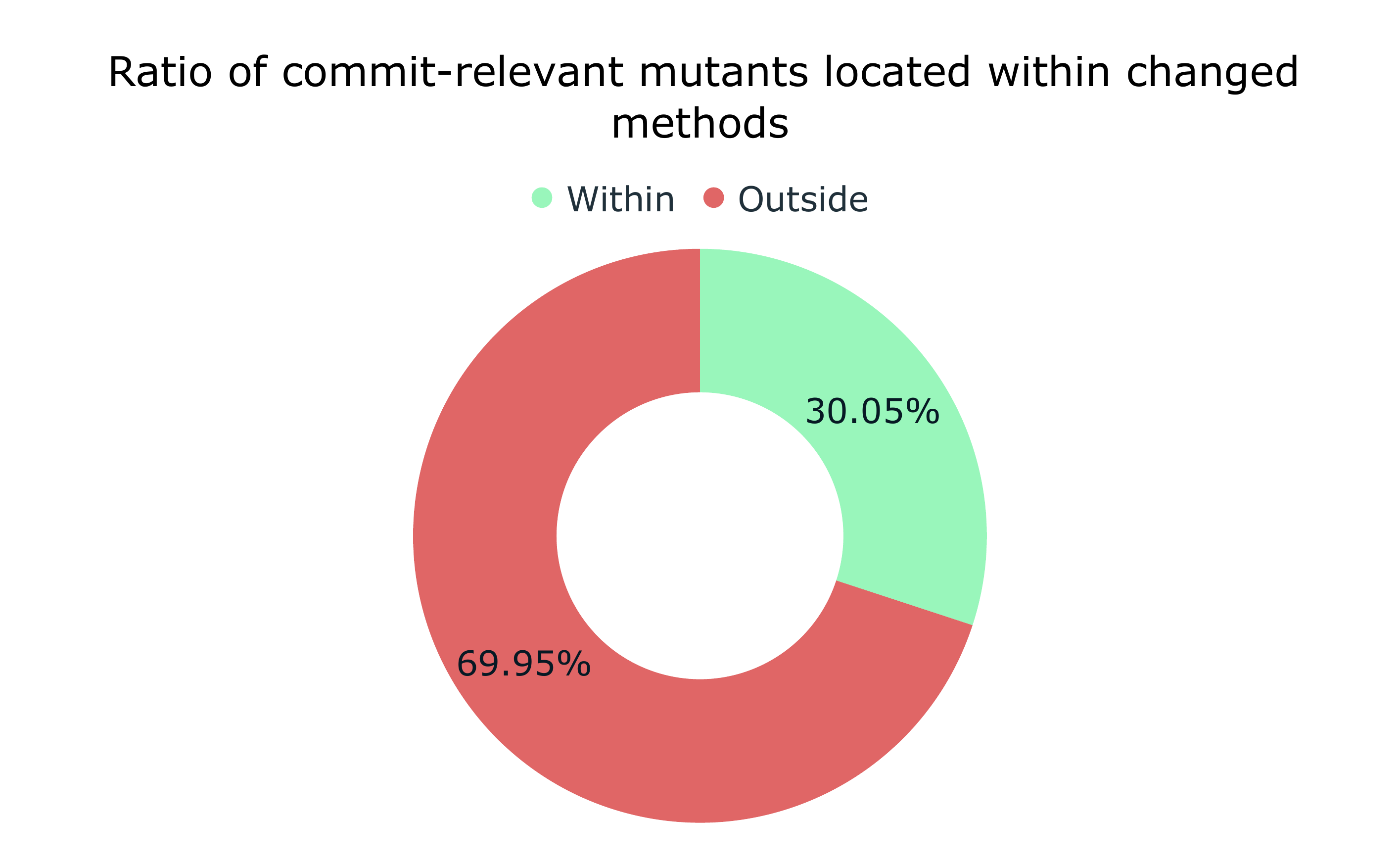}
      \captionof{figure}{Proportion of commit-relevant mutants within the change method (30.05\%) and the outside changed methods (69.95\%) }
      \label{fig:RQ1-ratio_ouside_methods_ii}
    \end{minipage}
    \end{figure}    
    

\subsubsection{\textit{\textbf{RQ1.1: }\revise{What is the proportion of commit-relevant mutants out of all mutants?}}}

\autoref{tab:prevalence-commit-relevant-mutants} and \autoref{fig:RQ1-Distribution_plot} illustrate the distribution of commit-relevant mutants among all mutants. 
In our evaluation, we found that \textit{only about one in three ($\approx$30\%) mutants are commit-relevant}, on average. In particular, we observed that only about 225 mutants are relevant to a commit out of 833 mutants, on average.  \revise{This implies that an \textit{effective commit-aware mutation testing technique can reduce significant mutation testing effort, both computational when executing mutants and manual when analysing mutants. }} In addition, we found some (21) outliers in our analysis of commit-aware mutants, see columns ``\# C. All R. M.'' (Number of Commits with all Relevant Mutants) and ``\# C. No R. M.'' (Number of Commits with No Relevant Mutants): In particular, we found that only 2.8\% of commits (8) had 100\% commit-relevant mutants, this portrays the importance of mutant selection for evolving software systems. On the other hand, our evaluation results show that \revise{in 4.5\% of the commits (13) we found no commit-relevant mutants outside the change}; this suggests that it is pertinent to develop commit-aware mutation testing techniques that discern relevant from non-relevant mutants. 
Overall, these findings demonstrate the importance of developing commit-aware test selection for evolving software systems, in particular, in selecting relevant mutants to reduce testing effort. 


\begin{result}
One in three (approximately 30\%) mutants are commit-relevant; hence, selecting commit-aware mutants can significantly reduce 
\revise{mutation testing cost}.
\end{result}

\subsubsection{\textit{\textbf{RQ1.2: }\revise{Where are commit-relevant mutants located in the program, i.e., how many commit-relevant mutants are within or outside the committed changes?}}} In our evaluation, most (81\%) commit-relevant mutants are outside of developers' committed changes (\textit{see} \autoref{fig:RQ1-ratio_ouside_methods_i}). Making only about one in five (19\%) commit-relevant mutants being within the committed changes of developers. For instance, a developer that tests \textit{all commit-relevant mutants within the changed method} will test only 30\% of commit-relevant mutants, and miss almost 70\% of commit-relevant mutants (\textit{see} \autoref{fig:RQ1-ratio_ouside_methods_ii}). This result suggests that to test the impact of developer changes on the program effectively, it is important to not only test within the committed changes. It is also highly pertinent to test the interaction of committed changes with the rest of the unmodified program.


\begin{result}
Most (81\% of) commit-relevant mutants are located outside of the commit, and \\only a few (19\% of) commit-relevant mutants are within the commit.  

\end{result}

\begin{figure}[htp]
    \centering
    \begin{minipage}{.3\textwidth}
      \centering
    \vspace{-0.4cm}
    \includegraphics[width=1.0\linewidth]{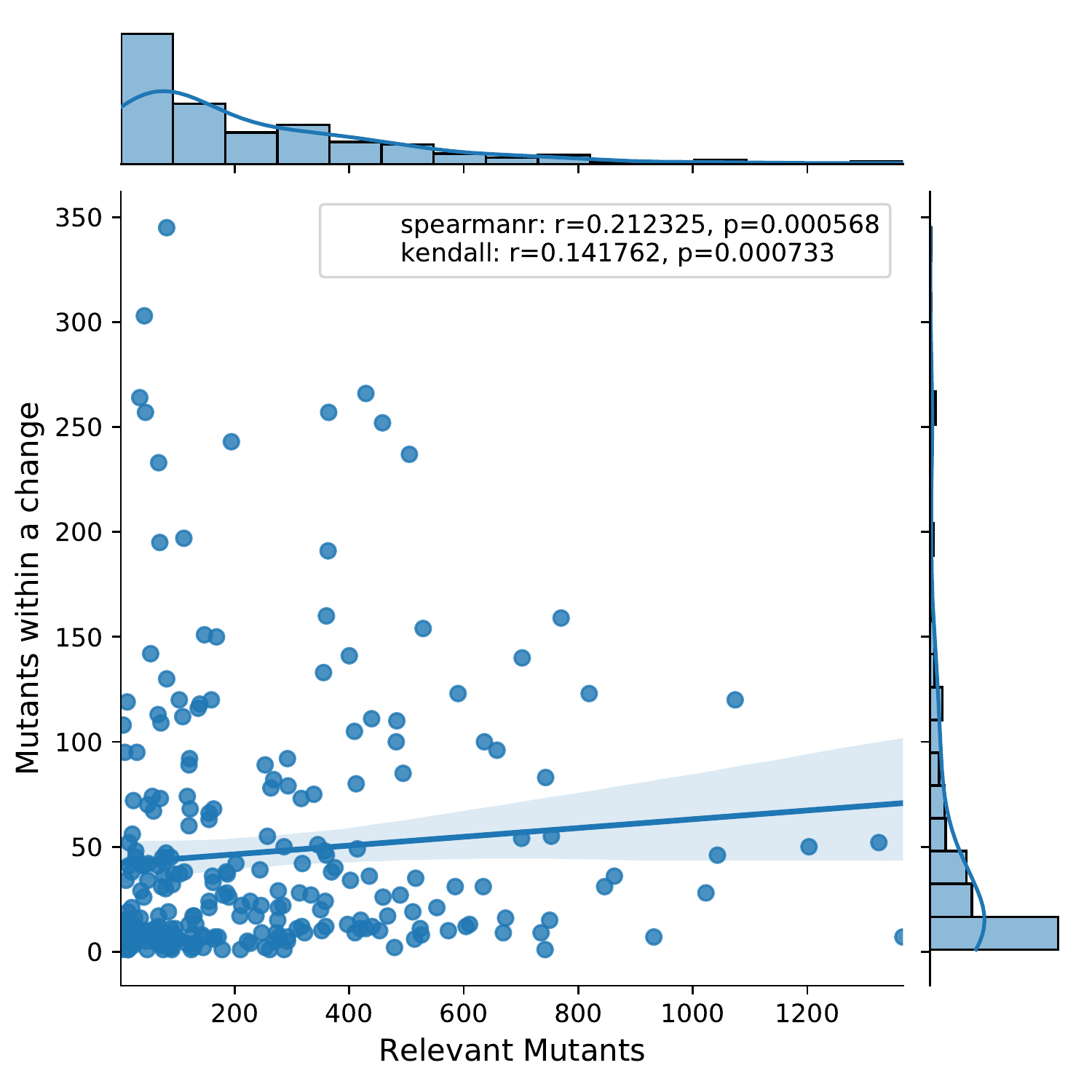}
      \captionof{figure}{
\revise{
      Correlation Analysis between the number \textit{mutants within a change} and the number of \textit{commit-relevant mutants}}
      }
      \label{fig:RQ1-correlation_plot_i}
    \end{minipage}%
    \hspace{0.2cm}
    \begin{minipage}{.3\textwidth}
      \centering
      \vspace{-0.4cm}
      \includegraphics[width=1.0\linewidth]{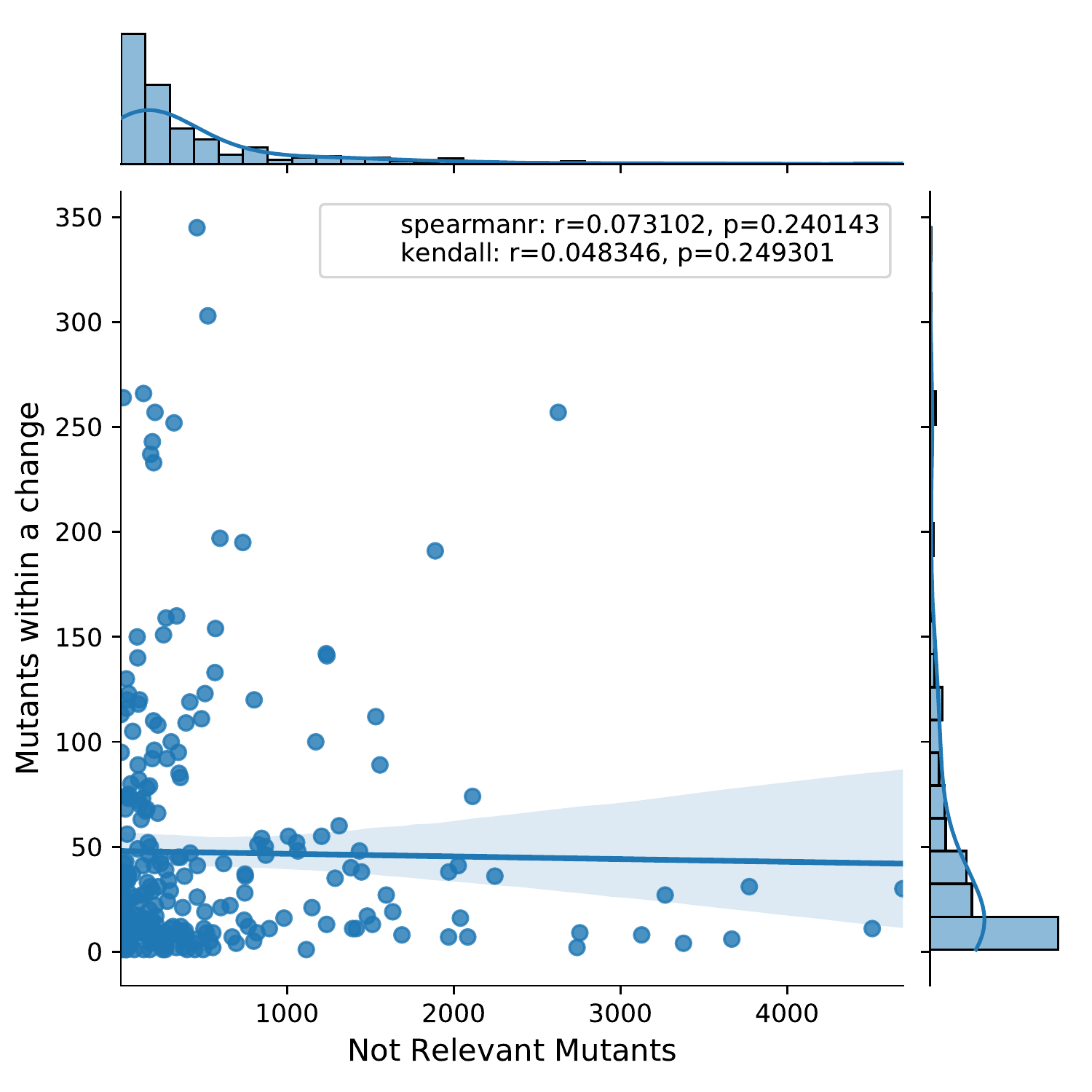}
      \captionof{figure}{
\revise{
      Correlation Analysis between the number of \textit{mutants within a change} and the number of \textit{non-relevant mutants}}
      }
      \label{fig:RQ1-correlation_plot_ii}
    \end{minipage}
        \hspace{0.2cm}
    \begin{minipage}{.3\textwidth}
      \centering
      \vspace{-0.4cm}
      \includegraphics[width=1.0\linewidth]{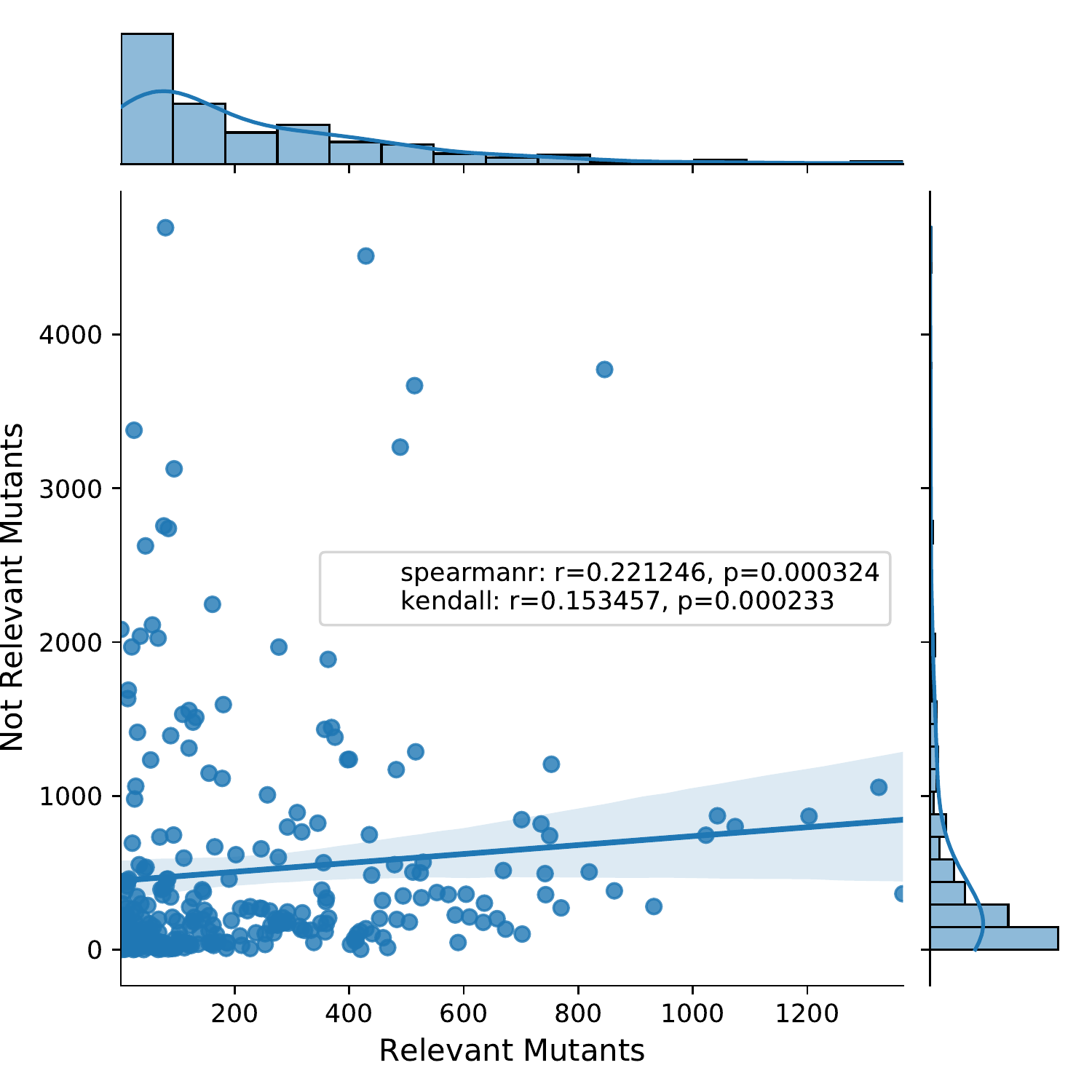}
      \captionof{figure}{
\revise{
      Correlation Analysis between the number of \textit{non-relevant mutants} and \textit{commit-relevant mutants}}
      }
      \label{fig:RQ1-correlation_plot_iii}
    \end{minipage}
    \end{figure}


\subsubsection{\textit{\textbf{RQ1.3: }\revise{Is there a correlation between the number of commit-relevant mutants and the number of mutants within the change?}}} Our evaluation results show that \textit{there is a weak trend between the number of commit-relevant mutants and the number of mutants within the commit}. Our statistical correlation analysis shows that there is a \textit{weak correlation} between both variables. \revise{In particular, we found a Spearman and Kendall correlation coefficients of 0.212 and 0.141, respectively. Indeed, both the Spearman and Kendall correlation coefficients are statistically significant (with p-values 0.0006 and 0.0007, respectively)}. \revise{Figures~\ref{fig:RQ1-correlation_plot_i}, \ref{fig:RQ1-correlation_plot_ii} and \ref{fig:RQ1-correlation_plot_iii} summarize the results of the different studied correlations.} 
These correlation results suggest that there is a weak relationship between the number of mutants within a change and the number of commit-relevant mutants, but \textit{no robust and predictable pattern or trend between both variables}. This implies that \textit{the number of mutants within the commit can not reliably predict the number of commit-relevant mutants} (in unmodified code regions), and vice versa. 

\begin{result}
There is a statistically significant weak positive correlation between the number of commit-relevant mutants and the number of mutants within the change (\revise{Spearman} and Kendall correlation coefficients of \revise{0.212} and 0.141, respectively). 
\end{result}

\smallskip\noindent
\subsection{\RQ2: Subsuming Commit Relevant Mutants} 

In this section, we investigate the prevalence of \emph{subsuming commit-relevant mutants} among \emph{commit-relevant mutants}. Estimating the proportion of subsuming commit-relevant mutants is important to demonstrate the further reduction (in number of mutants to analyse and test executions) achieved by ``selecting"  or ``optimizing'' for effectively identifying \emph{subsuming commit-relevant mutants}, in comparison to \emph{commit-relevant mutants}, \emph{subsuming mutants} and \emph{all mutants}. 
The two subsumption relations (i.e., one for the commit-relevant mutants and the other one for all mutants) are computed by following the definition introduced in Section~\ref{sec:subsuming_mutants}. 

Additionally, we examine the correlation between the number of subsuming commit-relevant mutants and the number of commit-relevant mutants within a change and subsuming mutants; this is important to determine if these variables are related can predict or serve as a proxy for determining subsuming commit-relevant mutants. 

\begin{figure*}[htp]
        \begin{center}
        \begin{minipage}{.45\textwidth}
      \centering
        \vspace{-0.55cm}
    \includegraphics[width=1.0\linewidth]{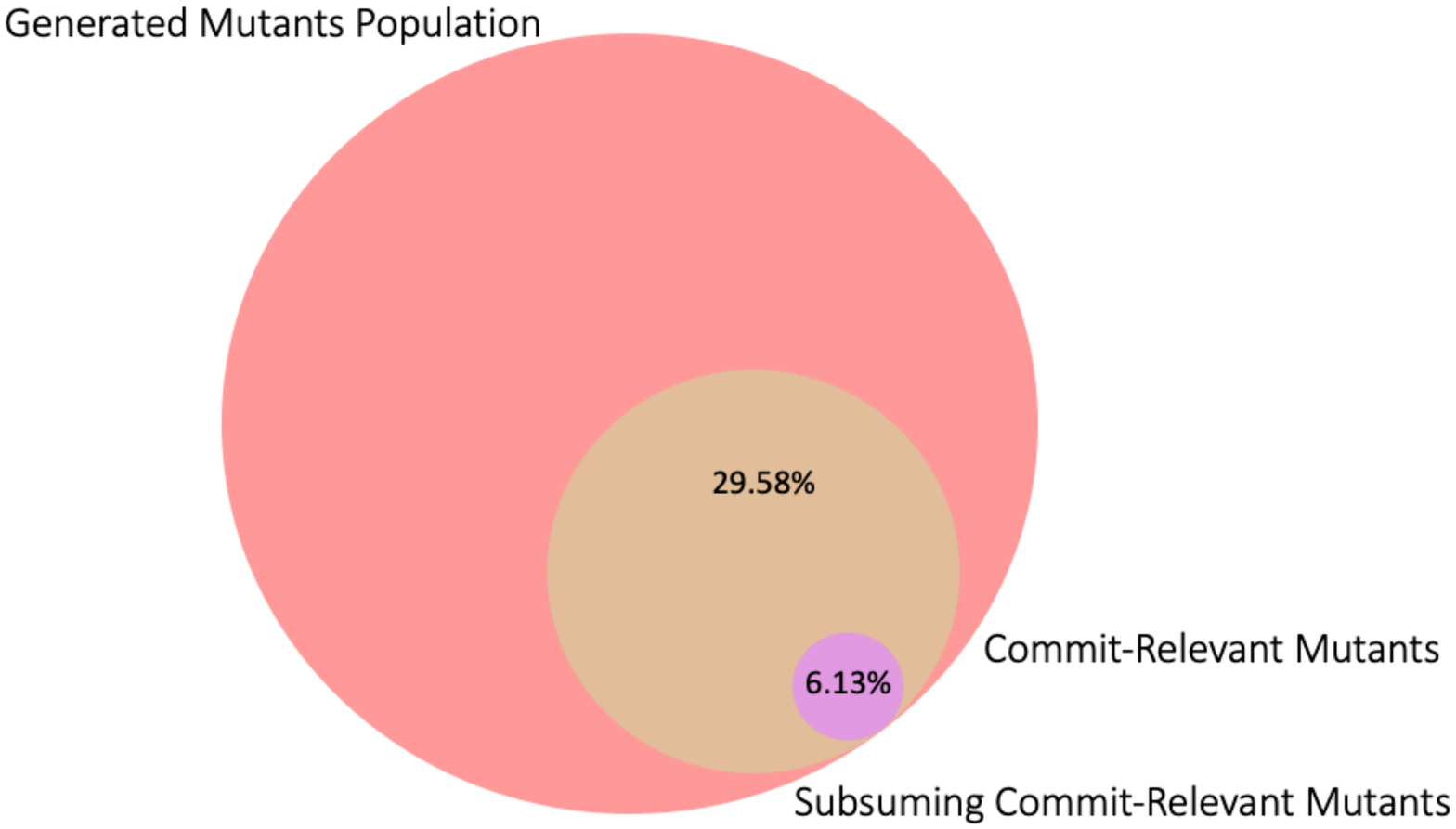}
      \captionof{figure}{Venn diagram showing the proportion of ``commit-relevant mutants''  (29.58\% in \textit{\color{orange} orange})) and ``subsuming commit-relevant mutants'' (6.13\% in \textit{\color{purple} purple}) among all mutants (in \textit{\color{pink} pink}).}
      \label{fig:venn_3a_categories_i}
    \end{minipage}%
    \hspace{0.2cm}
    \begin{minipage}{.45\textwidth}
      \centering
      \vspace{-0.4cm}
      \includegraphics[width=1.0\linewidth]{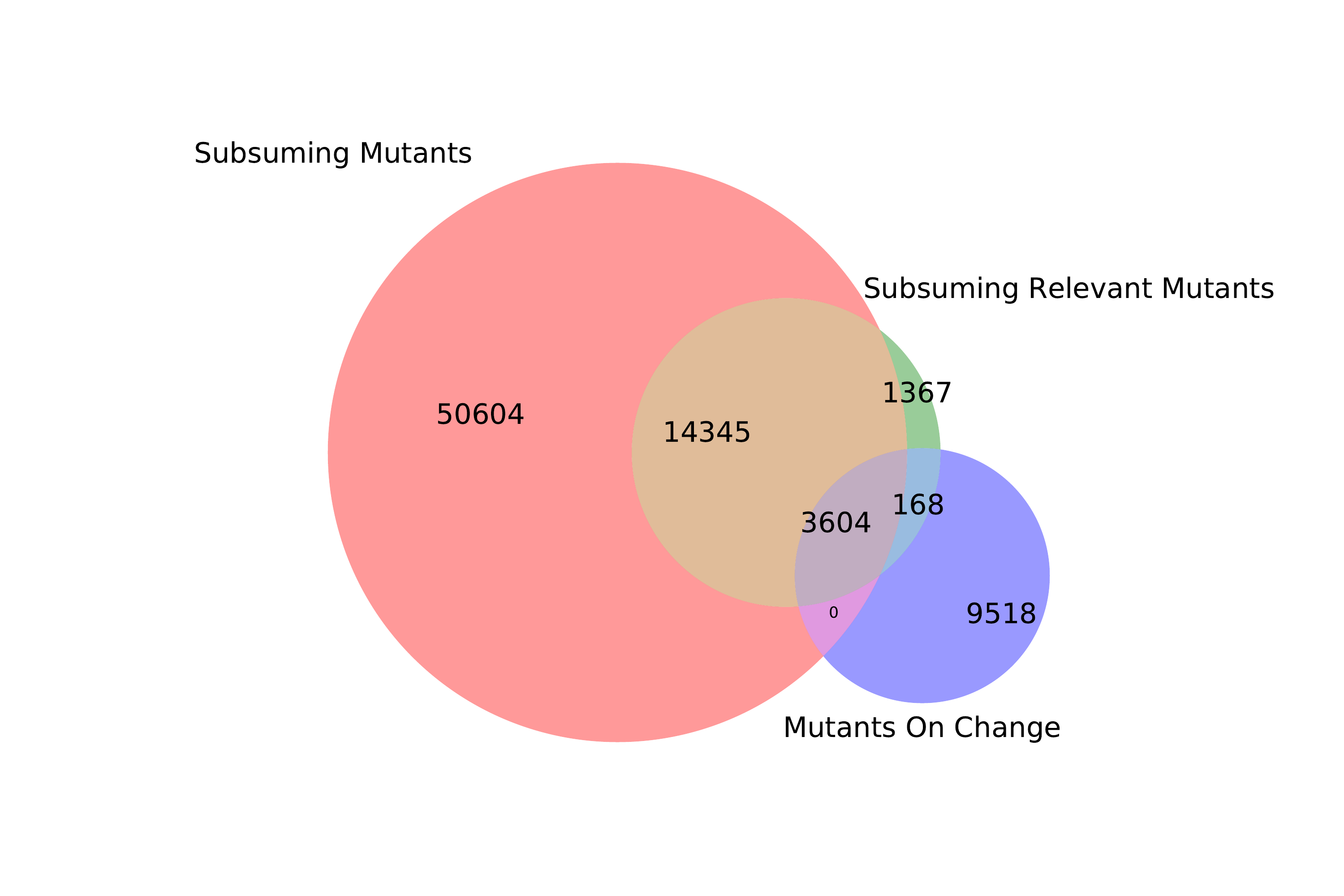}
      \captionof{figure}{Venn diagram showing the number and intersections among ``commit-relevant mutants within commit changes''  (in \textit{\color{blue} blue}),``subsuming commit-relevant mutants''  (in \textit{\color{orange} orange}) and ``subsuming mutants''  (in \textit{\color{pink} pink}). 
        }
      \label{fig:venn_3a_categories_ii}
    \end{minipage}
        \end{center}
    \end{figure*} 

\textit{What is the proportion of ``subsuming commit-relevant mutants'' among commit-relevant mutants, 
such that a test suite that distinguishes a(ll) subsuming commit-relevant mutant(s) covers (all) other commit-relevant mutants?} 
\autoref{fig:venn_3a_categories_i} illustrates the proportion of \emph{subsuming commit-relevant mutants} and their intersection with \emph{commit-relevant mutants} as well as all mutants. In our evaluation, we found that \textit{``subsuming commit-relevant mutants'' are significantly smaller than commit-relevant mutants and all mutants}. About one in 20 mutants is a \emph{subsuming commit-relevant mutant}, and about one in five (5) commit-relevant mutants is a subsuming commit-relevant mutant. Specifically, ``subsuming commit-relevant mutants'' represent 20.72\% and 6.13\% of all commit-relevant mutants and all mutants, respectively. This suggests it is worthwhile to identify and select subsuming relevant mutants from all (commit-relevant) mutants. Invariably, generating only subsuming commit-relevant mutants reduces the number of mutants to analyze by 79\% and 93\% compared to generating commit-relevant mutants and all mutants, respectively. This result implies that developing automated mutation testing methods that effectively identify, select or generate subsuming commit-relevant mutants can significantly reduce \revise{mutation testing cost}.  


\begin{result}
Selecting ``subsuming commit-relevant mutants'' can reduce the number of mutants to be considered by about 79\% and 93\%  in comparison to commit-relevant mutants and all mutants, respectively. 
\end{result}

\textit{What is the proportion of ``subsuming commit-relevant mutants'' among ``subsuming mutants'' and ``commit-relevant mutants within a change''?}
\autoref{fig:venn_3a_categories_ii} illustrates the intersections between all three types of mutants. Notably, \textit{most (92.98\% -- 18,117 out of 19,484) subsuming commit-relevant mutants are subsuming mutants as well, and they represent 26.42\% of all subsuming mutants (68,553)}. 
This implies that searching for subsuming commit-relevant mutants among subsuming mutants (instead of all mutants) is beneficial in reducing the search scope.   

We also observed that all subsuming commit-relevant mutants within committed changes are subsuming mutants. Meanwhile, about one in five (19.36\% -- 3,772 out of 19,484) subsuming commit-relevant mutants are within the developers' committed changes; they represent 28.38\% (3,772 out of 13,290) of all mutants within the change. This suggests that less than one in three mutants within the change are subsuming commit-relevant mutants. Hence, it is important to search for subsuming commit-relevant mutants outside of the committed changes since most subsuming commit-relevant mutants (81\%, 15,772) are outside the committed changes.

\begin{result}
  Most (92.98\% of) subsuming commit-relevant mutants are subsuming mutants, while a few (19.36\% of) subsuming commit-relevant mutants are located within committed changes.
\end{result}

    \begin{figure*}[htp]
        \begin{center}
        \begin{minipage}{.45\textwidth}
      \centering
        \vspace{-0.4cm}
    \includegraphics[width=1.0\linewidth]{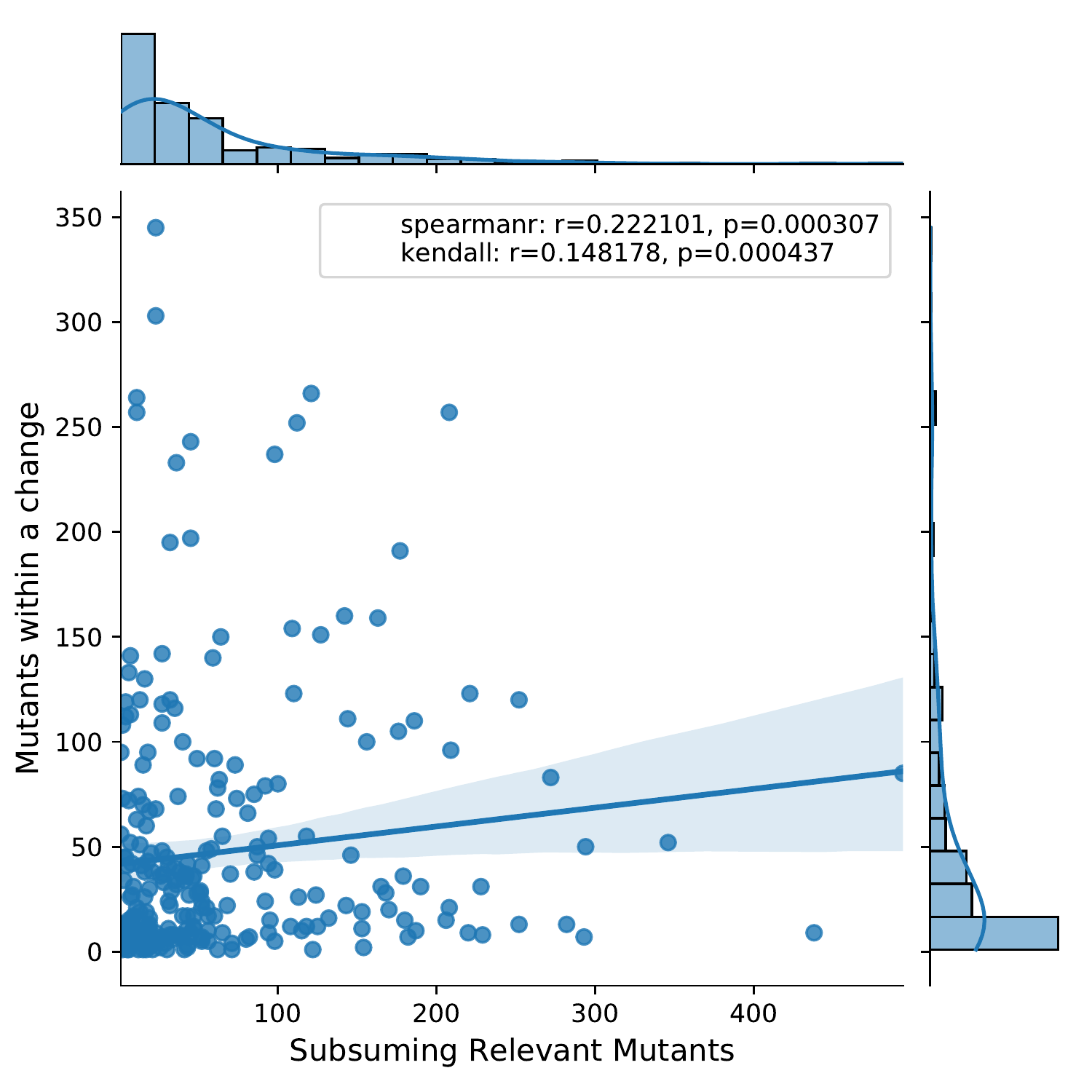}
      \captionof{figure}{ 
\revise{
      Correlation Analysis between the number mutants within a change and the number of subsuming commit-relevant mutants}
      }
      \label{fig:RQ2-correlation_plot_i}
    \end{minipage}%
    \hspace{0.2cm}
    \begin{minipage}{.45\textwidth}
      \centering
      \vspace{-0.4cm}
      \includegraphics[width=1.0\linewidth]{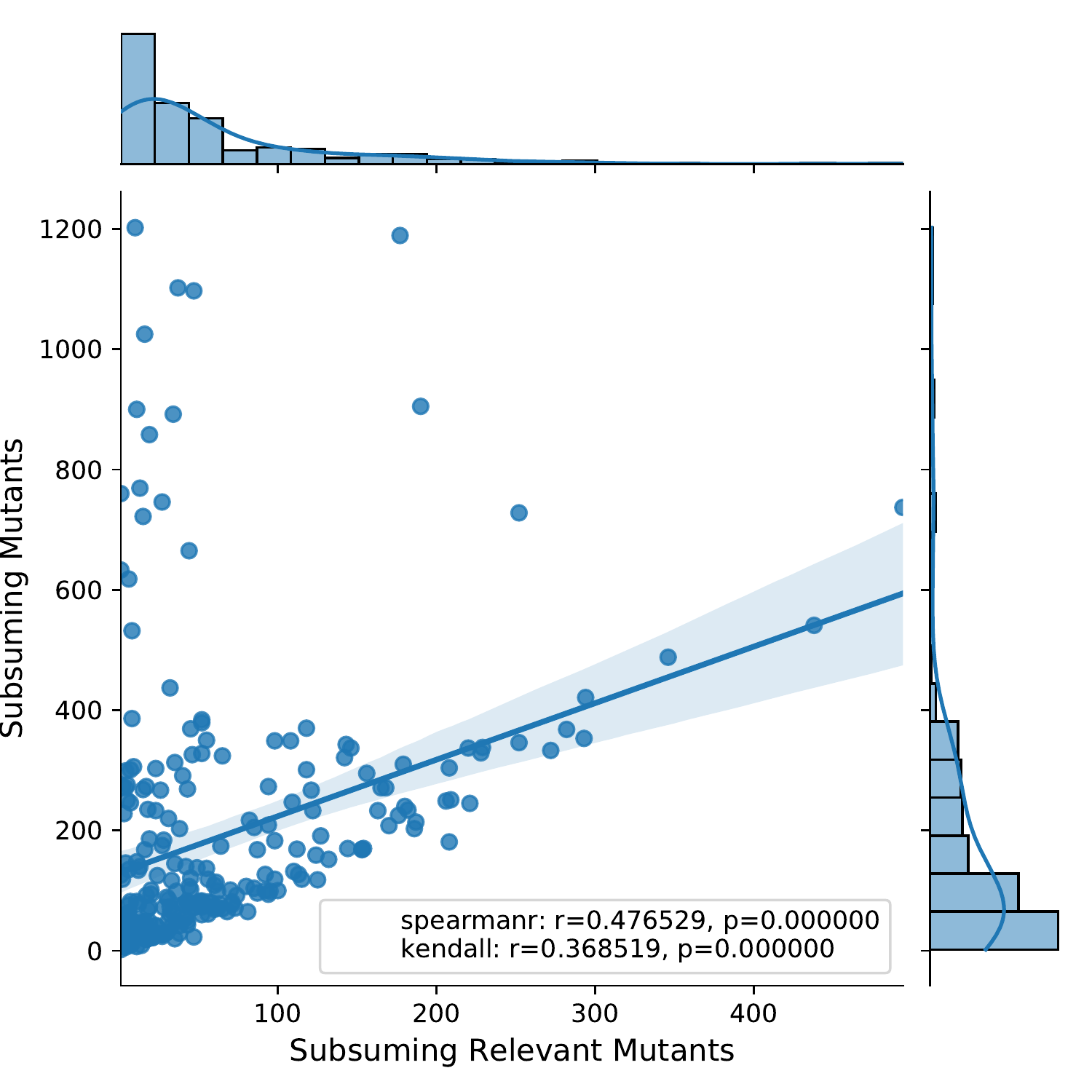}
      \captionof{figure}{ 
\revise{
      Correlation Analysis between the number of subsuming mutants and the number of subsuming commit-relevant mutants}
      }
      \label{fig:RQ2-correlation_plot_ii}
    \end{minipage}
        \end{center}
    \end{figure*}


\textit{Is there a correlation between the number of subsuming commit-relevant mutants and the number of mutants within a change?} Our correlation analysis shows that \textit{there is a weak positive correlation between the number of commit-relevant mutants within a change and the number of subsuming commit-relevant mutants} 
(\textit{see} \autoref{fig:RQ2-correlation_plot_i}). \revise{Both Spearman and Kendall correlation coefficients report a weak positive correlation, with correlation coefficients 0.222 and 0.148, respectively, (\textit{see} \autoref{fig:RQ2-correlation_plot_i}).} 
\revise{In particular, 
the correlation coefficients are statistically significant with p-values less than 0.05, specifically, 0.0003 and 0.0004 for Spearman and Kendall coefficients, respectively}. 
This result suggests that the number of mutants within a change can not strongly predict the number of \revise{subsuming commit-relevant mutants}; hence, it is important to identify all commit-relevant mutants that interact with the committed changes, and not only test the change itself.


\begin{result}
  The number of mutants within a change can not reliably predict the number of subsuming commit-relevant mutants 
  since there is only a weak positive correlation between both variables.  
\end{result}

    \begin{figure*}[bt!]
        \begin{center}
        \includegraphics[width=\textwidth]{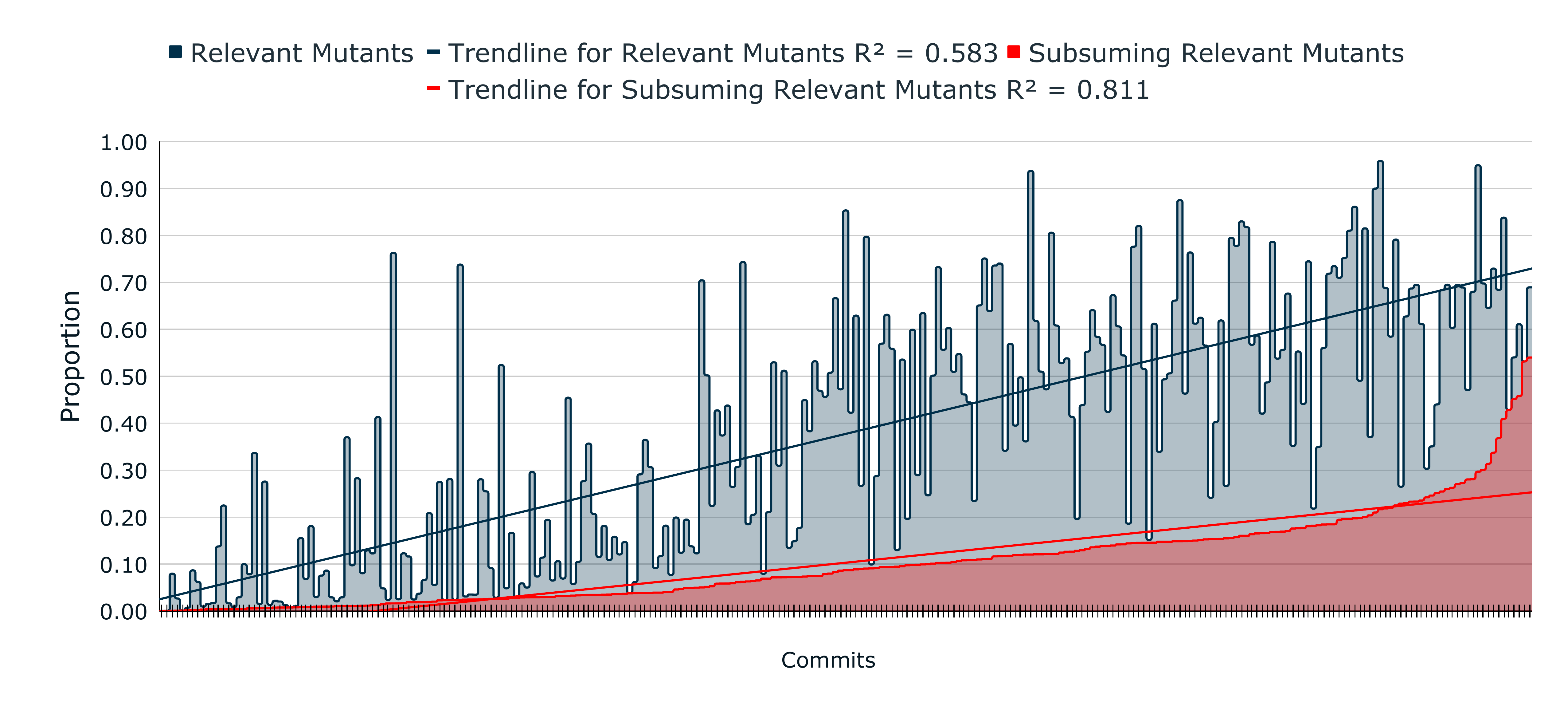}
        \caption{Distribution of the proportion of commit-relevant mutants (in \textit{\color{gray} gray}) and subsuming commit-relevant mutants (in \textit{\color{red} red}); Commits are sorted from left to right in ascending order of the proportion of subsuming relevant mutants
        }
        \label{fig:RQ2-Distribution_minimal_relevant}
        \end{center}
    \end{figure*}



\textit{What is the relationship between the number of subsuming commit-relevant mutants and the number of subsuming mutants?} \autoref{fig:RQ2-correlation_plot_ii} illustrates the distribution and correlation between the number of subsuming mutants and the number of subsuming commit-relevant mutants. In this figure, the trending line shows that \textit{there is a moderate positive correlation between both variables}. \revise{Indeed, both Spearman and Kendall correlation coefficients reports a \textit{moderate positive relationship} between both variables, with correlation coefficients 0.476 and 0.368, respectively, (\textit{see} \autoref{fig:RQ2-correlation_plot_ii}). The correlation coefficients also show that the positive relationship is statistically significant (p-value < 0.05)}. As expected, we observed that the proportion of subsuming relevant mutants per commit increases  (trendline $R^2$=0.881)  as the proportion of commit-relevant mutants increases (\textit{see} \autoref{fig:RQ2-Distribution_minimal_relevant}). 
Overall, this result implies that these variables can serve as a proxy to each other, hence predicting one variable could help identify the other. In particular, this implies that selecting subsuming mutants significantly increases the chances of selecting subsuming commit-relevant mutants. 

\begin{result}
There is a moderate positive relationship between the number of subsuming commit-relevant mutants and the number of subsuming mutants, such that one can predict the other and vice versa.
\end{result}

\smallskip\noindent
\subsection{\RQ3: Commit Size}     

In this section, we investigate if there is a relationship between the number of (subsuming) commit-relevant mutants and the size of the commit, measured in terms of the number of commit hunks. 

In particular, we pose the following question: \textit{Is there a relationship between the number of commit hunks and the number of (subsuming) commit-relevant mutants?}
 
\begin{figure*}[htp]
        \begin{center}
        \begin{minipage}{.45\textwidth}
      \centering
        \vspace{-0.4cm}
    \includegraphics[width=1.0\linewidth]{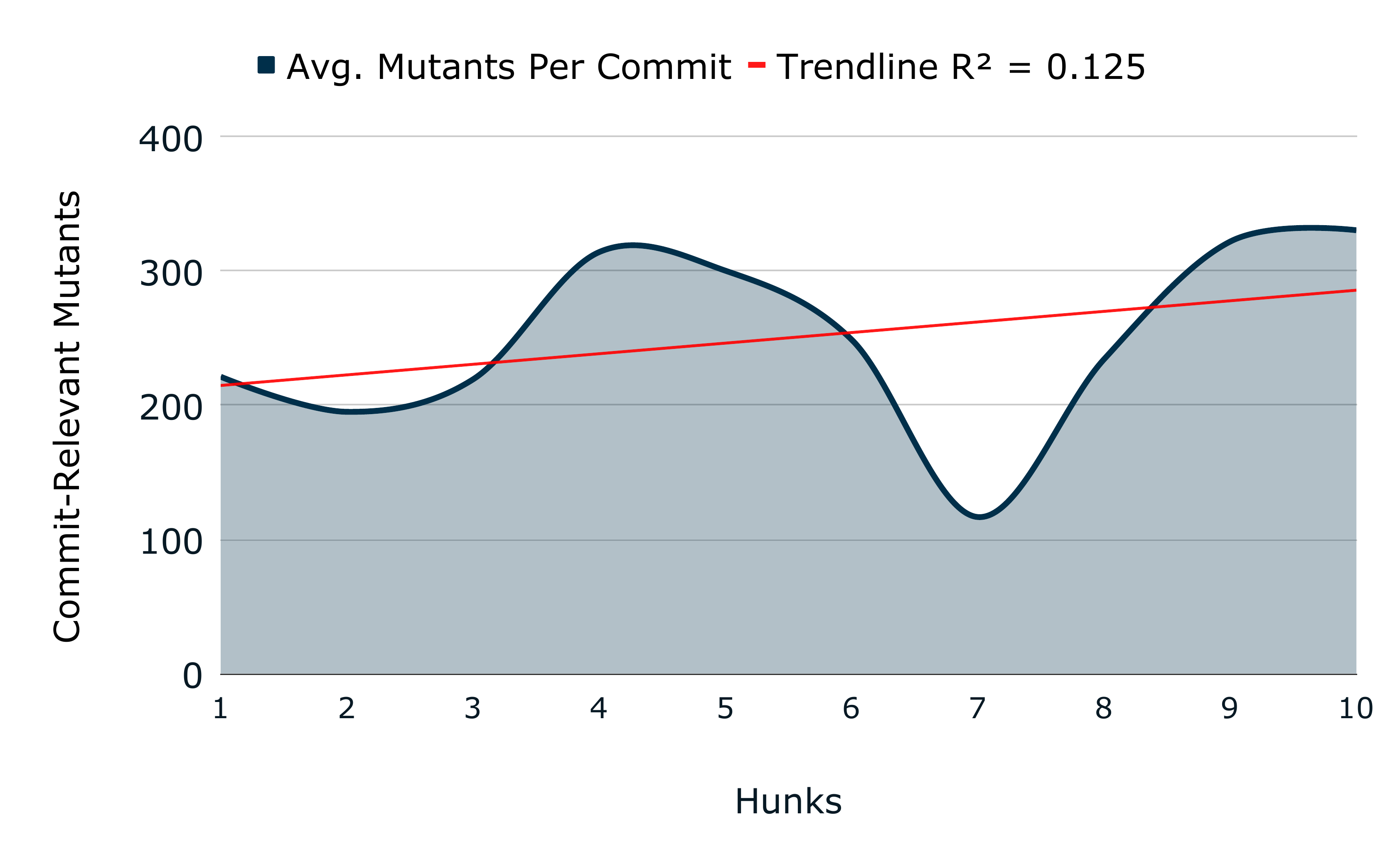}
      \captionof{figure}{Average Number of Commit-relevant mutants per commit}
      \label{fig:RQ3-distribution_hunks_i}
    \end{minipage}%
    \hspace{0.2cm}
    \begin{minipage}{.45\textwidth}
      \centering
      \vspace{-0.4cm}
      \includegraphics[width=1.0\linewidth]{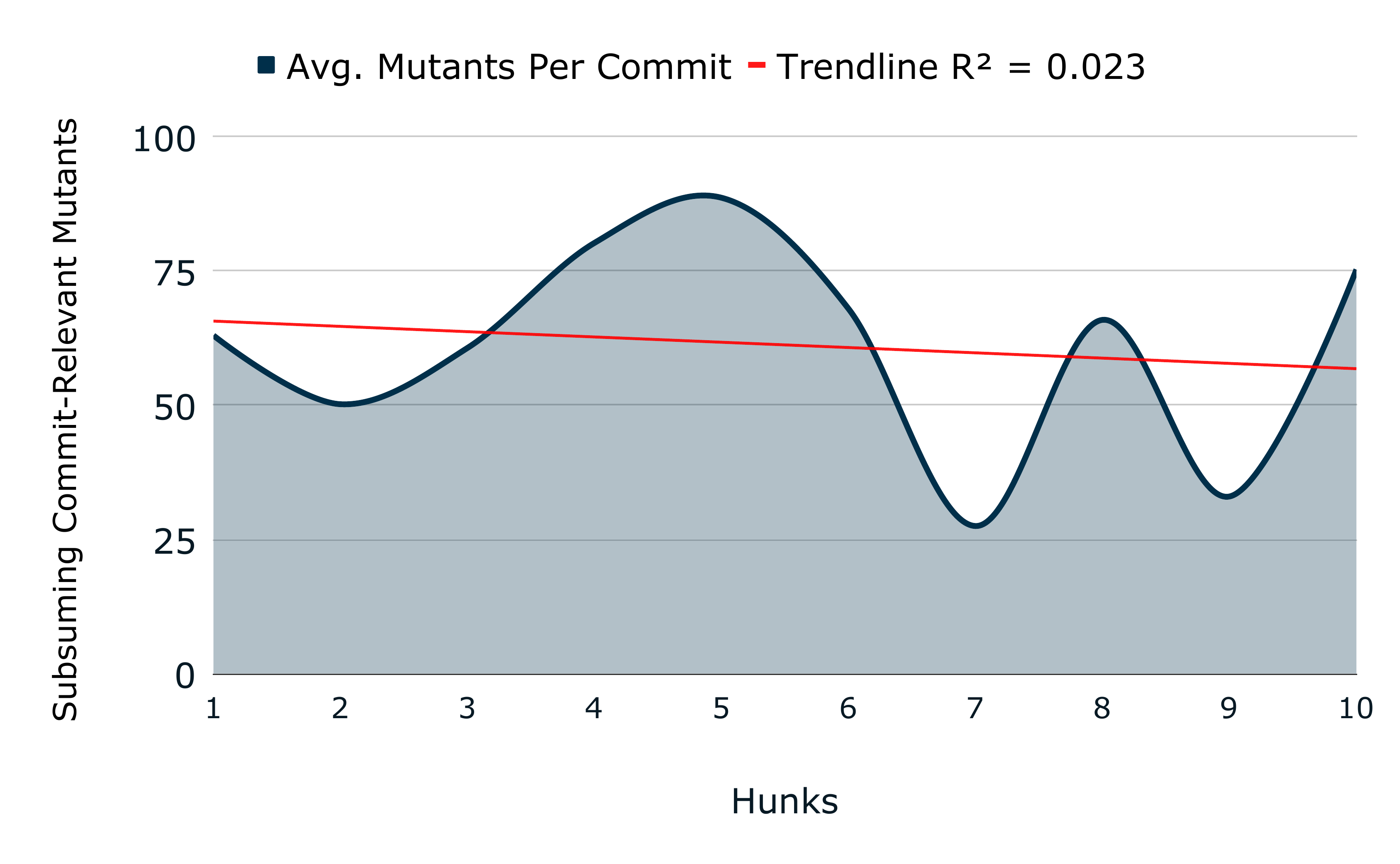}
      \captionof{figure}{Average Number of Subsuming Commit-relevant mutants per commit}
      \label{fig:RQ3-distribution_hunks_ii}
    \end{minipage}
        \end{center}
    \end{figure*}

\autoref{fig:RQ3-distribution_hunks_i} illustrates the relationship between the number of commit-relevant mutants and the number of commit hunks. For commit-relevant mutants, we found that the number of \emph{commit-relevant mutants} (moderately) \emph{increases} (trendline $R^2$=0.125) as the number of commit-hunks increases. This implies that there is \emph{positive direct relationship} between the size of the commit and the number of \emph{commit-relevant mutants}. However, \autoref{fig:RQ3-distribution_hunks_ii} shows that the number of \emph{subsuming commit-relevant mutants} (moderately) \emph{decreases} (trendline $R^2$=0.023) as the number of commit-hunks increases. 
These results suggest that there is an \emph{indirect relationship} between the size of the commit and the number of \emph{subsuming commit-relevant mutants}. The size of the commit does not directly predict the number of subsuming commit-relevant mutants. Indeed, the number of \emph{subsuming commit-relevant mutants} decreases as the average size of the commit increases. Overall, this result demonstrates the effectiveness and importance of \emph{subsuming commit-relevant mutants}  in reducing testing effort, even for large commit changes. 


\begin{result}
  The number of ``commit-relevant mutants'' increases as the size of the commit increases; however, the number of ``subsuming commit-relevant mutants'' decreases as the size of the commit increases. 
\end{result}
   
    \begin{figure*}[bt!]
        \begin{center}
        \begin{minipage}{.45\textwidth}
      \centering
        \vspace{-0.4cm}
    \includegraphics[width=1.0\linewidth]{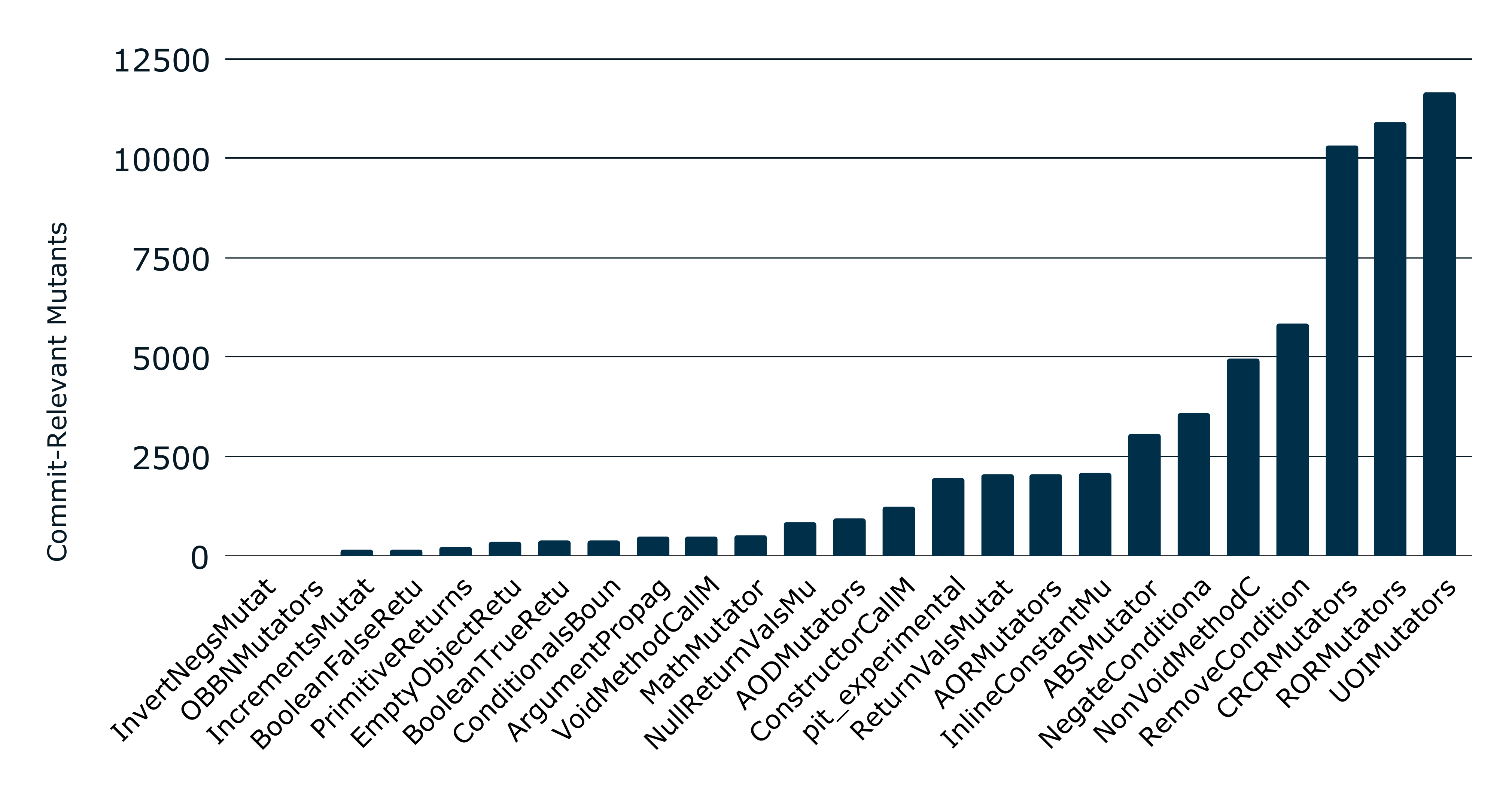}
      \captionof{figure}{Prevalence of Commit-relevant Mutant Types}
      \label{fig:RQ4-test_mutants_operators}
    \end{minipage}%
    \hspace{0.2cm}
    \begin{minipage}{.45\textwidth}
      \centering
      \vspace{-0.4cm}
      \includegraphics[width=1.0\linewidth]{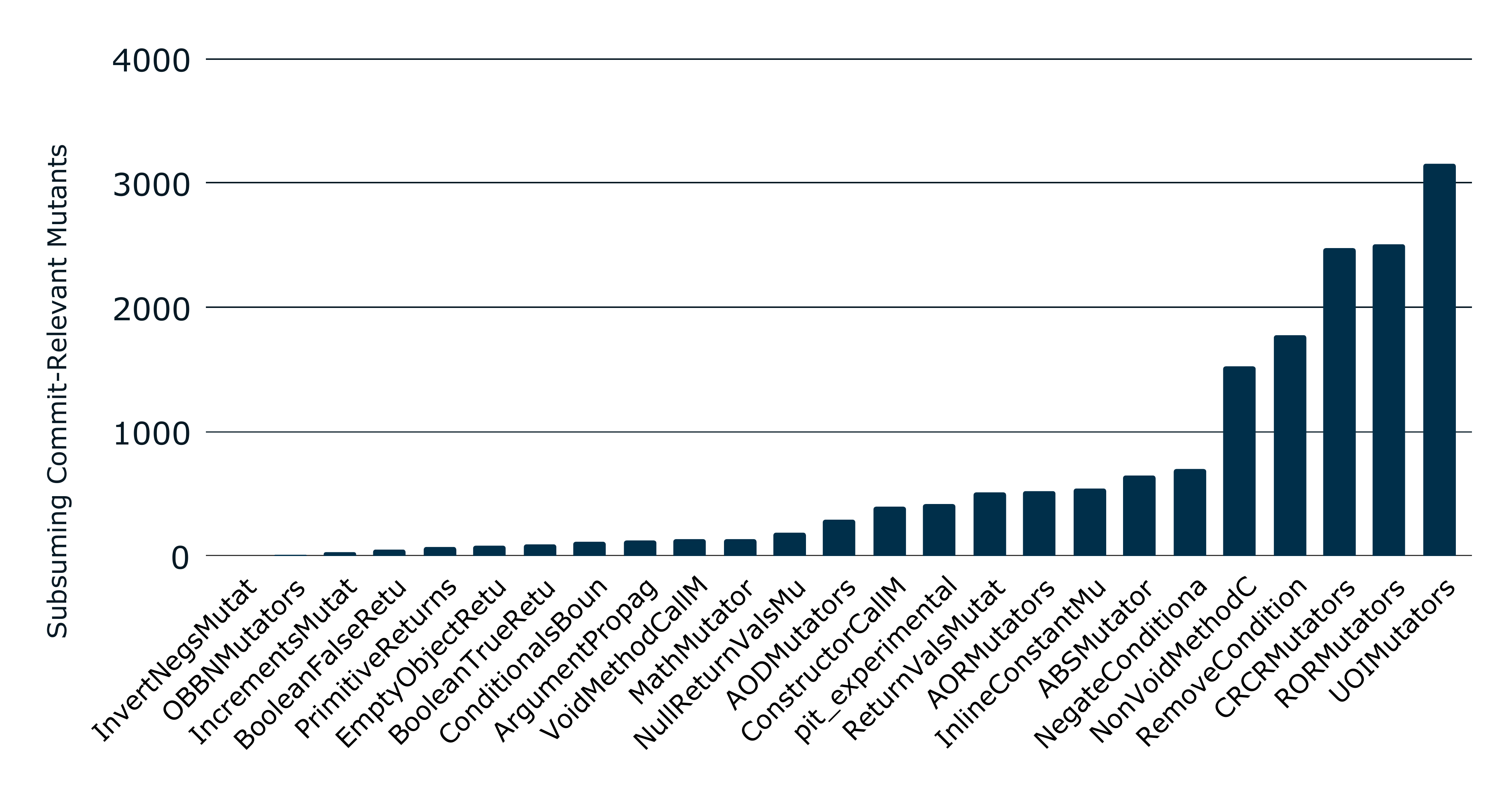}
      \captionof{figure}{Prevalence of Subsuming Commit-relevant Mutant Types}
      \label{fig:RQ4-test_sub_mutants_operators}
    \end{minipage}
        \end{center}
    \end{figure*}

    \begin{figure*}[bt!]
        \begin{center}
        \begin{minipage}{.45\textwidth}
      \centering
        \vspace{-0.4cm}
    \includegraphics[width=1.0\linewidth]{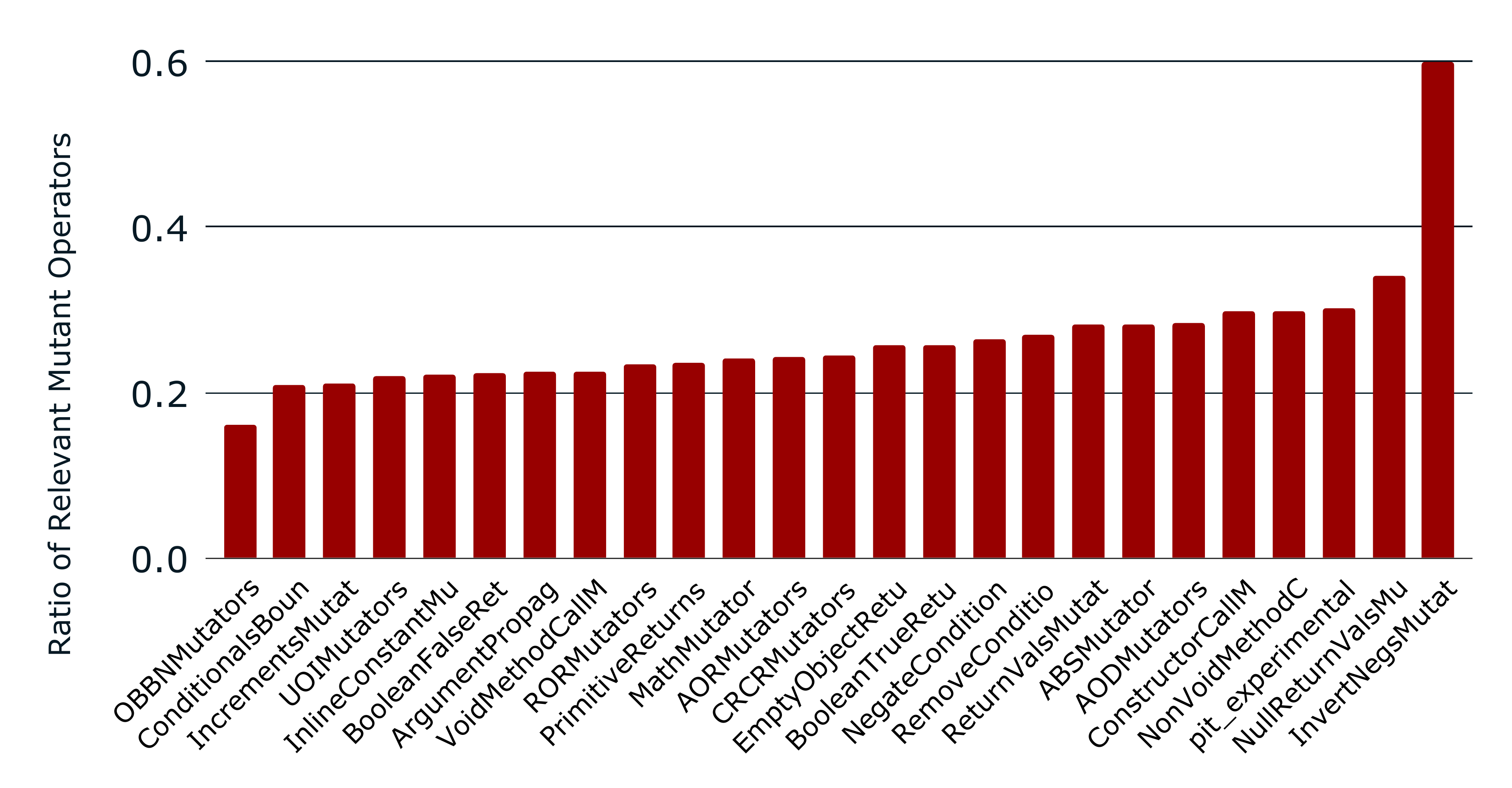}
      \captionof{figure}{Ratio of Commit-relevant Mutants over All Mutants per Mutant Type}
      \label{fig:RQ4-ratio_relevant_mutants_operators}
    \end{minipage}%
    \hspace{0.2cm}
    \begin{minipage}{.45\textwidth}
      \centering
      \vspace{-0.4cm}
      \includegraphics[width=1.0\linewidth]{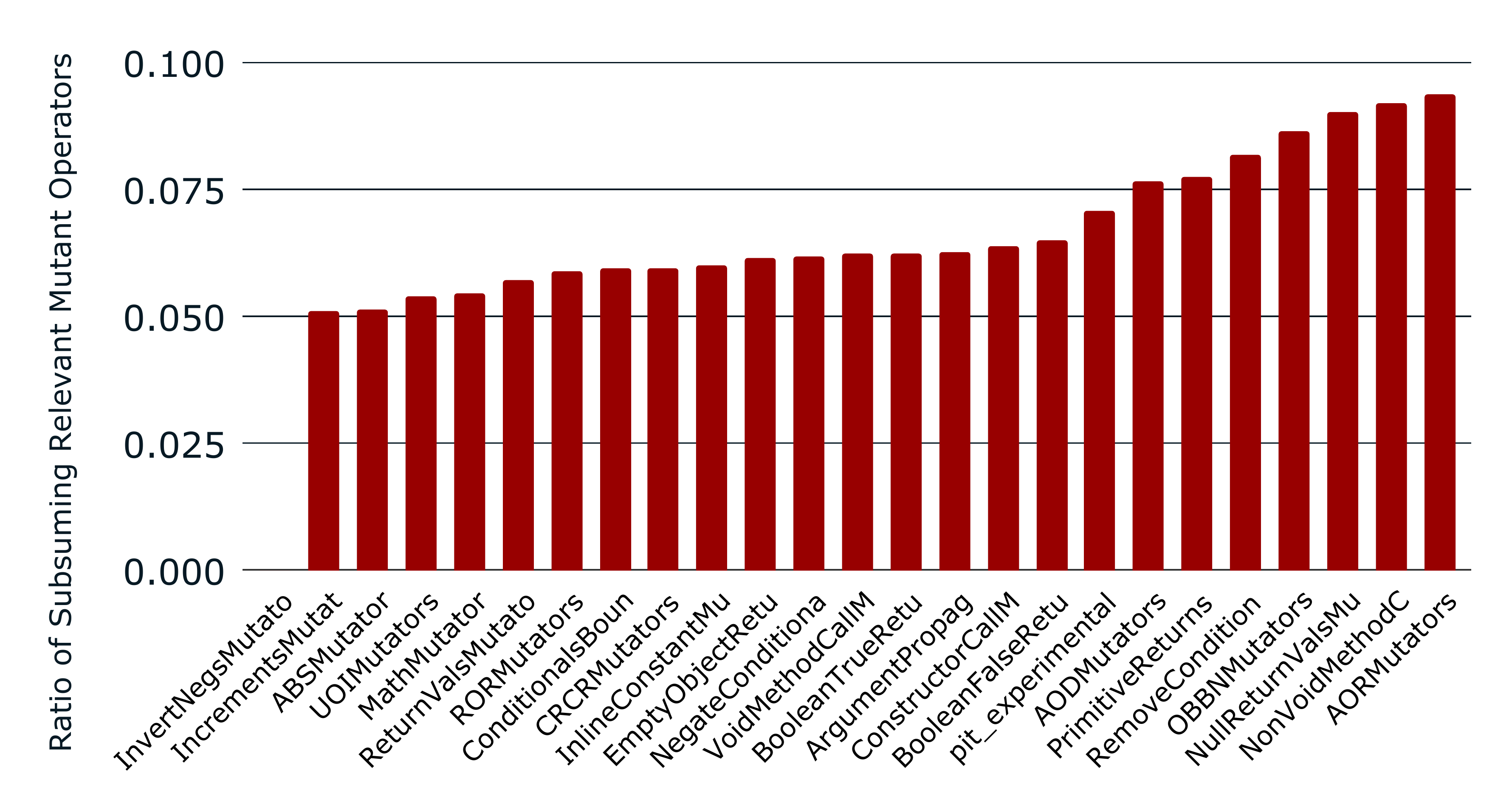}
      \captionof{figure}{Ratio of Subsuming Commit-relevant Mutants over All Mutants per Mutant Type}
      \label{fig:RQ4-ratio_sub_mutants_operators}
    \end{minipage}
        \end{center}
    \end{figure*}

\smallskip\noindent
\subsection{\RQ4: Commit-relevant Mutant Types}     


Let us investigate the prevalence of \emph{mutant types} among (subsuming) commit-relevant mutants, using 25 distinct mutant group types from Pitest~\cite{pitest}. This is important to determine whether the generation, selection or identification of commit-relevant mutants can be improved by focusing on specific mutant types.  

\textit{What is the prevalence of mutant types among (subsuming) commit-relevant mutants?} 
\autoref{fig:RQ4-test_mutants_operators} illustrates the prevalence of mutant types among commit-relevant mutants. 
Our evaluation results show that some mutant types are highly prevalent, such as \textit{Unary Operator Insertion Mutators (UOIMutators)}, \textit{Relational Operator Replacement Mutator (RORMutators)} and \textit{Constant Replacement Mutator (CRCRMutators)}. 
\revise{
On one hand, UOIMutators inject a unary operator (increment or decrement) on a variable, this may affect the values of local variables, arrays, fields, and parameters~\cite{pitest}, while RORMutators replace a relational operator with another one, e.g., ``$<$'' with ``$>$'' or ``$<=$'' with ``$<$''.  
On the other hand, CRCRMutators mutates inline constants. For further detauls about the mutant types, the table of constants and other mutation operators can be found in the official PiTest documentation\footnote{http://pitest.org/quickstart/mutators/}.}
Specifically, 50.77\% of the commit-relevant mutants are of one of these three mutant types. 
This is mainly related to the fact these three mutation operators produced the majority (54.5\%) of the mutants considered in our study. 
Precisely, \autoref{fig:RQ4-ratio_relevant_mutants_operators} shows that the \emph{distribution} of commit-relevant mutants is clearly \emph{uniform} per mutant type. That is, in general, between 20\% and 30\% of the mutants for each type result to be commit-relevant. This indicates that mutants type does not increase or reduce the chances for mutants of being commit-relevant. 
The outliers of \autoref{fig:RQ4-ratio_relevant_mutants_operators}, corresponding to mutant types \revise{Bitwise Operator Mutator} (\textit{OBBNMutators}) and \revise{Invert Negatives Mutator} (\textit{InvertNegsMutat}), are because of the low number of mutants for these types: 13 out of 81 (16\%) mutants are commit-relevant in the case of \textit{OBBNMutators} mutant type, while 3 out of 5 (60\%) mutants are commit-relevant for \textit{InvertNegsMutat} mutant type. 
\revise{
In particular, OBBNMutators mutates (i.e., reverses) bitwise ``AND'' (\&) and ``OR'' ($|$) operators, 
while InvertNegsMutat operators inverts the negation of integers and floating-point numbers.}

Similarly, Figures \ref{fig:RQ4-test_sub_mutants_operators} and \ref{fig:RQ4-ratio_sub_mutants_operators} show that the ratio of subsuming commit-relevant mutants per mutant type follows a uniform distribution as well. Typically, between 5-7\% of the mutants per mutant type turn to be subsuming commit-relevant. 
The outlier of \autoref{fig:RQ4-ratio_sub_mutants_operators} corresponds to \textit{InvertNegsMutat} mutant type, where none of the 3 commit-relevant mutants identified for this mutant type are subsuming (because of mutants of a different mutant type subsume them). 

\begin{result}
  The distribution of (subsuming) commit-relevant mutants per mutant type is uniform. Typically, between 20-30\% (5-7\%) of the mutants per mutant type are (subsuming) commit-relevant.
\end{result}

\smallskip\noindent
\subsection{\RQ5: Effectiveness of Commit-relevant Mutants Selection}

This section simulates a mutation testing scenario where the tester selects a mutant for analysis for which a test to kill it is developed. 
Note that a test case that is designed to kill a mutant may collaterally kill other mutants. 
Consequently, opening a space to examine the effectiveness of the test suites developed when guided by different mutant selection strategies. Accordingly, this study compares the following mutant selection strategies: “random mutants selection,” “mutants within a change,” and (subsuming) commit-relevant mutants.  
We measure their effectiveness in terms of the \textit{Relevant Mutation Score} (RMS) and \textit{Minimal-Relevant Mutation Score} (RMS*), which intuitively measures the number of  (subsuming) commit-relevant mutants killed by the different test suites. Specifically, we investigate the extent to which selecting and killing each aforementioned mutant types improves the test suite quality, in terms of the number of (subsuming) commit-relevant mutants killed by the test suite.  
Then we pose the question: \textit{How many (subsuming) commit-relevant mutants are killed if a developer or test generator selects and kills random mutants or only mutants within a change?} 

\begin{table}[]
    \caption{Comparative Effectiveness of selecting and killing (subsuming) commit-relevant mutants in comparison to ``\revise{all mutants}'' and ``mutants within a change'' by observing RMS (Relevant Mutation Score) and RMS* (Subsuming Relevant Mutation Score)}
    \resizebox{\textwidth}{!}{
    \begin{tabular}{c|cccccccccc|cccccccccc|}
    \cline{2-21}
     & \multicolumn{10}{c|}{\textbf{RMS}} & \multicolumn{10}{c|}{\textbf{RMS*}} \\ \hline
    \multicolumn{1}{|c|}{\textbf{Selection Strategy/Interval}} & \textit{\textbf{2}} & \textit{\textbf{4}} & \textit{\textbf{6}} & \textit{\textbf{8}} & \textit{\textbf{10}} & \textit{\textbf{12}} & \textit{\textbf{14}} & \textit{\textbf{16}} & \textit{\textbf{18}} & \textit{\textbf{20}} & \textit{\textbf{2}} & \textit{\textbf{4}} & \textit{\textbf{6}} & \textit{\textbf{8}} & \textit{\textbf{10}} & \textit{\textbf{12}} & \textit{\textbf{14}} & \textit{\textbf{16}} & \textit{\textbf{18}} & \textit{\textbf{20}} \\ \hline
    \multicolumn{1}{|c|}{\textbf{Random}} & 46.67 & 70.59 & 82.42 & 88.10 & 91.95 & 95.26 & 95.74 & 96.85 & 97.50 & 98.05 & 11.11 & 35.00 & 54.17 & 66.13 & 75.00 & 81.25 & 85.71 & 88.24 & 90.89 & 92.76 \\ \cline{1-1}
    \multicolumn{1}{|c|}{\textbf{Within a change}} & 46.48 & 59.52 & 65.91 & 67.48 & 68.42 & 69.47 & 69.96 & 70.18 & 70.40 & 71.09 & 11.95 & 25.00 & 28.95 & 32.31 & 33.33 & 33.67 & 34.38 & 34.88 & 35.23 & 35.29 \\\cline{1-1}
    \multicolumn{1}{|c|}{\textbf{Commit-Relevant}} & 75.00 & 95.05 & 100 & 100 & 100 & 100 & 100 & 100 & 100 & 100 & 40.74 & 83.72 & 100 & 100 & 100 & 100 & 100 & 100 & 100 & 100 \\ \cline{1-1}
    \multicolumn{1}{|c|}{\textbf{Subsuming Commit-Relevant}} & 80.00 & 98.51 & 100 & 100 & 100 & 100 & 100 & 100 & 100 & 100 & 65.75 & 95.35 & 100 & 100 & 100 & 100 & 100 & 100 & 100 & 100 \\ 
     \hline
    \end{tabular}
    }
    \label{tab:RQ5-median-developer_simulation}
    \end{table}

   \begin{figure*}[bt!]
        \centering
        \begin{subfigure}[t]{0.49\textwidth}
        \centering
        \includegraphics[width=\textwidth]{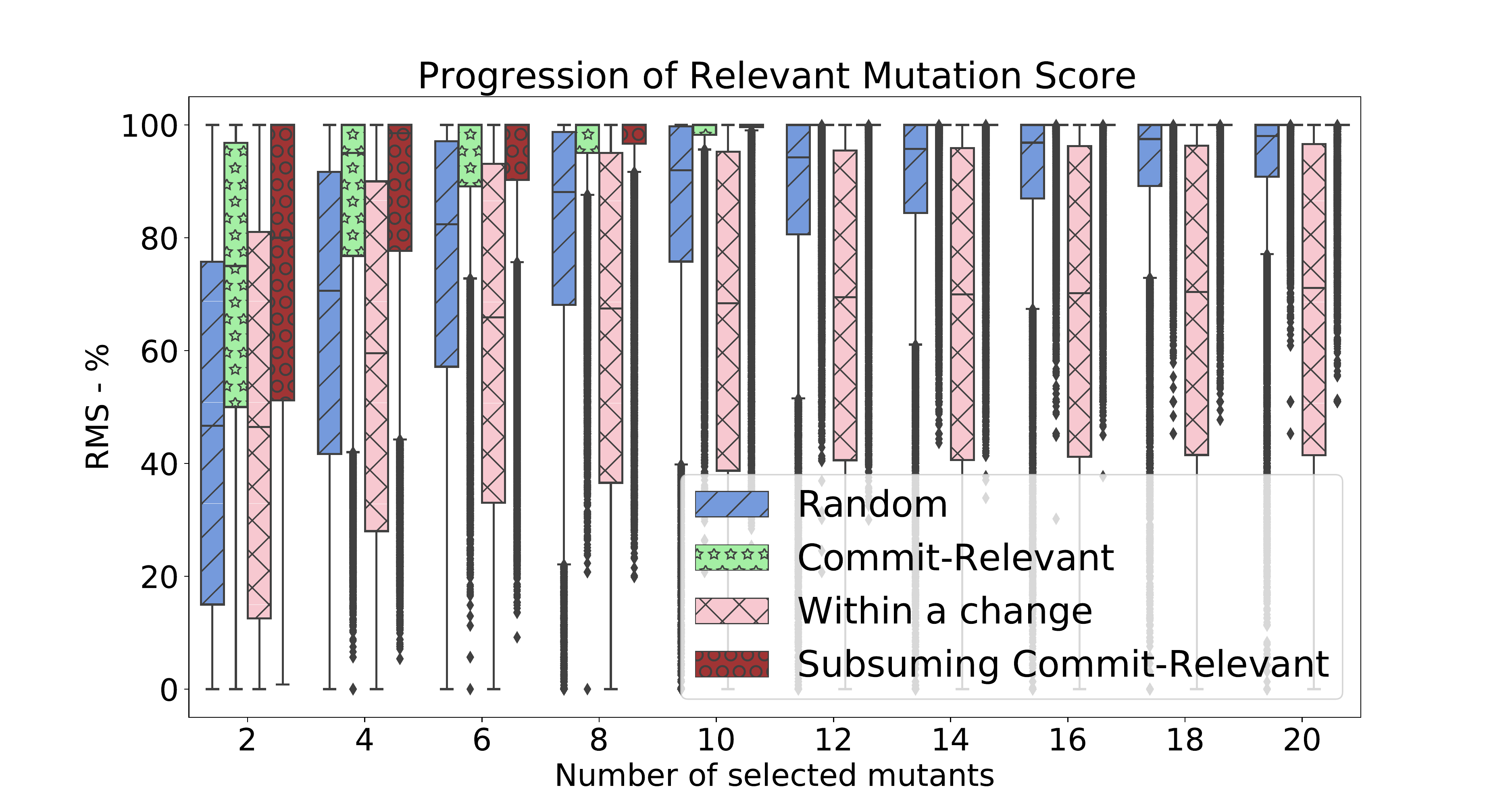}
        \caption{Relevant Mutants Progression}
        \end{subfigure}
        \begin{subfigure}[t]{0.49\textwidth}
        \centering
        \includegraphics[width=\textwidth]{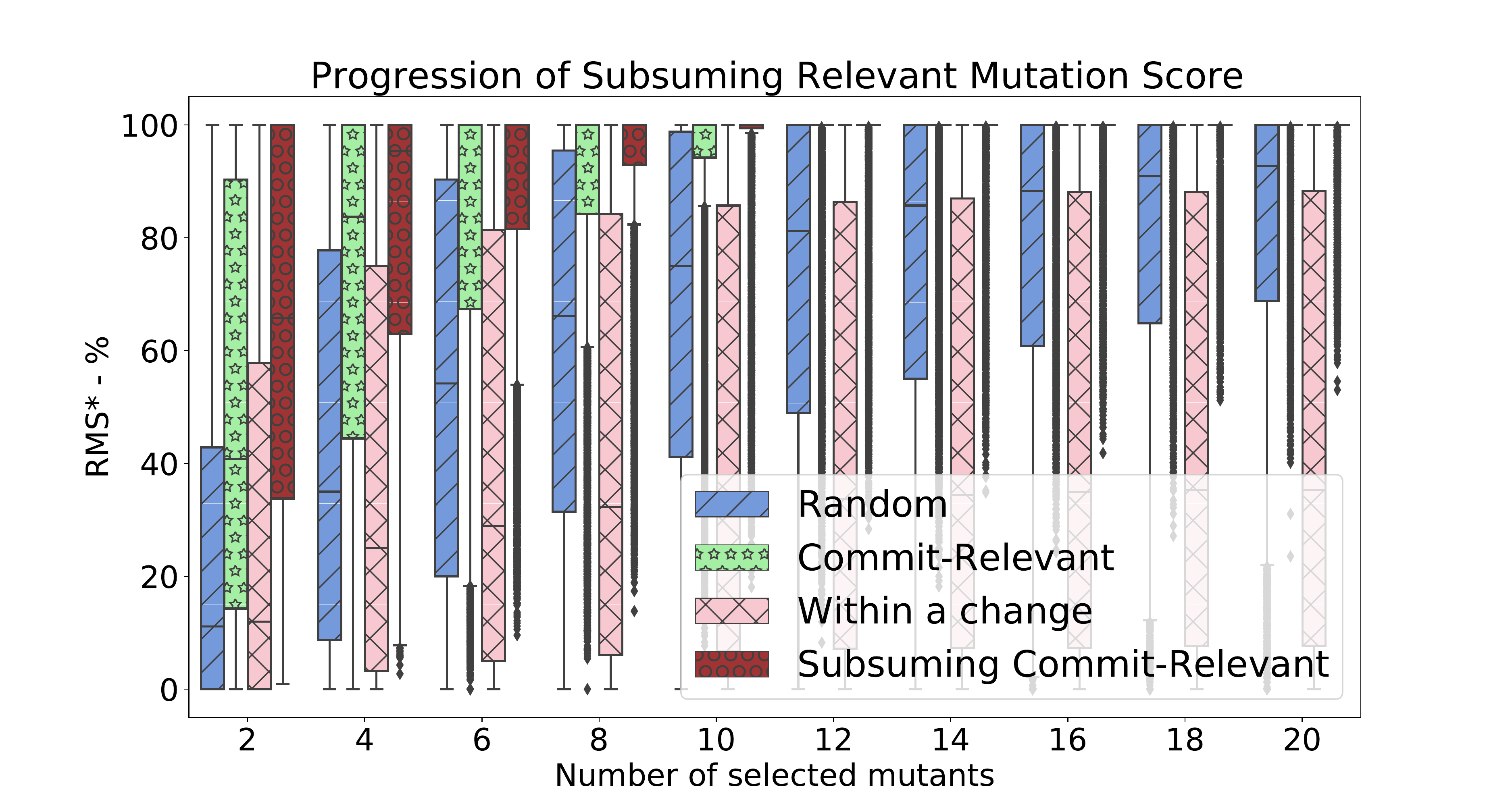}
        \caption{Subsuming Relevant Mutants Progression}
        \end{subfigure}
        \caption{
        Comparative Effectiveness of selecting and killing (subsuming) commit-relevant mutants in comparison to ``random mutants'' and ``mutants within a change''}
        \label{fig:RQ5-developer_simulation}
    \end{figure*}

\autoref{tab:RQ5-median-developer_simulation} and \autoref{fig:RQ5-developer_simulation} demonstrates how the effectiveness of the developed test suites progresses when we analyze up to 20 mutants from the different mutant pools. 
We observed that when the same number of mutants are selected from the different pools, better effectiveness is reached by test suites developed for killing (subsuming) commit-relevant mutants. 

For instance, a test suite designed to kill six (6) selected (subsuming) commit-relevant mutants will achieve 100\% of RMS and RMS*. However, a test suite designed to kill six randomly selected mutants will achieve 82.42\% RMS and 54.17\% RMS*, while a test suite that kills six mutants within a change will achieve 65.91\% RMS and 28.95\% RMS*, respectively. 
More precisely, even after selecting 20 mutants, neither random selection \revise{from all mutants} nor within a change selection achieved 100\% of RMS and RMS*. 
This result demonstrates the significant advantage achieved by selecting (subsuming) commit-relevant mutants. 

Moreover, we observed that random selection \revise{from all mutants} is up to 1.6 times more effective than selecting mutants within a change. 
For instance, selecting 20 random mutants achieves 98.05\% RMS and 92.76\% RMS*, while selecting 20 mutants within a change only achieves 71.09\% RMS and 35.29\% RMS*. 
This result demonstrates the importance of selecting mutants \emph{outside} developers' committed changes.

\begin{result}
  Selecting and killing (subsuming) commit-relevant mutants led to more effective test suites. They significantly reduced the number of mutants requiring analysis compared to random mutant selection and selecting mutants within a change.
\end{result}

\smallskip\noindent
\subsection{\RQ6: Test Executions} 
In this section, we study the \emph{efficiency} of the different mutant sets in terms of the number of \emph{test executions} \revise{required to run the tests resulting from the analysis of 2-20 mutants. 
We thus, approximate the \revise{computational} demands involved when using all mutants, relevant mutants, (subsuming) relevant mutants and mutants located within commit changes. }

\begin{figure*}[bt!]
        \centering
        \includegraphics[scale=0.3]{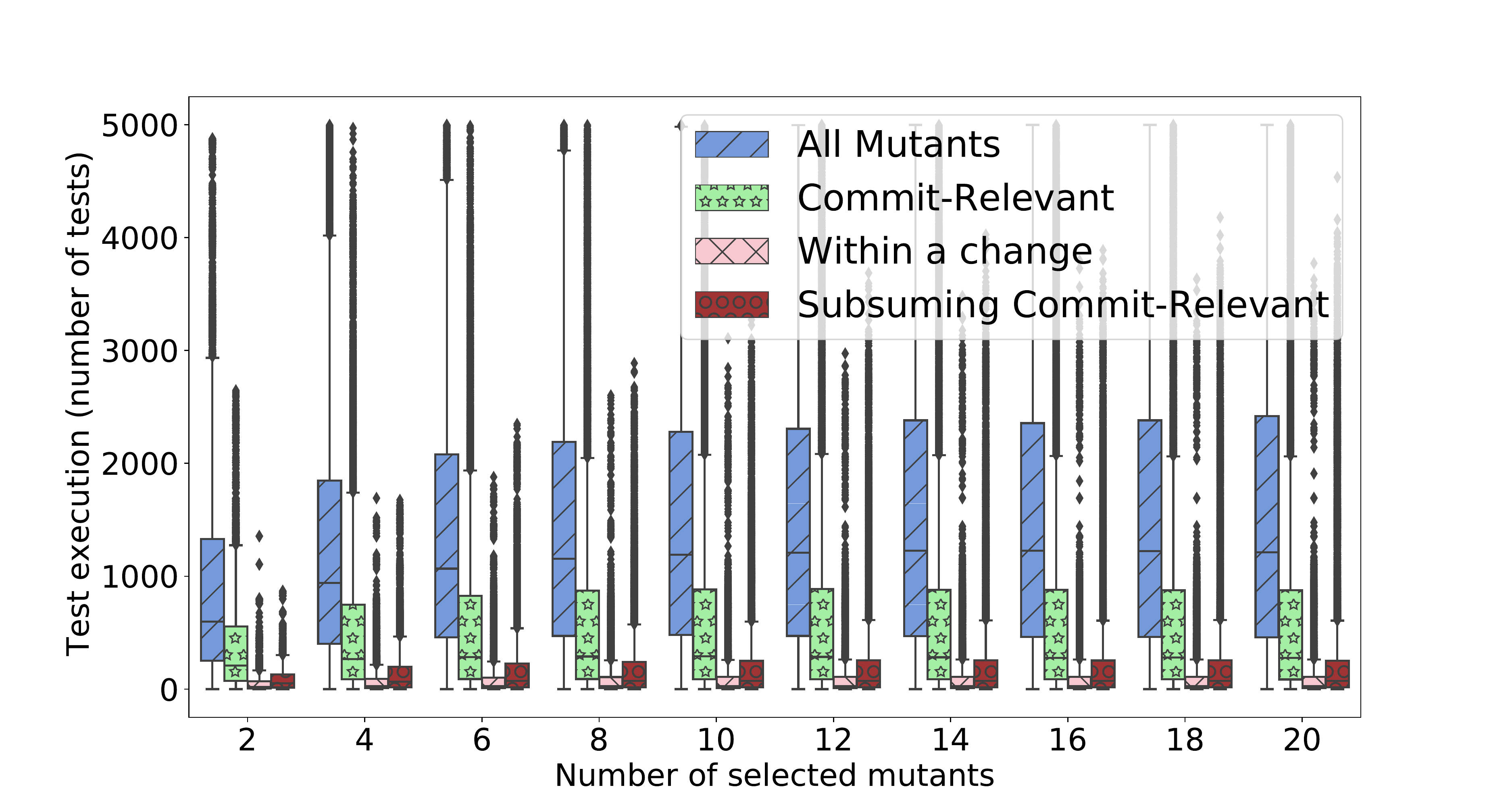}
        \caption{Efficiency, number of test executions required when deriving test suite sizes (in the range [2, 20]).}
        \label{fig:RQ6-developer_simulation}
    \end{figure*}

\revise{\autoref{fig:RQ6-developer_simulation} illustrates the number of test executions required by the test suites \revise{derived by the analysis of 2-20 mutants}.  
We found that the analysis of commit-relevant mutants significantly reduces the number of required test executions by \revise{4.28} times on average over different intervals of analysed mutants (and \revise{16} times when using subsuming commit-relevant mutants) in comparison to \revise{test execution required when analysing all mutants}. 
For instance, users will need to perform \revise{601} test executions when deriving tests based on the \revise{analysis of 2} mutants, from the set of all mutants, compared to \revise{185 or 52} test executions needed by the use of commit-relevant mutants or subsuming commit relevant mutants, respectively. }

\revise{The difference increases with the number of analysed mutants. Thus, for the 2 analysed mutants, the difference in test execution is 2.8 times. For 4 mutants, 3.65, and 6 mutants, the difference in test execution is 4 times comparing all mutants and the commit-relevant mutants.
We can also observe an increase in the difference between test executions needed by the use of subsuming commit relevant mutants and all mutants over different intervals. This difference is 11.55 times, 14.68 and 16 for analysed 2,4 and 6 mutants, respectively.
Overall, we can compare test execution needed by using commit-relevant and subsuming commit-relevant mutants and observe a 4 times difference on average, with no considerable differences between intervals. }

\begin{result}
  Selecting subsuming commit-relevant mutants reduces test execution cost \revise{(i.e., the number of test executions,}) by up to \revise{16} times compared to all mutants.
\end{result}

\section{Discussion}\label{Discussion}


\subsection{Summary of Findings}

Commit-relevant mutation testing allows developers to identify and select the mutants necessary for testing the program changes to avoid regression bugs and newly introduced failures. This paper presents an empirical study that examines the prevalence and characteristics of commit-relevant mutants and provides scientific insights concerning the mutation testing of evolving software systems. Our main empirical findings include the following:

\begin{enumerate} 

\item 
Commit-relevant mutants, at unit level, are \textit{highly prevalent} (30\%) and  most commit-relevant mutants (81\%) are \textit{located outside of program commit changes}. Hence, it is important to conduct mutation analysis of evolving systems to determine the influence of the program changes on the rest of the unmodified code.

\item 
Adequate selection of (subsuming) commit-relevant mutants significantly reduces \revise{the number of mutants involved (approximately 93\%)}; thus, there is a huge benefit to developing effective and practical techniques for the selection of (subsuming) commit-relevant mutants in evolving systems.  

\item
Predicting (subsuming) commit-relevant mutants is not a trivial task. In our evaluation, we studied several candidate \textit{proxy variables} that \emph{do not reliably predict} commit-relevant mutants, including the number of mutants within a change, mutant type, and commit size. Hence, we encourage the development of statistical or machine learning approaches and program analysis techniques to predict or identify commit-relevant mutants automatically.

\item 
Selecting commit relevant mutants is \textit{significantly more effective and efficient than random mutant selection and the analysis of only mutants within the program change}. Commit-relevant mutation testing can reduce testing effort \revise{(i.e., number of test executions)} by up to \revise{16} times, and by half, compared to random mutant selection and mutants within a change, respectively.

\end{enumerate}

Firstly, our evaluation results show that most commit-relevant mutants located outside of the commit changes due to the interaction of changes with the unmodified program code. However, in our evaluation, commit-relevant mutants that capture evolving software behavior are located all around the program changes. Besides, we observe that effective selection of commit-relevant mutants significantly reduces the number of mutants requiring analysis. Thus, we encourage researchers to investigate automated methods for identifying and selecting commit-relevant mutants, for instance, using statistical analysis or program analysis.

\begin{figure*}[bt!]
        \begin{center}
		\includegraphics[width=1.\textwidth]{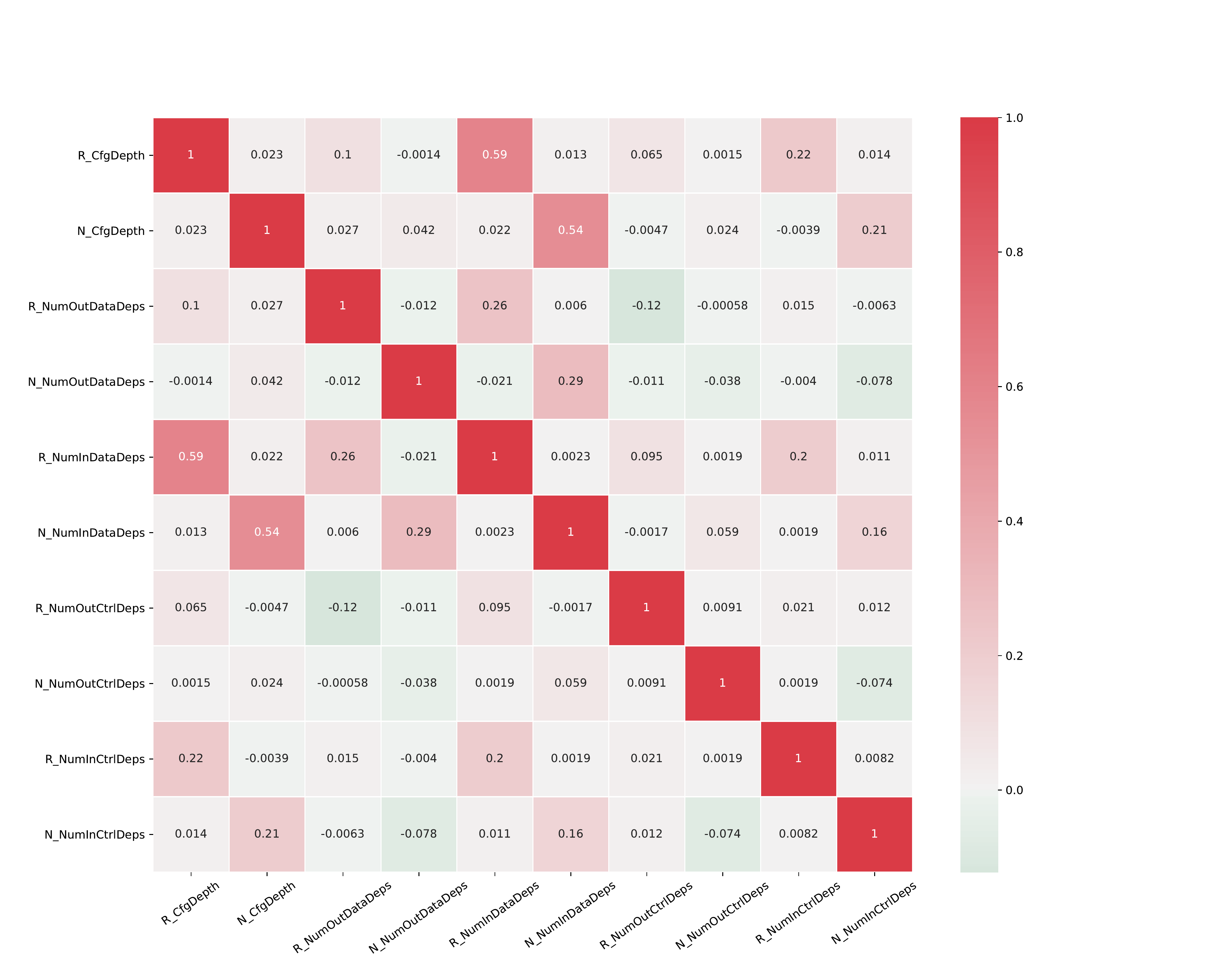}
        \caption{Correlation between features of relevant and non-relevant mutants labeled with suffixes ``R" and ``N", respectively. The features examined include the following: 
\texttt{CfgDepth} - \textit{Depth of a mutant in Control Flow Graph, i.e., the number of basic blocks to follow in order to reach the mutant}; 
\texttt{NumOutDataD} - \textit{Number of mutants on expressions data-dependent on a mutant expression}; 
\texttt{NumInDataD} - \textit{Number of mutants on expressions on which a mutant $m$ is data-dependent}; 
\texttt{NumOutCtrlD} - \textit{Number of mutants on expressions control-dependent on a mutant}; and 
\texttt{NumInCtrlD} - \textit{Number of mutants on expressions on which $m$ is control-dependent}.
        }
        \label{fig:RQ4-features_correlation}
        \end{center}
    \end{figure*}

In addition, we observed that commit-relevant prediction and selection is a challenging task. For example, many proxy variables could not reliably predict commit-relevant mutants in our analysis (\RQ2 to \RQ4). To buttress this,  we further conducted a correlation analysis of the features of commit-relevant and non-relevant mutants using control and data flow features selected from Chekam \etal~\cite{ChekamPBTS20}. 
The goal is to determine if mutants' features previously used for other prediction tasks, for instance, for selecting fault revealing mutants ~\citet{ChekamPBTS20}, can also distinguish commit-relevant mutants. 
\autoref{fig:RQ4-features_correlation} presents our findings using a heat map, where each map coordinate represents Spearman correlation coefficient calculated between two features on the coordinates. 
These features characterize relevant and not relevant mutants, labeled with suffix "R" or "N", respectively. 
Notably, we observe that there are no strong positive or negative correlations among these features. 
This implies that these features can not directly help in distinguishing between commit-relevant and non-relevant mutants. 
However, we can observe two cases of a medium positive correlation between the same class features, in particular, \textit{CfgDepth} and \textit{NumInDataD} between both classes show correlation.\footnote{\textit{CfgDepth} means the depth of a mutant in the control flow graph, i.e., the number of basic blocks to follow to reach the mutant, and \textit{NumOutDataD} refers to the number of mutants on expressions on which a mutant $m$ is data-dependent.} This phenomenon is expected since there will be more data-dependent expressions as the depth of a mutant in the control flow graph increases.

Furthermore, we found that commit-relevant mutant selection considerably improves the effectiveness and efficiency of testing evolving systems, especially in comparison to the random mutant selection, and using the mutants within the program changes (\RQ5 and \RQ6). Overall, these empirical findings shed more light on the challenge of mutation testing of evolving systems and provide directions for future research into the selection and prediction of commit-relevant mutants.

\subsection{Implications}


\revise{
The main insight of our study is 
\textit{the need to pay attention to the effective identification, selection or prioritization of commit-relevant mutants}. This is particularly important to reduce the effort required for mutation-based regression testing. Notably, an effective commit-aware mutant selection method can significantly reduce the number of mutants involved. We also shown that \textit{commit-relevant mutants are located both within and outside program changes}. Precisely, we demonstrate that beyond the committed changes, other program locations are also important for commit-aware mutation testing. Hence, it is important to identify the relevant program locations for commit-aware mutant injection. To achieve this, we encourage the use of program analysis techniques (e.g., slicing) that determines the program dependencies between changes and the rest of the program, such that mutant injection is focused on selecting such dependencies to effectively reduces the search space and cost for mutation testing. It is also pertinent to note that \textit{the subsumption relation of mutants can help in reducing considerably the effort during commit-aware mutation testing}. Indeed, it is important to prioritize subsuming mutants during mutation testing of evolving systems. }

\revise{
To achieve the aforementioned goals, i.e., automate the identification and selection of commit-relevant mutants to aid developers, 
we turn to the research community to develop and investigate the techniques required for effective commit-aware mutation testing. 
Thus, the takeaway of this study is the need to develop: a) novel techniques for \textit{selecting, prioritizing and predicting commit-relevant mutants}; and b) \textit{commit-aware test metrics} to determine the adequacy of commit-aware mutation testing. Although the problem of mutant selection/identification of the relevant mutants is active for traditional mutation testing, this is hardly well-studied for commit-aware mutation testing. 
This is an important problem since several studies~\cite{ma2020commit, ma2021mudelta} (including this study) have demonstrated that traditional (random) mutation testing 
is significantly costly for evolving software. 
}

\revise{
This paper has further illustrated that dynamic approaches (like observation slicing) can complement 
static or machine learning based 
 approaches in effectively identifying commit-relevant mutants. 
 We have also observed that commit-relevant mutants  
cannot be predicted using only the committed changes or program dependence properties. 
This implies that 
the current state-of-the-art is not generally applicable for commit-aware mutation testing in practice. Thus, for more effective approaches, we believe researchers need to consolidate the knowledge from several sources, including the commit difference, mutant properties, the semantic behavior of mutants, and the semantic divergence produced by the change. 
To this end, we encourage further investigation of the effectiveness of such techniques for commit-aware mutation testing, and the development of newer program analysis based approached (e.g., symbolic execution or search-based techniques) for identifying commit-aware mutants. 
}

\revise{
Finally, previous research~\cite{ma2021mudelta} has  shown that commit-aware mutation testing requires different test metrics from traditional mutation testing. Thus, we encourage the researchers to define new test metrics targeting the changes and their dependencies, and investigate their effectiveness for commit-aware mutation testing. Overall, we expect that addressing these challenges will reduce the performance gap between the state-of-the-art in traditional mutation testing and commit-aware mutation testing.  
}

\section{Threats to Validity}\label{ValidityThreats}

Our empirical study and findings may be limited by the following validity threats.

\smallskip \noindent  \textbf{\textit{External Validity:} }This refers to the generalizability of our findings. We have empirically evaluated the characteristics of commit-relevant mutants on a small set of open-source Java programs, test cases, and mutants. Hence, there is a threat that our experimental protocol and findings do not generalize to other mutants, programs, or programming languages. 
\revise{Additionally, there is the threat that our findings do not generalize to other Java projects, since our subject programs are all from the Apache Commons project and may share similar characteristics in terms of architecture, implementation, coding style and contributors.} 
We have mitigated these threats by conducting our experiments on five (5) matured Java programs with a varying number of tests and a considerably large number of mutants. In our experiments, we had 288 commits and 10,071,872 mutants with 25 different groups of mutant types. In addition, our subject programs have 216,489 KLOC and 17 years of maturity, on average. 
Hence, \revise{we are confident that our empirical findings hold for the tested (Java) projects, programs, commits, and mutants. } 
\revise{
    Furthermore, we encourage other researchers to replicate this study using other (Java) programs, projects and mutation tools. 
}
\revise{
    In our experiments we used  Pitest \cite{pitest} to perform our analysis. 
    However,  it is likely that the use of a different mutation tool may impact our findings, since it may contain different operators than Pitest. 
    While this is possible, recent empirical evidence \cite{KintisPPVMT18}  has shown that Pitest has one of the most complete sets of mutation operators that subsumes the operators of the most popular mutation testing tools in almost all cases. Nevertheless, we are confident on our results since Pitest includes a large sample of mutants the general results are unlikely to change with different types of simple mutations. 
}

\smallskip \noindent \textbf{ \textit{Internal Validity:}} This threat refers to the \textit{incorrectness} of our implementation and analysis, especially if we have correctly implemented/deployed our experimental tools (e.g., Evosuite, Pitest and Pitest assert), performed our experiment as described and accounted for randomness in our experiments. We mitigate the threat of incorrectness by (manually) testing our implementation, tools, and experimental protocol on few programs and commits to ensure our setup works as expected. \revise{Specifically, we performed manual testing by examining five (5) representative Apache programs containing about 500 LoC per commit on average. While we inspected in total about 20 commits with over 30 LoC in patch sizes, on average.} 
We also address the threat of randomness in our experiments by repeating our experiments 100 times to mitigate any random or stochastic effects. 


\smallskip \noindent \textbf{\textit{Construct Validity:}} This refers to the \textit{incompleteness} of our experimental approach, in terms of \textit{identifying all commit-relevant mutants}. Despite the soundness of our approach, it only provides an approximation of commit-relevant mutants, such that the set of identified commit-relevant mutants is only a subset of the total number of all commit-relevant mutants. This is due to the finite set of test cases and mutants employed in our experiments. We have mitigated this threat by ensuring we have a reasonably large set of mutants and test cases for our experiments. 
For instance, following the standards set up by previous studies \cite{KurtzAODKG16,ammann_establishing_2014,PapadakisK00TH19}, we augmented developers' written tests by automatically generating additional tests (using EvoSuite), to expand the observable input space for commit-relevant mutants. 
Our experimental findings are also threatened by the potential noise introduced by \textit{equivalent mutants}. First, notice that commit-relevant mutants come either from lines within the change or outside the change. On the one hand, considering our \autoref{algo:relevant} for identifying commit-relevant mutants outside the change, you can notice that in case that mutant $X$ is equivalent, then condition $Yval \neq XYval$ in Line 9 will evaluate to \emph{false}, since mutants $Y$ and $XY$ will be equivalent as well, then mutant $X$ will not be considered as commit-relevant. 
On the other hand, our approach selects by default all the mutants within the change as commit-relevant, so there is a potential threat in selecting some equivalent mutant, even though mutants within the change are a small fraction concerning the total number of mutants. To mitigate this threat, we employ standard methods in mutation testing to reduce the probability of generating equivalent mutants, for instance, by applying \texttt{Pitest} to ensure no common language frameworks are mutated. 

\revise{
Furthermore, our experimental approach is limited by our measure of mutant execution effort (i.e., efficiency), as well as the granularity of our test assertion checks. Firstly, in our experiments, we have estimated the efficiency of commit-aware mutation testing using the \textit{number of mutant test executions}. This measure is limited because it assumes that all tests have similar execution time (on average). 
Thus, there is a threat that our measure of efficiency may not be representative of actual execution time, especially if some mutants/tests have a longer execution time than others. Though, we argue that number of mutant test executions generalizes better than execution or CPU time because it is independent of the infrastructure, level of parallelization and test execution optimizations used. 
Consider the case of a test execution optimization that avoids issues caused by infinite loops. 
This optimization will result in significantly different execution times than if not employing them. 
Similarly, parallelization impacts the requested execution time if different strategies are used. 
Therefore, execution time measurements can be more accurate than the number of mutant executions that we use only if one uses the same infrastructure, parallelization, and mutant test execution optimizations. 
In our case, we ran our experiments in our University HPC\footnote{https://hpc.uni.lu/} with a heavy parallelization scheme. 
Therefore, we feel that its test execution results are hard to generalize to other environments. 
We also note, that there are many test execution optimizations \cite{PapadakisK00TH19, WangLXL21} that are not implemented yet by the existing mutation testing tools, fact that may reduce the generalization of our results. }

\revise{
To determine the interactions between mutants, we employ a coarse-grained assertion check in our experiments. 
Specifically, our assertion checks are at the assert parameter level. 
As an example, given a first-order mutant and a second-order mutant, we directly check the equality of the parameter values (i.e., the expected and actual outcomes) for both mutants. 
This raises the threat of missing more fine-grained assertion properties, especially
the effect of dependencies within assertions and test cases. 
Our approach may mask such dependencies, e.g., if there is a dependency between the expected and actual value within the assertion. 
Indeed, this assertion check may limit the number of observed commit-relevant mutants, as a more fine-grained approach (e.g. one that accounts for such dependencies) may reveal more commit-aware mutants. 
In the future, we plan to investigate the effect of assertion granularity on (commit-aware) mutation testing. 
Finally, we also encourage other researchers to investigate the effect of these issues (i.e., assertion granularity and test execution effort) on the performance of commit-aware mutation testing.  }

\section{Related Work}\label{related_work}

In this section, we discuss closely related work in the areas of change impact analysis, regression testing, test augmentation, and commit-aware mutation testing. 

\smallskip \noindent \textbf{\textit{Program slicing:}}
A related line of work regards the formulation of dynamic or observation-based slicing \cite{BinkleyH05, BinkleyHK07, BinkleyGHIKY14}. These techniques aim at identifying relevant program statements and not mutants. Though, they could be used in identifying relevant mutant locations, in which every located mutant could be declared as relevant. 
\revise{For instance, \citet{guimaraes2020optimizing} proposed the use of dynamic program slicing to determine the subsumption relations among mutants, in order to detect redundant mutants and reduce the number of tested mutants. In their evaluation, the authors demonstrate that using dynamic subsumption relation among mutants reduces mutation testing time by about 53\%, on average.} 
%
\revise{Similarly, \citet{delamaro2001interface} proposed interface mutation to reduce the mutation testing effort required during system integration. The goal of the paper is to apply interface mutants as a test adequacy criterion for system integration testing. The paper demonstrates that inter-procedural program slicing is applicable for mutation analysis, particularly for integration testing. Their approach leverages the data flow dependencies between two system units to determine the set of mutants that are relevant for the integration of both units.} 
 This is because many non-killable or irrelevant mutants located in dependent statements will be considered as relevant. This is evident from the previously reported results of Binkley \etal \cite {BinkleyH05, BinkleyHK07} that showing simple changed slices occupying 35-70\% of the entire programs.

\smallskip \noindent \textbf{\textit{Change-Impact Analysis:}}
It is important to analyze and test the impact of program changes on evolving software systems. To this end, researchers have proposed several automated methods to assess the impact of program changes on the quality of the software, e.g., in terms of correctness and program failures. For instance, researchers have employed \textit{program analysis} techniques (such as program slicing) to identify relevant coverage-based test requirements, specifically,  
by analyzing the impact of all control and data dependencies affected by the changed code to determine all tests that are affected by the change~\cite{RothermelH94,Binkley97}. 
Unlike these works, in this paper, we focus on performing change impact using commit-aware mutation testing, in particular, we empirically evaluate the properties, distribution and prevalence of (subsuming) commit-relevant mutants.

\smallskip \noindent \textbf{\textit{Regression Testing:}}
The field of regression testing investigates how to automatically generate test cases for evolving software systems to avoid regression bugs. Researchers have proposed several approaches in this field for decades~\cite{YooH12}. The closest work to ours is in the area of regression mutation testing~\cite{ZhangMZK12} and predictive mutation testing~\cite{0050ZHH0019, MaoCZ19}. The goal of regression mutation testing is to identify change-affected mutants (i.e., mutants on execution trace affected by changes), and incrementally calculate mutation score. Meanwhile, predictive mutation testing seeks to estimate the mutation score without mutant execution using machine learning classification models trained on different static features~\cite{MaoCZ19}. These approaches are focused on speeding up test execution and mutation score computation while testing evolving software systems. 
In contrast, in this paper, we focus on identifying the test requirements (mutants) relevant to the program changes. i.e., the mutants that need to be analyzed and tested by the engineer, and we provide a more refined and precise mutation testing score, specially adapted for commit-relevant mutants. 

\smallskip \noindent \textbf{\textit{Test Augmentation:}}
This line of research aims to automatically generate additional test cases to improve the (fault revealing) quality of existing test suites. This is particularly important when a software system changes (often); hence, it is vital to generate new tests that exercise the program changes. Researchers have proposed several test augmentation approaches to trigger program output differences~\cite{QiRL10}, increase coverage~\cite{XuKKRC10} and increase mutation score~\cite{SmithW09JSS,SmithW09EMSE}. Some test augmentation approaches have been developed to address code coverage problems using propagation-based techniques~\cite{ApiwattanapongSCOH06,SantelicesCAOH08,SantelicesH10,SantelicesH11}.  Other approaches employ symbolic execution for test augmentation by generating tests that exercise the semantic differences between program versions by incrementally searching the program path space from the changed locations and onwards, this includes approaches such as differential symbolic execution \cite{PersonDEP08}, KATCH \cite{MarinescuC13} and Shadow symbolic execution \cite{KuchtaPC18}.  These techniques rely on dependency analysis and symbolic execution to decide whether changes can propagate to a user-defined distance by following the predefined propagation conditions. Hence they are considerably complex and computationally expensive, for instance, because they are limited by the state explosion problem of symbolic execution. These papers are complementary to our work since the aim is to generate additional tests to improve existing test suites. However, our work is focused on test augmentation to exercise code changes, albeit using mutation testing.

%
%

\smallskip \noindent \textbf{\textit{Commit-Aware Mutation Testing:}} 
The goal of commit-aware mutation testing is to select mutants that exercise program changes in evolving software systems. The problem of commit-relevant test requirements has not been studied in depth by the mutation testing literature \cite{PapadakisK00TH19}. The closest work to ours in this area includes the formalization of commit-aware mutation testing~\cite{ma2020commit}, 
diff-based commit-aware mutant selection (i.e., mutants within program changes only)~\cite{petrovic2018}, and a machine learning-based mutant selection method~\cite{ma2021mudelta}. Petrovic \etal~\cite{petrovic2018} presents a diff-based probabilistic mutation testing approach that is focused on selecting mutants within committed program changes only. Unlike this paper, this approach ignores the dependencies between program changes and the unmodified code. \texttt{Mudelta}~\cite{ma2021mudelta} presents a machine learning-based approach for selecting commit-relevant mutants.  Ma \etal~\cite{ma2020commit} defines commit-relevant mutants and evaluates their relationship with traditional mutation test criteria, emphasizing the importance of commit-aware mutation testing. Unlike these works, we conducted an in-depth empirical study to understand the characteristics of commit-relevant mutants to shed more light on their properties and provide scientific insights for future research in commit-aware mutation testing. In particular, in comparison to Ma \etal~\cite{ma2020commit}, our work impose more generalizable and easy to fulfill requirements on the programs and test contracts. For instance, Ma \etal~\cite{ma2020commit}, determines commit-aware mutants by comparing mutants test suites from both pre- and post-commits, under the assumption that the test contract is intact and remains the same across both program versions. 
However, this critical requirement considerably limits the application and adoption of the approach of Ma \etal~\cite{ma2020commit} in practice. Our empirical study observed that test contracts frequently change across versions in evolving software systems (>60\%). 

\section{Conclusion}\label{sec:conclusion}
We presented an empirical evaluation of the characteristics of commit-relevant mutants. In particular, we have studied the prevalence, location, effectiveness, and efficiency of commit-relevant mutants. We have also examined the comparative advantage of commit-relevant mutants compared to two baseline methods, i.e., random mutant selection and selecting mutants within program changes. Notably, we found that commit-relevant mutants are \textit{highly prevalent} (30\%), and primarily located outside of program changes (81\%). 
 In addition, we observed that effective selection of commit-relevant mutants affords a significant testing advantage. Specifically, it has the potential of significantly reducing the cost of mutation, and it is significantly more effective and efficient than random mutant selection and analysis of only mutants within the program change. 
We also investigate the predictability of commit-relevant mutants by considering typical proxy variables (such as the number of mutants within a change, mutant type, and commit size) that may correlate with commit-relevant mutants. However, our empirical findings show that these candidate proxy features do not reliably predict commit-relevant mutants, indicating that more research is required to develop tools that successfully detect this kind of mutants. 

\bibliographystyle{ACM-Reference-Format}
\bibliography{bibfile}

\end{document}